\documentclass[10pt, a4paper, onecolumn, twoside]{article}

\usepackage{graphicx, xcolor}

\usepackage[top=35.2truemm,bottom=35.2truemm,left=34.5truemm,right=34.5truemm]{geometry}

\title{\bf \Large 
Relaxing in Warped Spaces: \\ 
Generalized Hierarchical and Modular \\ 
Dynamical Neural Network  
}

\author{
Kazuyoshi Tsutsumi
\thanks{
Ryukoku University, 
67 Fukakusa-Tsukamoto-Cho, Fushimi-Ku
Kyoto-Shi, Kyoto-Fu 612-8577, JAPAN}
\and 
Ernst Niebur
\thanks{
The Zanvyl Krieger Mind/Brain Institute, 
Johns Hopkins University, 
3400 N. Charles Street, Baltimore, MD 21218, USA} 
}

\date{}

\begin{document}

\maketitle 
\thispagestyle{empty}

\begin{abstract} 
\noindent
We propose a dynamical neural network model 
with a hierarchical and modular structure. 
The network architecture can be derived by minimizing 
an energy function that is originally designed 
based on two kinds of neurons with quite different time constants. 
It has multiple subspaces 
that are spanned by neural parameters 
employed in the energy function, 
and adjacent subspaces are related to each other 
by a layered ``internetwork" with modifiable connections. 
Each internetwork further consists of 
a pair of a forward subnet and a backward one, 
and signals flowing through these subnets 
determine total dynamics of the network. 
The model can operate in either a learning mode or an association mode. 
In the learning mode, when periodic signals 
equivalent to repetitive neuronal bursting 
are adequately applied to input ports existing in all subspaces, 
mapping relationships corresponding to those input signals 
are eventually formed in internetworks between subspaces. 
If the periodic input signals applied to the subspaces are deformed, 
various non-linear mapping relationships can be obtained in the internetworks. 
Two-dimensional mapping relationships between subspaces can be shaped 
by employing an appropriate set of periodic input signals 
with different frequencies 
based on the same mechanism as a Lissajous curve. 
Due to the cooperation of the two kinds of neurons, 
the model in the association mode provides an overall framework 
such that state variables inside the network individually 
relax in warped spaces, each of which has been designed 
as favorable for a (or some) state variable(s). 
The association mode is further classified 
into two modes; unconstrained and constrained. 
In the former mode, a ``goal" point is applied 
to an input port located in a subspace, 
and then the convergence speed of the model's output toward the goal point 
varies depending on how the mapping relationships 
in the internetworks are deformed. 
Therefore, in the two-dimensional model, 
output trajectories are not necessarily straight but can curve, 
since the convergence speed on each axis changes 
in accordance with the degree of space warping in the internetworks. 
In the latter mode,  
a time-varying ``trajectory," rather than a fixed goal point, 
is given to an input port in a subspace. 
For instance, 
when a fast (high speed) periodic trajectory, 
relative to the time constant of a dynamical neuron, 
is applied to an input port in the outermost subspace, 
a periodic output trajectory with only a small radius 
is generated in any subspace. 
When a sufficiently slow (low speed) periodic trajectory is set as an input, 
however, a warped output trajectory appears in each subspace 
as if ``imaginary" layered networks 
with the ``inverse" mapping relationships to existing forward subnets' 
were located hierarchically from outside to inside. 
These results suggest that a certainty/uncertainty relation exists 
between ``the speed of an input trajectory" applied from the outside 
and ``the quality of an output trajectory" generated in a subspace. 

\vspace{2.00mm}
\end{abstract}

\markboth{\rm Relaxing in Warped Spaces}{\rm Kazuyoshi Tsutsumi \& Ernst Niebur}

\section{Introduction}

A brain composed of huge number of neurons 
is definitely not homogeneous but does have heterogeneous structures 
between the neuron as a component and the brain as a whole. 
In order to understand the brain functioning 
and construct its mathematical model, 
it is significant not to overlook 
microscopic through macroscopic aspects 
that strikingly emerge owing to such heterogeneous structures. 
Various fundamental principles and models 
for the brain collectively and the neural circuits as subsystems 
have been proposed so far. 
If they have their own plausibility, 
it is then crucial to seek the upper frameworks 
into which those primal principles are integrated.

In terms of mathematically understanding a brain 
that is an aggregation of neurons, 
two important components are 
``input-output characteristics of the neuron itself" 
and ``functioning of a synapse as a connection between the neurons," 
both of which are based on electrophysiological variation. 
With regard to the former, 
the first mathematical model of the neuron 
is the one proposed by W. S. McCulloch and W. Pitts 
from a historical perspective
\cite{McCulloch1943}. 
The output was determined 
by applying a threshold function to 
the product-sum of ``input signals" and ``connections for weighting" 
on the simple assumption of no time delay 
in the input-output relation. 
As for the latter, an idea for synaptic plasticity 
suggested by D. O. Hebb is the first one
\cite{Hebb1949}. 
After that, a series of learning machines and layered neural networks 
composed of such neuronal and synaptic models was proposed, 
and separability of input patterns was vigorously discussed
\cite{Rosenblatt1958}\cite{Widrow1960}\cite{Steinbuch1963}. 
Those learning machines and neural network models 
at the dawn of a new era were thought to be very promising. 
However, it was pointed out thereafter 
that they could not recognize and/or classify pixel patterns 
which were not linearly separable, 
and further developments had been long-awaited
\cite{Minsky1969}. 
In addition to those situations, 
research for symbolic AI (Artificial Intelligence) 
started to become active in the 1970s. 
Consequently, a quiet period in neural network research 
followed from the late 1960s to the early 1980s, 
although proposed even during this era were 
various important ideas and models 
that influenced later studies
\cite{Amari1967}\cite{Kohonen1972}\cite{Nakano1972}
\cite{Marlsburg1973}\cite{Werbos1974}
\cite{Grossberg1976}\cite{Amari1977}. 
In the early research stage for learning machines, 
a piece-wise linear function was often employed 
as the input-output characteristics for the conversion 
from membrane potential to nerve impulse density
\cite{Block1962}. 
In the middle of the 1980s, 
presented with many suggestive simulation studies 
was a new learning algorithm named the generalized delta rule, 
in which a linear learning algorithm proposed in the early days 
based on a gradient descent method was expanded and united 
with a continuous and differentiable non-linear function 
for the neuron's input-output property; 
this proposal succeeded in overcoming 
the limitation of linear separability after a couple of decades
\cite{Rumelhart1986a}. 
In addition to the improvement 
in logical relationship between input and output 
to be acquired through learning process, 
it was clarified that a layered neural network 
composed of neurons with a non-linear input-output property 
described by such a sigmoid function 
had excellent interpolation and generalization capabilities
\cite{Rumelhart1986b}. 
Owing to those characteristics, 
great expectations had risen in the engineering field
\cite{Hect-Nielsen1988}\cite{Izui1990}.

In the vision system, well-known is a pioneering work 
in which the existence of a feature extraction mechanism 
with the highly hierarchical structure was revealed 
based on precise electrophysiological experiments
\cite{Hubel1959}. 
Inspired by this study, 
some mathematical models of layered neural networks 
with not full but limited connections between layers 
have been proposed so far 
along with their excellent processing capability 
of visual information
\cite{Fukushima1975}\cite{Fukushima1980}
\cite{Fukushima1983}. 
A layered neural network with non-linear hidden units 
has a great ability for static mapping 
even if it has only one hidden layer
\cite{Funahashi1989}. 
That with multiple non-linear hidden layers is expected 
to have even better mapping characteristics in principle 
due to the effectiveness of 
feature extraction and information integration 
as can be seen in biological neural networks. 
The generalized delta rule can practically be applied 
to a layered neural network 
with an arbitrary number of non-linear hidden layers. 
As the number of hidden layers increases, however, 
the values of back-propagated errors sometimes 
become too small to modify synaptic connections near the input layer 
depending on the mapping relationship to be gained for the network; 
under such circumstances, 
the synaptic modification (i.e., the learning) 
does not proceed successfully.
\footnote{The experimental results in Sections 3 \& 4 
of \cite{Tsutsumi2022} are useful.} 
In recent years, it is clarified that ingenious and smooth learning 
in a super multi-layered static neural network comes to be possible 
through the introduction of limited connections between layers
\cite{Hinton2006a}\cite{Hinton2006b}. 
Such type of neural networks with excellent performance 
as never before gathers attention again 
from a viewpoint of various technological applications
\cite{Markoff2012}.

With regard to discussion about those learning machines, 
modifications of connections between neurons 
have been treated as learning dynamics since the early stages, 
even if differential equations were not explicitly shown 
but only simple difference equations 
based on a first-order approximation were given. 
On the other hand, 
as for input-output characteristics of a neural network, 
the acquisition of a non-linear mapping 
and the generalization capability 
were mainly focused on, 
and dynamics for neuron's activity had not been considered 
in most cases. 
A mathematical approach to dynamical characteristics of a neuron 
was originated from an ionic mechanisms' model 
of the squid giant axon 
employing a set of non-linear differential equations
\cite{Hodgkin1952}. 
The research on simplified models 
extracting the essence of the characteristics 
and more detailed models based on clinical trials 
has continued since then
\cite{FitzHugh1955}\cite{Nagumo1962}
\cite{Niebur1988}\cite{Candel2012}. 
In a real neuron, a minimum unit of signal information passing through an axon 
is a nerve impulse, and its density varies moment by moment. 
In this meaning, the density of nerve impulses should play an important role 
also when considering neural activities as a network, 
and it is of validity to expect that 
the information is expressed using the density of nerve impulses. 
Repetitive bursts of nerve impulses 
that are generally observed in a biological neural network, 
i.e., sinusoidal or quasi-sinusoidal signals 
in the short-term average density of nerve impulses, 
should further be taken into account.

In principle, neural network's behavior can be regarded 
as a dynamical system 
because it is based on time-varying electrophysiological process. 
Possible is the direction in which 
we construct a complex network 
combining strict and precise neuron models. 
On the contrary, when we discuss large-scaled neural systems, 
it is reasonable to simplify a neuron model 
as long as it does not lose its essence, 
depending on what part of the dynamical behavior 
we pay attention to. 
It is a quite important choice to determine 
how precisely and faithfully a biological neuron 
should be reproduced as a model 
when studying dynamical neural networks
\cite{Millman2010}\cite{Plenz2014}. 
When only a first-order time delay between input and output 
is implemented into the above-mentioned simple neuron model 
mainly used in static learning machines, 
a feedback network with a lot of such neuron models 
becomes a higher-order system 
described by a set of non-linear differential equations; 
as a result, it offers rich dynamical features. 
In fact, complex neural dynamics including oscillatory and chaotic phenomena 
are being studied employing those simple types of dynamical neuron models
\cite{Morishita1972}\cite{Tsutsumi1984a}\cite{Tsutsumi1984b}
\cite{Matsuoka1985}\cite{Tsutsumi1988}
\cite{Aihara1999}\cite{Tsujita2003}. 
From the perspective of acquiring a mapping relationship 
based on learning, 
one of the expectations to such dynamical neural networks 
is the expansion from static mapping to dynamic one. 
Lively discussed are various learning methods 
how to obtain a set of connections for a network 
that can generate an arbitrary output trajectory
\cite{Williams1990}\cite{Williams1995}.

The framework for an energy function 
introduced by J. J. Hopfield is essential 
for research on neural networks
\cite{Hopfield1982}\cite{Hopfield1984}\cite{Carpenter1987}. 
This is available as a design technique 
for dynamical neural networks. 
Based on activity dynamics of the derived network 
in which the values of feedback connections 
were preliminarily determined from multiple constraints 
necessary for problem solving, 
we can search a convergence point satisfying all of those constraints
\cite{Hopfield1985}\cite{Tank1986}. 
Various real-world applications are being explored, 
and in parallel with such approaches, 
their network dynamics are being studied in detail
\cite{Tsutsumi1987}\cite{Tsutsumi1988b}\cite{Tsutsumi1988c}. 
There exists some rationality in network architecture 
derived from an evaluation function 
in which each term has a dimension of energy, 
and the existence of such an upper criterion as the energy function 
seems to have validity. 
Symmetry (or bidirectionality) in connectivity is needed 
as a necessary condition for deriving a network; 
taking the relation between an upper criterion and a derived network 
(e.g., the relation between power consumption 
and conductance (or resistance) in electric circuitry) 
into consideration, 
such symmetric (or bidirectional) connections 
should be natural and essential. 
In a series of the Bidirectional Associative Memory (BAM) models, 
``activity dynamics based on dynamical neurons" 
and ``learning dynamics assuming Hebb type synapses" 
are derived from one energy function; 
this means that the derivation of 
``(bidirectional) connections necessary for solving a problem" 
in and of itself is subject to learning
\cite{Kosko1987}\cite{Kosko1988}. 
It is interesting that the model has an architecture 
in which two ensembles of dynamical neurons 
are coupled by bidirectional connections. 
In the early stage of neural network research, 
``activity dynamics of a network with fixed feedback connections" 
and ``learning dynamics for modifying interconnections between neuron models" 
had been separately discussed. 
Therefore, the integration or the fusion of 
activity dynamics and learning dynamics 
that are put forward in those models 
seems to be a major progress in neural network exploration. 
In the framework proposed by one of the authors, 
``activity dynamics in a modular network 
composed of neural elements with different time constants" 
and ``learning dynamics based on the generalized delta rule" 
are integrated
\cite{Tsutsumi1990}\cite{Tsutsumi1991}\cite{Ozawa1998}. 
Derived from one energy function (one evaluation function) 
is an architecture in which 
ensembles of dynamical neurons are mutually connected 
not directly but via layered networks 
composed of small neurons with tiny time constants 
(regarded as static ones). 
It is further crucial that 
a higher mechanism in neural processing is discussed there 
on the basis of the mutual interaction between modules 
due to the cooperation of neurons with different time constants.

In the cerebellum, 
existence of the spread-wise modular structure 
has been known for a long time
\cite{Szentagothai1968}. 
With respect to the cerebrum, 
the depth-wise modular structure has precisely been revealed 
owing to state-of-the-art inspection technologies 
\cite{Meunier2010}\cite{Nicolini2016}. 
Although the sizes of modules and the relations 
between modules seem to be multifarious, 
modularity in neural architecture can fundamentally 
be classified into spread-wise (horizontal) one 
and depth-wise (vertical or hierarchical) one, 
and both of them must be essential
\cite{Tsutsumi1989}\cite{Tsutsumi2003}. 
As well as local bidirectional connections between adjacent modules, 
we should pay attention to global/detoured bidirectional routes 
such as centrifugal and centripetal paths
\cite{Fukushima1986}
\cite{Tsutsumi1999a}\cite{Tsutsumi1999b}\cite{Tsutsumi2001}
\cite{Fukushima2013}. 
Thus, it comes to be a vital viewpoint in brain modeling
to consider in detail how neural modules are formed 
(i.e., modularity, hierarchy of modules, 
and bidirectionality between modules, etc.). 
On what principle and through what process 
are those particular structures built in the brain?
These points are quite important in terms of modeling, 
and it seems to be reasonable to expect that 
a higher level criterion such as 
an evaluation function or an energy function
creates those structures.

Because of this situation, 
we proposed, in a previous paper, 
a dynamical neural network model 
that incorporated the above-mentioned frameworks 
such as nerve impulse's periodic bursts, 
integration of activity and learning dynamics, 
modular structure, and hierarchy of the modules
\cite{Tsutsumi2022}. 
The network architecture was not constructed 
in a building block manner, 
but it was able to be derived in a mathematical way 
from a specifically designed energy function 
composed of plural neuronal parameters. 
The obtained network had the internal and external spaces 
that were related to each other 
by a pair of ``{\sl Forward Subnet}" and ``{\sl Backward Subnet}," 
i.e., ``{\sl Internetwork}" collectively. 
The model could operate in either a Learning Mode or an Association Mode. 
In the Learning Mode, when sinusoidal or quasi-sinusoidal signals 
(equivalent to repetitive neuronal bursting or semi-bursting 
in the nerve impulse level) 
were adequately applied to all input ports, 
the {\sl Internetwork} was able to acquire 
a linear or non-linear mapping relationship between both spaces. 
Various two-dimensional mapping relationships could also be obtained 
by employing sinusoidal or quasi-sinusoidal signals 
with different frequencies 
according to the same mechanism as a well-known Lissajous curve. 
The Association Mode offered an entire framework 
in which some of various constraints 
represented by synaptic connections in the model 
were satisfied/minimized in the internal warped space 
that was designed as favorable 
for a (or some) network parameter(s). 
In a particular Association Mode 
named the Constrained Association Mode 
in which not a goal point but a trajectory 
was applied to a unique input port, 
if the input's speed was very low, 
it was revealed that 
a warped trajectory was generated in the internal space 
in such a manner that an ``imaginary" mapper 
with the ``inverse" mapping relationship 
to the real {\sl Forward Subnet}'s 
existed in the direction from outside to inside.

However, there were some points 
that we could not refer in our basic model, 
where the number of {\sl Internetworks} was limited to one, 
and the number of spaces (or {\sl Subspaces}) 
spanned by state variables inside the model 
were only two; internal and external. 
In this paper, we propose the generalized model 
that is able to have 
arbitrary number of {\sl Internetworks} and {\sl Subspaces}, 
and examine its dynamical behavior in detail 
regarding what changes and effects are produced 
due to the expansion. 
In Section 2, we specifically derive the model's architecture 
to which our basic model is extended 
so that there is no upper bound 
on the number of {\sl Internetworks} and {\sl Subspaces}. 
It is predicted that signals flowing through the network 
congestedly interfere with each other 
because of complicated signaling routes. 
In Section 3, taking notice of that point in particular, 
we explain, by means of detailed simulation studies, 
how the {\sl Internetworks} inside the model can be trained. 
In the following section, 
with respect to the Unconstrained Association Mode 
in which a fixed goal point is given 
to an input port in a {\sl Subspace}, 
we describe the primary dynamics with some simulation results. 
In Section 5, on the basis of precise simulation studies, 
we illustrate the dynamics for the Constrained Association Mode 
in which periodic target signals 
with different frequencies and amplitudes 
are applied to a unique input port. 
In the last section, 
we summarize the main points of the entire and state our future work.

\section{Model}

\subsection{Modeling Concept}

\vspace{1.00mm} 
\noindent 
We put \( U_{i} \) and \( V^{<0>}_{i} \) 
``the inner state (the membrane potential) 
of the \( i \)-th dynamical neuron" 
and 
``its output (the short-term average density 
of nerve impulses)" respectively. 
We then suppose that both are functions of time 
and related to each other with the input-output function given by Eq. (1):

\vspace{1.00mm}
\begin{equation}
V^{<0>}_{i} = g(~U_{i}~) ~.
\hspace*{20.00mm}
(~ i = 0, 1, 2, ... ~) 
\hspace*{-10.00mm}
\end{equation}

\vspace{2.00mm}
\noindent
Based on this fundamental relation, 
we construct a dynamical neural network described by Eq. (2) 
that is according to Kirchhoff's Current Low (KCL):

\vspace{2.00mm} 
\begin{equation}
- ~c_{i} \frac{dU_{i}}{dt} ~=~ 
~\sum_{j} T^{<0>}_{ij} V^{<0>}_{j} ~- ~\sum_{k} S^{<0>}_{ik} Z^{<0>}_{k} 
~- ~J_{i} ~+ ~\frac{U_{i}}{r_{i}} ~.
\end{equation}

\vspace{-4.00mm} 
\[
\hspace*{56.00mm}
(~ i = 0, 1, 2, ... ~) 
\]

\vspace{2.00mm} 
\noindent
Here, \( T^{<0>}_{ij} \) indicates the direct fixed-valued connection 
(conductance) with symmetry 
from the \( j \)-th dynamical neuron to the \( i \)-th one. 
Due to this, the output of a dynamical neuron 
is converted from voltage to current 
and fed back to other dynamical neurons and/or itself. 
\( Z^{<0>}_{k} \) means the input voltage 
(the short-term average density of nerve impulses) 
applied from the outside of the model, 
and it is added to the time-constant circuit consisting of dynamical neurons 
after being converted from voltage to current 
via the fixed-valued connection (conductance) \( S^{<0>}_{ik} \). 
\( J_{i} \) is the bias current injected from the outside. 
\( c_{i} \) and \( r_{i} \) are respectively capacitance and resistance, 
and they determine the dynamical response of the \( i \)-th neuron. 
Note that we redundantly use such indices 
as \( i \), \( j \), and \( k \) 
without their maximum values to reduce the number of indices.

By changing each term's sign in Eq. (2), 
we can control the model's basic dynamical behavior 
into various forms including whether it is convergent or divergent. 
From the viewpoint of circuit designing, 
we can utilize an energy function as stated below. 
Now let us consider the following function 
in which each term has a dimension of electric power:

\begin{eqnarray} 
E & \stackrel{\triangle}{=} & 
\frac{1}{2} \sum_{i} V^{<0>}_{i} ~( \sum_{j} T^{<0>}_{ij} V^{<0>}_{j} )
~- ~\sum_{i} V^{<0>}_{i} ~( \sum_{k} S^{<0>}_{ik} Z^{<0>}_{k} )
\nonumber \\ 
  &                         & 
\hspace*{10.00mm}
~- ~\sum_{i} V^{<0>}_{i} J_{i} 
~+ ~\sum_{i} \frac{1}{r_{i}} \int_{0}^{V^{<0>}_{i}} \hspace*{-3.00mm} g^{-1}(V) dV ~.
\end{eqnarray}

\vspace{1.00mm} 
\noindent 
By differentiating this energy function with respect to time, 
we obtain the following equation 
on the assumption that the connection \( T^{<0>}_{ij} \) is symmetric, 
i.e., \( T^{<0>}_{ij} = T^{<0>}_{ji} \):

\begin{equation} 
\frac{dE}{dt} ~=~ \sum_{i} \frac{dV^{<0>}_{i}}{dt} 
~\mbox{\Large [} 
~\sum_{j} T^{<0>}_{ij} V^{<0>}_{j} ~- \sum_{k} S^{<0>}_{ik} Z^{<0>}_{k} 
~- J_{i} ~+ \frac{U_{i}}{r_{i}} 
~\mbox{\Large ]} ~.
\end{equation}

\vspace{1.00mm} 
\noindent 
The terms inside the brackets in Eq. (4) 
are the same as the ones of the right-hand side in Eq. (2).
So substituting Eq. (2) for Eq. (4), 
Eq. (4) can be represented as Eq. (5) through the use of the chain rule:

\begin{eqnarray} 
\frac{dE}{dt} & = & \sum_{i} ~\mbox{\Large (} 
                    \frac{\partial V^{<0>}_{i}}{\partial U_{i}}
                    ~\frac{dU_{i}}{dt} \mbox{\Large )} 
                    ~\mbox{\Large (} - c_{i} \frac{dU_{i}}{dt} \mbox{\Large )} 
                    \nonumber \\ 
              & = & \sum_{i} ~\mbox{\Large (} 
                    \frac{\partial }{\partial U_{i}} ~g(~U_{i}~)
                    \mbox{\Large )} ~\frac{dU_{i}}{dt} 
                    ~\mbox{\Large (} - c_{i} \frac{dU_{i}}{dt} \mbox{\Large )} 
                    \nonumber \\  
              & = & - ~\sum_{i} ~c_{i} ~g^{\prime} (U_{i}) 
                    ~\mbox{\Large (} \frac{dU_{i}}{dt} \mbox{\Large )}^{2} 
                    ~.
\end{eqnarray}

\vspace{1.00mm} 
\noindent 
For instance, the following monotonously increasing function 
can be assumed as \( g(x) \) in Eq. (1);

\begin{equation} 
g(x) ~=~ \alpha x
\hspace{22.20mm} (~ \alpha ~> ~0 ~: ~\mbox{Const.} ~) 
\end{equation}

\vspace{-0.50mm} 
\noindent 
or

\begin{equation} 
g(x) ~=~ \frac{1}{1 + e^{- \alpha x}}
\hspace{15.00mm} (~ \alpha ~> ~0 ~: ~\mbox{Const.} ~)~. 
\end{equation}

\vspace{3.00mm} 
\noindent 
If we employ Eq. (6) or (7) as \( g(x) \), 
the following inequality is always established 
by taking account of \( g'(~U_{i}~) > 0 \) in Eq. (5):

\begin{eqnarray} 
\frac{dE}{dt} & \leq & 0 ~.~
\end{eqnarray}

\vspace{2.00mm} 
\noindent 
In the network described by Eq. (2), it is then assured 
that the value of the energy function given by Eq. (3) 
continues to be minimized toward a global or local minimum. 
The symbol \( < \hspace{-0.5mm} 0 \hspace{-0.5mm} > \) 
placed on the upper right of \( T \), \( V \), \( S \), and \( Z \) 
means that the corresponding variable is located 
in {\sl Subspace \#0} (i.e., the \( 0 \)-th {\sl Subspace}) 
having dynamical neurons.

Now let us suppose that 
the output variables, \( V^{<0>}_{0}, V^{<0>}_{1}, V^{<0>}_{2}, ... \), 
in {\sl Subspace \#0} (i.e., the innermost {\sl Subspace}) 
are converted in turn to the output variables 
\( V^{<p>}_{i_{p}} \) in {\sl Subspace \#\( p \)} 
(i.e., the \( p \)-th {\sl Subspace} 
counting from the right-hand neighbor of the innermost one) 
based on Eq. (9), 
where \( p = 1, 2, ..., P, ~~i_{p} = 0, 1, 2, ... \) :

\begin{equation} 
V^{<p>}_{i_{p}} = f^{<p>}_{i_{p}}( V^{<p-1>}_{0}, V^{<p-1>}_{1}, V^{<p-1>}_{2}, ... ~) ~. 
\end{equation}

\vspace{-3.00mm} 
\[
\hspace*{40.00mm}
(~ p = 1, 2, ..., P, ~~~i_{p} = 0, 1, 2, ... ~) 
\]

\vspace{1.00mm} 
\noindent 
On that basis, we expand the energy function defined by Eq. (3) 
to the following one in which \( P + 1 \) {\sl Subspaces} are supposed in general:
\footnote{As for the energy function defined in our basic model 
corresponding to the generalized model with \( P = 1 \), 
refer to Eq. (9) in \cite{Tsutsumi2022}.
Note here that the term for \( J_{i} \) appearing in Eq. (3) 
is omitted in Eq. (10) for simplicity 
since it can be included in the second term.}

\vspace{-1.00mm} 
\begin{eqnarray} 
E & \stackrel{\triangle}{=} & \sum_{p=1}^{P} ~\Bigl[
                              ~\frac{1}{2} ~\sum_{i_{p}} V^{<p>}_{i_{p}}
                              ~\bigl( ~\sum_{j} T^{<p>}_{i_{p}j} ~V^{<p>}_{j} ~\bigr)
                              \nonumber \\
  &                         & \hspace{30mm}
                              - ~\sum_{i_{p}} V^{<p>}_{i_{p}}
                              ~\bigl( ~\sum_{k} S^{<p>}_{i_{p}k} ~Z^{<p>}_{k} ~\bigr)
                              ~\Bigr] \nonumber \\
  &                         & + ~\Bigl[ ~\frac{1}{2} ~\sum_{i} V^{<0>}_{i}
                              ~\bigl( ~\sum_{j} T^{<0>}_{ij} ~V^{<0>}_{j} ~\bigr)
                              ~- ~\sum_{i} V^{<0>}_{i}
                              ~\bigl( ~\sum_{k} S^{<0>}_{ik} ~Z^{<0>}_{k} ~\bigr) ~\Bigl]
                              \nonumber \\
  &                         & \hspace{50mm}
                              + ~\sum_{i} \frac{1}{r_{i}}
                              \int_{0}^{V^{<0>}_{i}} \hspace{-3.00mm} g^{-1}(V) dV ~.
\end{eqnarray}

\vspace{1.00mm} 
\noindent 
In the same way as the case for the variable \( V \) in Eq. (9), 
the numerical value in the first superscript with \( < \) and \( > \) 
on \( T \), \( S \), and \( Z \) 
means the {\sl Subspace} number 
to which the corresponding variable belongs inside the model. 
Differentiating Eq. (10) with respect to time, 
we get the following Eq. (11) 
on the assumption of the symmetry in \( T^{<p>}_{i_{p}j} \):

\vspace{-1.50mm} 
\begin{eqnarray} 
\frac{dE}{dt} & = & \sum_{p=1}^{P} ~\Bigl[
                    ~\sum_{i_{p}} \frac{dV^{<p>}_{i_{p}}}{dt} 
                    ~\bigl(
                    ~\sum_{j} T^{<p>}_{i_{p}j} V^{<p>}_{j} 
                    ~- ~\sum_{k} S^{<p>}_{i_{p}k} Z^{<p>}_{k}
                    ~\bigr)
                    ~\Bigr]
                    ~~~~~~~\nonumber \\
              &   &
                    ~\nonumber \\
              &   & + \sum_{i} \frac{dV^{<0>}_{i}}{dt} 
                    ~\bigl(
                    ~\sum_{j} T^{<0>}_{ij} V^{<0>}_{j} 
                    ~- ~\sum_{k} S^{<0>}_{ik} ~Z^{<0>}_{k} 
                    ~+ ~\frac{U_{i}}{r_{i}} 
                    ~\bigr) ~.
\end{eqnarray}

\vspace{1.00mm} 
\noindent 
\( dV^{<p>}_{i_{p}} / dt \)~ in the first term of Eq. (11) 
can be deformed with the chain rule as follows:

\vspace{1.50mm}
\begin{equation}
  \left \{
  \begin{array}{lll}
    {\displaystyle
      \frac{ d V^{<p>}_{i_{p}} }{dt}
    } & = &
    {\displaystyle
      \sum_{i_{p-1}} \frac{ \partial V^{<p>}_{i_{p}} }{ \partial V^{<p-1>}_{i_{p-1}} } 
      \frac{ d V^{<p-1>}_{i_{p-1}} }{dt} ~, 
    } \\
      &   &
    \hspace{22.00mm} 
    {\displaystyle
      (~ p = P, P - 1, ..., 2, ~~i_{p} = 0, 1, 2, ... ~)
    } \\
    \\
    {\displaystyle
      \frac{d V^{<1>}_{i_{1}}}{dt}
    } & = &
    {\displaystyle
      \sum_{i} \frac{\partial V^{<1>}_{i_{1}}}{\partial V^{<0>}_{i}} 
      \frac{d V^{<0>}_{i}}{dt} ~. 
      \hspace{22.00mm} 
      (~ i_{1} = 0, 1, 2, ... ~) 
      \hspace{-10.00mm} 
    } \\
  \end{array}
  \right. 
\end{equation}

\vspace{3.00mm}
\noindent 
Associating all of 
\( d V^{<P>}_{i_{P}} / dt \), \hspace{1mm} 
\( d V^{<P-1>}_{i_{P-1}} / dt \), \hspace{1mm} 
..., \hspace{1mm}
\( d V^{<2>}_{i_{2}} / dt \), \hspace{1mm} and 
\( d V^{<1>}_{i_{1}} / dt \) 
with \( d V^{<0>}_{i} / dt \) 
based on Eq. (12) yields:

\vspace{-1.50mm}
\begin{eqnarray} 
\frac{d V^{<p>}_{i_{p}}}{dt} & = & 
          \sum_{i_{p-1}} \frac{\partial V^{<p>}_{i_{p}}}{\partial V^{<p-1>}_{i_{p-1}}} 
          ~\biggl(
          ~\sum_{i_{p-2}} \frac{\partial V^{<p-1>}_{i_{p-1}}}{\partial V^{<p-2>}_{i_{p-2}}} 
          ~\Bigl(
          ~\sum_{i_{p-3}} \frac{\partial V^{<p-2>}_{i_{p-2}}}{\partial V^{<p-3>}_{i_{p-3}}} 
          ~\times
          \nonumber \\
                             &   &
          \nonumber \\
                             &   &
          \cdot \cdot \cdot \times 
          ~\bigl(
          ~\sum_{i_{1}} \frac{\partial V^{<2>}_{i_{2}}}{\partial V^{<1>}_{i_{1}}} 
          ~(
          ~\sum_{i} \frac{\partial V^{<1>}_{i_{1}}}{\partial V^{<0>}_{i}} 
          ~\frac{dV^{<0>}_{i}}{dt} 
          ~)
          ~\bigr)
          \cdot \cdot \cdot
          ~\Bigr)
          ~\biggr) ~. 
          \hspace{10mm}
\end{eqnarray}

\vspace{-4.00mm} 
\[
\hspace*{42.80mm}
(~ p = P, P - 1, ..., 2, 1, ~~i_{p} = 0, 1, 2, ... ~) 
\]

\vspace{2.00mm}
\noindent 
Substituting Eq. (13) for Eq. (11), 
Eq. (11) can be expressed as Eqs. (14) and (15) 
owing to its nested structure:

\vspace{-1.50mm}
\begin{eqnarray}
\frac{dE}{dt} & = & \sum_{i} \frac{dV^{<0>}_{i}}{dt}
                    ~\times \nonumber \\
              &   & ~\Bigl[
                    ~\sum_{j} T^{<0>}_{ij} V^{<0>}_{j}
                    - ~\sum_{k} S^{<0>}_{ik} ~Z^{<0>}_{k}
                    ~+ ~D^{<1>}_{i} ~+ ~\frac{U_{i}}{r_{i}}
                    ~\Bigr] ~,
\end{eqnarray}

\noindent 
where

\vspace{-2.00mm}
\begin{equation}
  \left \{
  \begin{array}{lll}
    {\displaystyle
      D^{<P>}_{i_{P-1}}
    }
    & = & 
    {\displaystyle
      \sum_{i_{P}} 
      \frac{\partial V^{<P>}_{i_{P}}}{\partial V^{<P-1>}_{i_{P-1}}} 
      ~\bigl(
      ~\sum_{j} T^{<P>}_{i_{P}j} V^{<P>}_{j} 
      - \sum_{k} S^{<P>}_{i_{P}k} ~Z^{<P>}_{k} 
      ~\bigr) ~, 
    } \\

    &   & \\

    &   &
    \hspace{40.0mm} (~ i_{P - 1} = 0, 1, 2, ... ~) \\

    &   & \\

    {\displaystyle
      D^{<p>}_{i_{p-1}}
    }
    & = & 
    {\displaystyle
      \sum_{i_{p}} 
      \frac{\partial V^{<p>}_{i_{p}}}{\partial V^{<p-1>}_{i_{p-1}}}
      ~\Bigl[
      ~\bigl(
      ~\sum_{j} T^{<p>}_{i_{p}j} V^{<p>}_{j} 
      - \sum_{k} S^{<p>}_{i_{p}k} ~Z^{<p>}_{k} 
      ~\bigr)
    } \\
    &   &
    {\displaystyle
      \hspace{69.5mm} ~+ ~D^{<p+1>}_{i_{p}}
      ~\Bigr] ~, 
    } \\

    &   & \\

    &   &
    \hspace{25.0mm} (~ p = P-1, P-2, ..., 3, 2, ~~i_{p - 1} = 0, 1, 2, ... ~) \\

    &   & \\

    {\displaystyle
      D^{<1>}_{i}
    }
    & = & 
    {\displaystyle
      \sum_{i_{1}}
      \frac{ \partial V^{<1>}_{i_{1}} }{ \partial V^{<0>}_{i} }
      ~\Bigl[
      ~\bigl(
      \sum_{j} T^{<1>}_{i_{1}j} V^{<1>}_{j} 
      - \sum_{k} S^{<1>}_{i_{1}k} ~Z^{<1>}_{k} 
      \bigr)
      + D^{<2>}_{i_{1}}
      ~\Bigr] ~.
    } \\

    &   & \\

    &   &
    \hspace{40.0mm} (~ i = 0, 1, 2, ... ~) \\
  \end{array}
  \right .
\end{equation}

\vspace{2.00mm}
\noindent
Here, let us construct a dynamical neural network described by Eq. (16) 
taking Eq. (15) into account. 
The only difference from Eq. (2) is the third term of the right-hand side 
\( D^{<1>}_{i} \):

\vspace{1.00mm} 
\begin{equation}
- ~c_{i} \frac{dU_{i}}{dt} ~= 
          ~\sum_{j} T^{<0>}_{ij} V^{<0>}_{j} 
          ~- ~\sum_{k} S^{<0>}_{ik} ~Z^{<0>}_{k} 
          ~+ ~D^{<1>}_{i} ~+ ~\frac{U_{i}}{r_{i}} ~.
          \hspace{5mm}
\end{equation}

\vspace{-4.00mm} 
\[
\hspace*{40.00mm}
(~ i = 0, 1, 2, ... ~) 
\]

\vspace{2.00mm} 
\noindent 
As is clear from substituting Eq. (16) 
for the part between \hspace{0.5mm} [ \hspace{0.5mm} 
and \hspace{0.5mm} ] \hspace{0.5mm} in Eq. (14), 
it always holds that \( dE/dt \leq 0 \) in the same way as in Eqs. (5)-(8); 
therefore the energy function defined by Eq. (10) is minimized 
with the network given by Eqs. (9), (15), and (16).

Figure 1 illustrates a block diagram of the dynamical neural network 
described by Eqs. (9), (15), and (16).
Regarding the relation between {\sl Subspace \#\( (p - 1) \)} 
and {\sl Subspace \#\( p \)}, 
the conversion 
from \( V^{<p-1>}_{0} \), \( V^{<p-1>}_{1} \), \( V^{<p-1>}_{2} \), ... 
to \( V^{<p>}_{0} \), \( V^{<p>}_{1} \), \( V^{<p>}_{2} \), ... 
based on Eq. (9) is called {\it Mapper \( M^{<p>} \)}, 
where \( p = 1, 2, ..., P \); 
in addition, the product-sum operation 
with \( \partial V^{<1>}_{i_{1}} / \partial V^{<0>}_{i} \) 
or \( \partial V^{<p>}_{i_{p}} / \partial V^{<p-1>}_{i_{p-1}} \) 
according to Eq. (15) is called 
{\it Mapper \( N^{<1>} \)} 
or 
{\it Mapper \( N^{<p>} \)}, 
where \( p = 2, 3, ..., P \). 
The derived network basically includes, in {\sl Subspace \#0}, 
the direct feedback circuit described by Eq. (2). 
As a whole, it is of ladder-shaped architecture 
in which a pair of {\it Mapper \( M^{<p>} \)} and {\it Mapper \( N^{<p>} \)} 
is placed in all the {\sl Subspaces} except for {\sl Subspace \#0}, 
where \( p = 1, 2, ..., P \), 
and also a pair of a feedback path and an input port 
exists in all of \( P + 1 \) {\sl Subspaces}.

\begin{figure}[!t]

    \hspace*{-6.50mm}
    \includegraphics[scale=0.84]{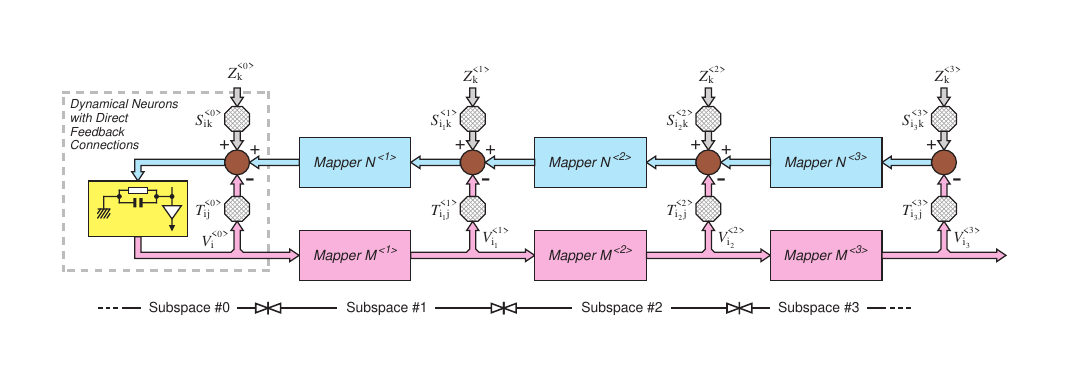}

    \vspace*{-5.00mm}
    \caption{
      A block diagram of the network described by 
      Eq. (16) (for the activity dynamics) 
      and Eqs. (9) and (15) (for the processing 
      by {\it Mapper \( M^{<p>} \)} and {\it Mapper \( N^{<p>} \)}, 
      where \( p = 1, 2, ..., P \)). 
      The whole state-space spanned by the network parameters, 
      \( V^{<0>}_{i} \), \( V^{<1>}_{i_{1}} \), ..., 
      \( V^{<P>}_{i_{P}} \), 
      is divided into multiple {\sl Subspaces} as shown in the figure, 
      where \( i = 0, 1, 2, ... \), \( i_{1} = 0, 1, 2, ... \), \( ... \), 
      and \( i_{P} = 0, 1, 2, ... \)\hspace*{0.50mm}. 
      Regarding the relation between {\sl Subspace \#\( (p - 1) \)} 
      and {\sl Subspace \#\( p \)}, 
      the conversion 
      from \( V^{<p-1>}_{0} \), \( V^{<p-1>}_{1} \), \( V^{<p-1>}_{2} \), ... 
      to \( V^{<p>}_{0} \), \( V^{<p>}_{1} \), \( V^{<p>}_{2} \), ... 
      based on Eq. (9) is called {\it Mapper \( M^{<p>} \)}, 
      where \( p = 1, 2, ..., P \). 
      In addition, the product-sum operation 
      with \( \partial V^{<1>}_{i_{1}} / \partial V^{<0>}_{i} \) 
      or \( \partial V^{<p>}_{i_{p}} / \partial V^{<p-1>}_{i_{p-1}} \) 
      according to Eq. (15) is called 
      {\it Mapper \( N^{<1>} \)} 
      or 
      {\it Mapper \( N^{<p>} \)}, 
      where \( p = 2, 3, ..., P \). 
      The network has multiple feedback paths 
      that can be divided into two kinds.
      One is inside {\sl Subspace \#0} and is included in Eq. (2). 
      The others are via outer {\sl Subspaces} 
      {\sl \#1}, {\sl \#2}, ..., {\sl \#\( P \)}; 
      output of dynamical neurons in {\sl Subspace \#0} 
      is returned back to {\sl Subspace \#0} with those dynamical neurons 
      based on detoured paths through {\it Mapper \( M^{<p>} \)}, 
      the connections \( T^{<p>}_{i_{p}j} \), 
      and {\it Mapper \( N^{<p>} \)}, where \( p = 1, 2, ..., P \). 
      The network also has an input port in every {\sl Subspace}. 
    }

\end{figure}

\vspace{1.00mm} 
\subsection{Employment of Static Neurons}

\vspace{1.00mm} 
\noindent 
In the previous subsection, we assumed that 
output of {\sl Subspace \#\( (p - 1) \)} 
and output of {\sl Subspace \#\( p \)} 
were related to each other with the function given by Eq. (9), 
where \( p = 1, 2, ..., P \). 
Here, instead of that, 
we employ layered neural networks described by Eqs. (17), (18), and (19); 
each of them has \( L \) layers and 
is composed of neurons with a tiny time constant 
that can be regarded as static 
compared with the time constant of a dynamical neuron 
placed in {\sl Subspace \#0} (i.e., the innermost {\sl Subspace}):

\vspace{-3.00mm}
\begin{eqnarray} 
V^{<p>}_{i}      & \stackrel{\triangle}{=} & O^{<p><L-1>}_{i} ~, \\
                 &                         & \nonumber \\
O^{<p><l>}_{i}   & =                       & g ( ~net^{<p><l>}_{i} ~) ~, \\
                 &                         & \nonumber \\
net^{<p><l>}_{i} & =                       & \sum_{j} W^{<p><l-1>}_{ij} O^{<p><l-1>}_{j} ~. 
\end{eqnarray}

\vspace{-4.50mm}
\[
\hspace*{5.00mm}
(~ p = 1, 2, ..., P-1, P, ~~l = 1, 2, ..., L-2, L-1, ~~i = 0, 1, 2, ... ~)
\]

\vspace{2.00mm} 
\noindent 
The numerical value in the first superscript with \( < \) and \( > \) 
on such variables as \( O \) , \( net \) , and \( W \) 
means the {\sl Subspace} number 
or the {\sl Mapper} (layered neural network) number 
to which the corresponding variable is directly related inside the model; 
it is limitedly assumed here to be a positive integer 
(i.e., except for \( 0 \)) 
since these variables are what become necessary 
after supposing {\sl Subspaces} outside the \( 0 \)-th (innermost) one. 
In the formulation, the layer number of each layered network 
is assigned from \( 0 \) to \( L - 1 \), 
and its number is placed on the upper right of a relevant variable 
as the second superscript with \( < \) and \( > \). 
\( O^{<p><l>}_{i} \) and \( net^{<p><l>}_{i} \) 
in Eqs. (18) and (19) respectively indicate 
``the output emitted from the \( i \)-th neuron in the \( l \)-th layer 
of the layered network inside {\sl Subspace \#\( p \)}" 
and ``the sum of inputs applied to its neuron." 
The modifiable connection 
from ``the \( j \)-th neuron in the \( (l - 1) \)-th layer" 
to ``the \( i \)-th neuron in the \( l \)-th layer" 
inside {\sl Subspace \#\( p \)} 
is written as  \( W^{<p><l>}_{ij} \). 
Equation (17) means that 
``the output emitted from the \( i \)-th neuron 
in the outermost ( output / \( (L - 1) \)-th ) layer 
of the layered network inside {\sl Subspace \#\( p \)}" 
\hspace{0.5mm} \( O^{<p><L-1>}_{i} \) is regarded 
as ``the corresponding output of the very {\sl Subspace}" 
\hspace{0.5mm} \( V^{<p>}_{i} \) 
in order to maintain consistency with Eq. (9). 
We presume that a neuron in the input layer of each layered network 
has linear input-output characteristics described by Eq. (6), 
and also that the input layer functions as a simple relaying unit 
for sending the output of inner {\sl Subspace \#\( (p - 1) \)} 
\hspace{0.5mm} \( V^{<p - 1>}_{i} \) 
efficiently to the upper layers of the layered network 
in outer {\sl Subspace \#\( p \)}; 
that is, the following relation holds:

\vspace{-1.00mm} 
\begin{equation} 
O^{<p><0>}_{i} = V^{<p-1>}_{i} ~. 
\end{equation}

\vspace{-3.00mm} 
\[
\hspace*{40.00mm} (~ p = 1, 2, ..., P - 1, P, ~~i = 0, 1, 2, ... ~) 
\]

\vspace{1.00mm} 
\noindent 
In the hidden layers, we assume a non-linear neuron 
whose input-output characteristics are given by Eq. (7). 
In the output layer, we employ a linear or non-linear neuron. 
For simplicity, we suppose here that 
a neuron has common linearity or non-linearity in each layer, 
although it is possible to make each neuron 
have different input-output characteristics in general.

Differentiation of Eq. (17) in association with Eqs. (18) and (19) 
can be written as follows:

\begin{eqnarray} 
\frac{d V^{<p>}_{i}}{dt} & = & 
        g^{\prime} ( ~net^{<p><L-1>}_{i} ~) 
        ~\biggl(
        ~\sum_{j} \frac{d W^{<p><L-2>}_{ij}}{dt} O^{<p><L-2>}_{j} 
        \hspace{5mm} \nonumber \\
                         &   &
        \hspace{15mm}
        + ~\sum_{j} W^{<p><L-2>}_{ij} \frac{d O^{<p><L-2>}_{j}}{dt} 
        ~\biggr) ~. 
\end{eqnarray}

\vspace{-4.00mm} 
\[
\hspace*{40.00mm} (~ p = 1, 2, ..., P - 1, P, ~~i = 0, 1, 2, ... ~) 
\]

\vspace{2.00mm} 
\noindent 
By particularly extracting the terms for {\sl Subspace \#\( P \)}
on the right-hand side of Eq. (11), 
we define \( H \) as shown below:

\vspace{1.00mm} 
\begin{eqnarray} 
H & \stackrel{\triangle}{=} & 
\sum_{i} \frac{d V^{<P>}_{i}}{dt}
~\mbox{\Large (}
~\sum_{j} T^{<P>}_{ij} V^{<P>}_{j} 
~- ~\sum_{k} S^{<P>}_{ik} Z^{<P>}_{k} 
~\mbox{\Large )} ~. 
\end{eqnarray}

\vspace{1.00mm} 
\noindent 
Substituting Eq. (21) with \( p = P \) for Eq. (22) 
and then doing some rearrangements, 
we obtain the following equation:

\vspace{1.00mm} 
\begin{eqnarray} 
H & = & \sum_{i} \sum_{j} \frac{d W^{<P><L-2>}_{ij}}{dt}
        ~O^{<P><L-2>}_{j} ~\delta^{<P><L-1>}_{i} \nonumber \\ 
  &   & + \sum_{j} \frac{d O^{<P><L-2>}_{j}}{dt}
        ~\mbox{\Large (} 
        ~\sum_{i} W^{<P><L-2>}_{ij} ~\delta^{<P><L-1>}_{i}
        ~\mbox{\Large )} ~. 
\end{eqnarray}

\vspace{1.00mm} 
\noindent 
We suppose here that 
\( \delta^{<P><L>}_{i} \) is defined as Eq. (24) 
for the sake of convenience. 
On that basis, 
 \( \delta^{<P><L-1>}_{i} \) appearing in Eq. (23) 
can be denoted as Eq. (25). 
The variable \( \delta \) also becomes to be needed 
after supposing two and more {\sl Subspaces} in the model, 
and we place the corresponding {\sl Subspace} number 
with \( < \) and \( > \) as the first superscript 
on the upper right of the variable. 
We further append the currently processing layer number 
with \( < \) and \( > \) as the second superscript 
on the upper right of the variable:

\begin{eqnarray} 
\delta^{<P><L>}_{i}   & \stackrel{\triangle}{=} & 
        ~\sum_{j} T^{<P>}_{ij} V^{<P>}_{j} 
        ~- ~\sum_{k} S^{<P>}_{ik} Z^{<P>}_{k} ~, \\
                      &                         &
        \nonumber \\
\delta^{<P><L-1>}_{i} & \stackrel{\triangle}{=} & 
        g^{\prime} ( ~net^{<P><L-1>}_{i} ~) ~\delta^{<P><L>}_{i} ~. 
\end{eqnarray}

\vspace{-3.00mm} 
\[
\hspace*{25.00mm}
(~ i = 0, 1, 2, ... ~) 
\]

\vspace{3.00mm} 
\noindent 
Furthermore calculating \( d O^{<P><L-2>}_{j} / dt \) in Eq. (23) 
based on Eqs. (18) and (19), we obtain:

\begin{eqnarray} 
H & = & \sum_{i} \sum_{j} \frac{d W^{<P><L-2>}_{ij}}{dt}
        ~O^{<P><L-2>}_{j} ~\delta^{<P><L-1>}_{i} \nonumber \\ 
  &   & + ~\sum_{j} \sum_{k} \frac{d W^{<P><L-3>}_{jk}}{dt}
        ~O^{<P><L-3>}_{k} ~\delta^{<P><L-2>}_{j} \nonumber \\ 
  &   & + ~\sum_{k} \frac{d O^{<P><L-3>}_{k}}{dt}
        ~\mbox{\Large (} 
        ~\sum_{j} W^{<P><L-3>}_{jk} ~\delta^{<P><L-2>}_{j}
        ~\mbox{\Large )} ~, 
\end{eqnarray}

\vspace{1.00mm} 
\noindent 
where

\begin{eqnarray} 
\delta^{<P><L-2>}_{j} & \stackrel{\triangle}{=} & 
g^{\prime} ( ~net^{<P><L-2>}_{j} ~)
~\mbox{\Large (}
~\sum_{i} W^{<P><L-2>}_{ij} \delta^{<P><L-1>}_{i} 
~\mbox{\Large )} ~. ~~~~~ 
\end{eqnarray}

\vspace{2.00mm} 
\noindent 
Repeating these calculations in the direction 
from the output layer to the input one in {\sl Subspace \#\( P \)}, 
we get the following equation:

\begin{eqnarray} 
H & = & \sum_{l=0}^{L-2} 
        ~\mbox{\Large [} 
        ~\sum_{i} \sum_{j} \frac{d W^{<P><l>}_{ij}}{dt}
        ~\mbox{\Large (}
        ~O^{<P><l>}_{j} ~\delta^{<P><l+1>}_{i} 
        ~\mbox{\Large )}
        ~\mbox{\Large ]} \nonumber \\ 
  &   & \hspace{20mm}
        + ~\sum_{i} \frac{d O^{<P><0>}_{i}}{dt}
        ~\delta^{<P><0>}_{i} ~. 
\end{eqnarray}

\vspace{2.00mm} 
\noindent 
Here, including Eqs. (25) and (27), 
the variables \( \delta^{<P><l>}_{i} \) 
for \( l = L - 1, L - 2, ..., 2, 1, 0 \)
in {\sl Subspace \#\( P \)} can be written as follows:

\begin{eqnarray}
\delta^{<P><l>}_{i} & = & g^{\prime} ( ~net^{<P><l>}_{i} ~)
                          ~\mbox{\Large (}
                          ~\sum_{j} W^{<P><l>}_{ji} \delta^{<P><l+1>}_{j} 
                          ~\mbox{\Large )} ~, \\
                    &   & \nonumber \\
                    &   & \hspace{20.00mm}
                          ( ~l = L-1, L-2, ..., 2, 1, ~~i = 0, 1, 2, ... ~) 
                          \nonumber \\
                    &   & \nonumber \\
\delta^{<P><0>}_{i} & = & \sum_{j} W^{<P><0>}_{ji} \delta^{<P><1>}_{j} ~. \\
                    &   & \hspace{20.00mm} 
                          (~ i = 0, 1, 2, ... ~) 
                          \nonumber 
\end{eqnarray}

\vspace{1.00mm}
\noindent
As stated above, 
linear neurons in the input layer of an {\sl Internetwork} 
are assumed to have a simple relaying functionality 
for sending the output of the {\sl Subspace} 
on the inner side of the {\sl Internetwork} 
to the upper layers in the very {\sl Internetwork}. 
So, with regard to \( dO^{<P><0>}_{i} / dt \) in Eq. (28), 
we can utilize the following relation 
according to Eq. (20):

\vspace{1.00mm} 
\begin{equation} 
\frac{dO^{<p><0>}_{i}}{dt} = \frac{dV^{<p-1>}_{i}}{dt} ~. 
\hspace{10.00mm}
\end{equation}

\vspace{-3.00mm} 
\[
\hspace*{40.00mm}
(~ p = 1, 2, ..., P - 1, P, ~~i = 0, 1, 2, ... ~) 
\]

\vspace{2.00mm} 
\noindent 
Remind here that Eq. (22) can be represented by Eq. (28). 
Hence, putting Eq. (28) back to Eq. (11) 
and rearranging those equations 
by means of Eq. (31) with \( p = P \), we obtain

\vspace{-1.00mm} 
\begin{eqnarray} 
\frac{dE}{dt} & = & \sum_{l=0}^{L-2}
                    ~\mbox{\Large [}
                    ~\sum_{i} \sum_{j} \frac{d W^{<P><l>}_{ij}}{dt}
                    ~\mbox{\Large (}
                    ~O^{<P><l>}_{j} ~\delta^{<P><l+1>}_{i}
                    ~\mbox{\Large )}
                    ~\mbox{\Large ]} \nonumber \\
              &   & \hspace{-5mm}
                    + ~\sum_{i} \frac{d V^{<P-1>}_{i}}{dt}
                    ~\mbox{\Large (} 
                    ~\sum_{j} T^{<P-1>}_{ij} V^{<P-1>}_{j} 
                    ~- ~\sum_{k} S^{<P-1>}_{ik} Z^{<P-1>}_{k} \nonumber \\
              &   & \hspace{80mm}
                    + ~\delta^{<P><0>}_{i} ~\mbox{\Large )} \nonumber \\
              &   & \hspace{-5mm}
                    + ~\sum_{p=1}^{P-2} ~\Bigl( 
                    ~\sum_{i} \frac{dV^{<p>}_{i}}{dt} 
                    ~\mbox{\Large (} 
                    ~\sum_{j} T^{<p>}_{ij} V^{<p>}_{j} 
                    ~- ~\sum_{k} S^{<p>}_{ik} Z^{<p>}_{k}
                    ~\mbox{\Large )} ~\Bigr) \nonumber \\ 
              &   & \hspace{-5mm}
                    + ~\sum_{i} \frac{dV^{<0>}_{i}}{dt} 
                    ~\mbox{\Large (}
                    ~\sum_{j} T^{<0>}_{ij} V^{<0>}_{j} 
                    ~- ~\sum_{k} S^{<0>}_{ik} ~Z^{<0>}_{k} 
                    ~+ ~\frac{U_{i}}{r_{i}} 
                    ~\mbox{\Large )} ~.
\end{eqnarray}

\vspace{1.00mm} 
\noindent 
Extracting the second term of Eq. (32), 
we newly define it as \( H \) shown blow:

\vspace{-1.00mm} 
\begin{eqnarray} 
H & \stackrel{\triangle}{=} & \sum_{i} \frac{d V^{<P-1>}_{i}}{dt}
                              ~\mbox{\Large (}
                              ~\sum_{j} T^{<P-1>}_{ij} V^{<P-1>}_{j}
                              ~- ~\sum_{k} S^{<P-1>}_{ik} Z^{<P-1>}_{k} \nonumber \\
  &                         & \hspace{70mm}
                              + ~\delta^{<P><0>}_{i} ~\mbox{\Large )} ~.
\end{eqnarray}

\vspace{2.00mm} 
\noindent 
Substituting Eq. (21) with \( p = P - 1 \) for Eq. (33), 
we repeat, for {\sl Subspace \#(P - 1)}, 
the same sort of calculations as the ones from Eq. (22) to Eq. (28). 
Rearranging the result in view of Eq. (31) with \( p = P - 1 \), 
we get

\vspace{-1.00mm} 
\begin{eqnarray} 
H & = & \sum_{l=0}^{L-2} 
        ~\mbox{\Large [} 
        ~\sum_{i} \sum_{j} \frac{d W^{<P-1><l>}_{ij}}{dt}
        ~\mbox{\Large (}
        ~O^{<P-1><l>}_{j} ~\delta^{<P-1><l+1>}_{i} 
        ~\mbox{\Large )}
        ~\mbox{\Large ]} \nonumber \\ 
  &   & \hspace{20mm}
        + ~\sum_{i} \frac{d V^{<P-2>}_{i}}{dt}
        ~\delta^{<P-1><0>}_{i} ~. 
\end{eqnarray}

\vspace{2.00mm} 
\noindent 
Here, the variables \( \delta \)s 
for {\sl Subspace \#\( (P - 1) \)} 
appearing in Eq. (34) are as shown by Eqs. (35), (36), and (37):

\vspace{-1.00mm} 
\begin{eqnarray}
\delta^{<P-1><L>}_{i} & = & \sum_{j} T^{<P-1>}_{ij} V^{<P-1>}_{j} 
                            - \sum_{k} S^{<P-1>}_{ik} Z^{<P-1>}_{k} 
                            + \delta^{<P><0>}_{i} ~, \\
                      &   & \nonumber \\
                      &   & \hspace{30.00mm}
                            ( ~i = 0, 1, 2, ... ~) \nonumber \\
                      &   & \nonumber \\
\delta^{<P-1><l>}_{i} & = & g^{\prime} ( ~net^{<P-1><l>}_{i} ~)
                            ~\mbox{\Large (}
                            ~\sum_{j} W^{<P-1><l>}_{ji} \delta^{<P-1><l+1>}_{j} 
                            ~\mbox{\Large )} ~, \\
                      &   & \nonumber \\
                      &   & \hspace{10.00mm}
                            ( ~l = L-1, L-2, ..., 2, 1, ~~i = 0, 1, 2, ... ~) 
                            \nonumber \\
                      &   & \nonumber \\
\delta^{<P-1><0>}_{i} & = & \sum_{j} W^{<P-1><0>}_{ji} \delta^{<P-1><1>}_{j} ~. \\
                      &   & \nonumber \\
                      &   & \hspace{30.00mm} 
                            ( ~i = 0, 1, 2, ... ~) 
                            \nonumber 
\end{eqnarray}

\vspace{2.00mm} 
\noindent 
We put Eq. (34) back to Eq. (32). 
Then, regarding the third term on the right-hand side of Eq. (32), 
we perform the same sort of calculations as the ones from Eq. (33) to Eq. (34) 
for \( p = P - 2, P - 3, ..., 2, 1 \) in turn. 
After these procedures, we finally obtain Eq. (38) 
instead of Eqs. (11) and (32):

\begin{eqnarray} 
\frac{dE}{dt} & = & \sum_{i} \frac{d V^{<0>}_{i}}{dt}
                    ~\mbox{\Large (} 
                    ~\sum_{j} T^{<0>}_{ij} V^{<0>}_{j} 
                    ~- ~\sum_{k} S^{<0>}_{ik} Z^{<0>}_{k} \nonumber \\
              &   & \hspace{55mm}
                    ~+ ~\delta^{<1><0>}_{i} 
                    ~+ \frac{U_{i}}{r_{i}} 
                    ~\mbox{\Large )} \nonumber \\
              &   & \hspace{-10mm} + ~\sum_{p=1}^{P} \sum_{l=0}^{L-2} 
                    ~\mbox{\Large [} 
                    ~\sum_{i} \sum_{j} \frac{d W^{<p><l>}_{ij}}{dt}
                    ~\mbox{\Large (} 
                    ~O^{<p><l>}_{j} ~\delta^{<p><l+1>}_{i} 
                    ~\mbox{\Large )} 
                    ~\mbox{\Large ]} ~. 
\end{eqnarray}

\vspace{2.00mm} 
\noindent 
Including Eqs. (24), (29), (30), (35), (36), and (37), 
the variables \( \delta \)s in Eq. (38) 
can totally be denoted as follows:

%
\begin{equation}
\delta^{<p><L>}_{i} =
\left \{
\begin{array}{l}
  {\displaystyle
  \sum_{j} T^{<p>}_{ij} V^{<p>}_{j} 
  ~- ~\sum_{k} S^{<p>}_{ik} Z^{<p>}_{k} } \\
  {\displaystyle \hspace{52.50mm} (~ p = P, ~~i = 0, 1, 2, ... ~)} \hspace{-2.20mm} \\
  \\
  {\displaystyle
  \sum_{j} T^{<p>}_{ij} V^{<p>}_{j} 
  ~- \sum_{k} S^{<p>}_{ik} Z^{<p>}_{k} + \delta^{<p><0>}_{i} } ~, \\
  {\displaystyle \hspace{24.50mm} (~ p = P-1, P-2, ..., 2, 1, ~~i = 0, 1, 2, ... ~)} \hspace{-2.50mm} \\
\end{array}
\right .
\end{equation}

\begin{eqnarray}
\delta^{<p><l>}_{i} & = & g^{\prime} ( ~net^{<p><l>}_{i} ~)
                          ~\mbox{\Large (}
                          ~\sum_{j} W^{<p><l>}_{ji} \delta^{<p><l+1>}_{j} 
                          ~\mbox{\Large )} ~, \\
                    &   & \hspace{-8.00mm}
                          (~ p = P, P-1, ..., 2, 1,  
                          ~~l = L-1, L-2, ..., 2, 1, 
                          ~~i = 0, 1, 2, ... ~) \nonumber \\
                    &   & \nonumber \\
\delta^{<p><0>}_{i} & = & \sum_{j} W^{<p><0>}_{ji} \delta^{<p><1>}_{j} ~. \\
                    &   & \hspace{31.30mm} 
                          (~ p = P, P-1, ..., 2, 1, 
                          ~~i = 0, 1, 2, ... ~) \nonumber
\end{eqnarray}

\vspace{2.00mm} 
\noindent 
Constructing the dynamical network described by Eq. (42) 
in conjunction with Eqs. (39), (40), and (41) 
and assuming that the synaptic connections 
in the layered neural networks are modified according to Eq. (43), 
that is, defining

\begin{equation} 
- ~c_{i} \frac{dU_{i}}{dt} ~= 
        \sum_{j} T^{<0>}_{ij} V^{<0>}_{j}
        ~- ~\sum_{k} S^{<0>}_{ik} Z^{<0>}_{k}
        ~+ ~\delta^{<1><0>}_{i}
        ~+ ~\frac{U_{i}}{r_{i}} 
\end{equation}

\vspace{-2.00mm} 
\[
\hspace*{40.00mm}
(~ i = 0, 1, 2, ... ~) 
\]

\noindent
and

\begin{equation} 
- ~\eta_{ij}^{<p><l>} ~\frac{d W^{<p><l>}_{ij}}{dt} ~= 
        ~O^{<p><l>}_{j} ~\delta^{<p><l+1>}_{i} ~, 
\end{equation}

\vspace{-2.00mm} 
\[
\hspace{0.00mm}
(~ p = 1, 2, ..., P-1, P, ~~l = 0, 1, 2, ..., L-3, L-2, 
~~i = 0, 1, 2, ..., ~~j = 0, 1, 2, ... ~) 
\]

\vspace{2.00mm}
\noindent
then we can understand, 
through the process of substituting Eqs. (42) and (43) for Eq. (38), 
that the energy function defined by Eq. (10) 
is minimized in the manner of Eq. (44) 
under the condition that Eq. (6) or (7) is used as \( g(x) \):

\begin{eqnarray} 
\frac{dE}{dt} & = & \sum_{i} 
                    ~\frac{dV^{<0>}_{i}}{dt} 
                    ~\mbox{\Large (} - c_{i} \frac{dU_{i}}{dt} \mbox{\Large )} 
                    \nonumber \\ 
              &   & + ~\sum_{p=1}^{P} ~\sum_{l=0}^{L-2} 
                    ~\mbox{\LARGE [~} \sum_{i} \sum_{j} 
                    \frac{dW_{ij}^{<p><l>}}{dt} 
                    \mbox{\Large (} - \eta_{ij}^{<p><l>} 
                    \frac{dW_{ij}^{<p><l>}}{dt} \mbox{\Large )}
                    \mbox{\LARGE ~]} 
                    \nonumber \\ 
              &   & \nonumber \\ 
              & = & - ~\sum_{i} ~c_{i} ~g^{\prime} (U_{i}) 
                    ~\mbox{\Large (} \frac{dU_{i}}{dt} \mbox{\Large )}^{2} 
                    \nonumber \\ 
              &   & - ~\sum_{p=1}^{P} ~\sum_{l=0}^{L-2} 
                    ~\mbox{\LARGE [~} \sum_{i} \sum_{j} \eta_{ij}^{<p><l>} 
                    \mbox{\Large (} \frac{dW_{ij}^{<p><l>}}{dt} \mbox{\Large )}^{2} 
                    \mbox{\LARGE ~]} 
                    ~~\leq 0 ~.
\end{eqnarray}

\vspace{3.00mm}
\noindent
Here, \( \eta_{ij}^{<p><l>} \) is a positive constant 
determining the learning speed.

The total network described by Eqs. (42) and (43) 
can be illustrated as a block diagram of Fig. 2. 
The model consists of \( P + 1 \) {\sl Subspaces} 
from {\sl Subspace \#\( 0 \)} to {\sl Subspace \#\( P \)} 
as indicated in the figure. 
In the same way as our basic model, 
we name ``a pair of the layered networks between adjacent {\sl Subspaces}" 
an ``{\sl Internetwork}" collectively 
in the sense that it interconnects the inner and outer {\sl Subspaces}. 
We further call the paired layered networks in each {\sl Internetwork}, 
a ``{\sl Forward Subnet}" and a ``{\sl Backward Subnet}," 
as shown in the figure; 
both of them complementarily work for the total dynamics. 
The {\sl Forward Subnets} operate 
according to Eqs. (17), (18), and (19). 
On the other hand, 
\( \delta_{i}^{<1><0>} \) appearing in Eq. (42) 
can be calculated in turn from \( \delta_{i}^{<P><L>} \) 
in accordance with Eqs. (39), (40), and (41), 
and the processing based on those equations 
is performed in the {\sl Backward Subnets}; 
note that their computational processes for \( \delta \)s 
are expressed explicitly in the form of layered networks. 
Also note that, obviously from a comparison with Fig. 1, 
all sets of {\sl Mapper \( M^{<p>} \)} and {\sl Mapper \( N^{<p>} \)} 
in Fig. 1 are replaced 
here by pairs of static layered neural networks, 
where \( p = 1, 2, ..., P \).

\begin{figure}[!t]

    \hspace*{-6.50mm}
    \includegraphics[scale=0.84]{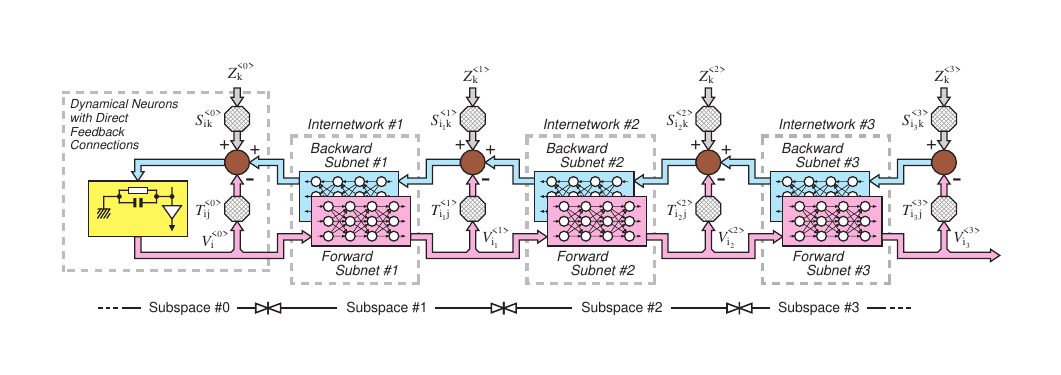}

    \vspace*{-5.00mm}
    \caption{
      A block diagram of the proposed network model 
      described by Eq. (42) (for the activity dynamics) 
      and Eq. (43) (for the learning dynamics). 
      The network has multiple {\sl Subspaces}, 
      each of which has a pair of a feedback path and an input port. 
      Two adjacent {\sl Subspaces} are connected 
      by an ``{\sl Internetwork}" composed of a pair of layered networks 
      with static neurons and modifiable connections; 
      the paired networks in each {\sl Internetwork} 
      are named a ``{\sl Forward Subnet}" and a ``{\sl Backward Subnet}" 
      as indicated in the figure. 
      The {\sl Forward Subnets} operate 
      according to Eqs. (17), (18), and (19), 
      and the processing in the {\sl Backward Subnets} 
      is based on Eqs. (39), (40), and (41); 
      both of them complementarily work for the total dynamics. 
      Note that the computational processes of 
      \( \delta^{<P><l>}_{i} \), \( \delta^{<P-1><l>}_{i} \), ..., 
      \( \delta^{<1><l>}_{i} \) 
      are expressed explicitly in the form of layered networks, 
      where \( l = L, L-1, ..., 1, 0, ~~i = 0, 1, 2, ... \) . 
      Also note that all sets of 
      {\sl Mapper \( M^{<p>} \)} and {\sl Mapper \( N^{<p>} \)} 
      in Fig. 1 are replaced 
      here by pairs of static layered neural networks, 
      where \( p = 1, 2, ..., P \). 
    } 

\end{figure}

\vspace{2.00mm} 
\subsection{Learning Mode and Association Mode}

\vspace{1.00mm} 
\noindent
The derived network has a pair of one feedback path and one input port 
in each {\sl Subspace} as depicted in Fig. 1 or Fig. 2. 
The fixed connections related to this particular architecture 
correspond to 
``\( T^{<0>}_{ij} \) and \( S^{<0>}_{ik} \)" 
and 
``\( T^{<p>}_{i_{p}j} \) and \( S^{<p>}_{i_{p}k} \)" 
in the first and second terms on the right-hand side 
of the energy function defined by Eq. (10); 
changing their values and signs, 
the potential field's shape can be designed in various forms, 
and it is reasonable, in this sense, 
that 
``\( T^{<0>}_{ij} \) and \( S^{<0>}_{ik} \)" 
or 
``\( T^{<p>}_{i_{p}j} \) and \( S^{<p>}_{i_{p}k} \)" 
in each {\sl Subspace} are treated as a pair, 
where \( p = 1, 2, ..., P - 1, P \). 
In the architecture shown by Fig. 1 or Fig. 2, 
let us consider, for instance, a situation in which 
only a pair of \( T^{<P>}_{i_{P}j} \) and \( S^{<P>}_{i_{P}k} \) 
in {\sl Subspace \#\( P \)} exists 
and all other pairs of 
``\( T^{<0>}_{ij} \) and \( S^{<0>}_{ik} \)" 
and 
``\( T^{<p>}_{i_{p}j} \) and \( S^{<p>}_{i_{p}k} \)" 
are removed, where \( p = 1, 2, ..., P - 1 \). 
The output \( V^{<0>}_{i} \) in the innermost {\sl Subspace} 
is converted to \( V^{<P>}_{i_{P}} \) in the outermost {\sl Subspace} 
through all {\sl Forward Subnets}. 
Its signal in {\sl Subspace \#\( P \)} 
is combined with the input signal from the outside \( Z^{<P>}_{k} \) 
via \( T^{<P>}_{i_{P}j} \) and \( S^{<P>}_{i_{P}k} \). 
These mixed signals are returned through all {\sl Backward Subnets}
to the innermost {\sl Subspace}, where there is no internal feedback loop 
based on the connections \( T^{<0>}_{ij} \) and \( S^{<0>}_{ik} \). 
Therefore, the network dynamics are determined only 
by the longest detoured feedback loop 
with all {\sl Forward Subnets}, 
the fixed connections \( T^{<P>}_{i_{P}j} \) and \( S^{<P>}_{i_{P}k} \), 
and all {\sl Backward Subnets}. 
It is of great interest to analyze how time-course properties 
of the whole network with such an architecture are. 
Before studying the model's behavior in the following sections, 
we topologically group the model into two types (two modes), 
the ``Learning Mode" and the ``Association Mode," as illustrated in Fig. 3, 
depending on the presence or absence of a set of the fixed connections 
in each {\sl Subspace}.

\begin{figure}[!t]

    \vspace*{-0.50mm}
    \hspace*{8.45mm}
    \includegraphics[scale=0.94]{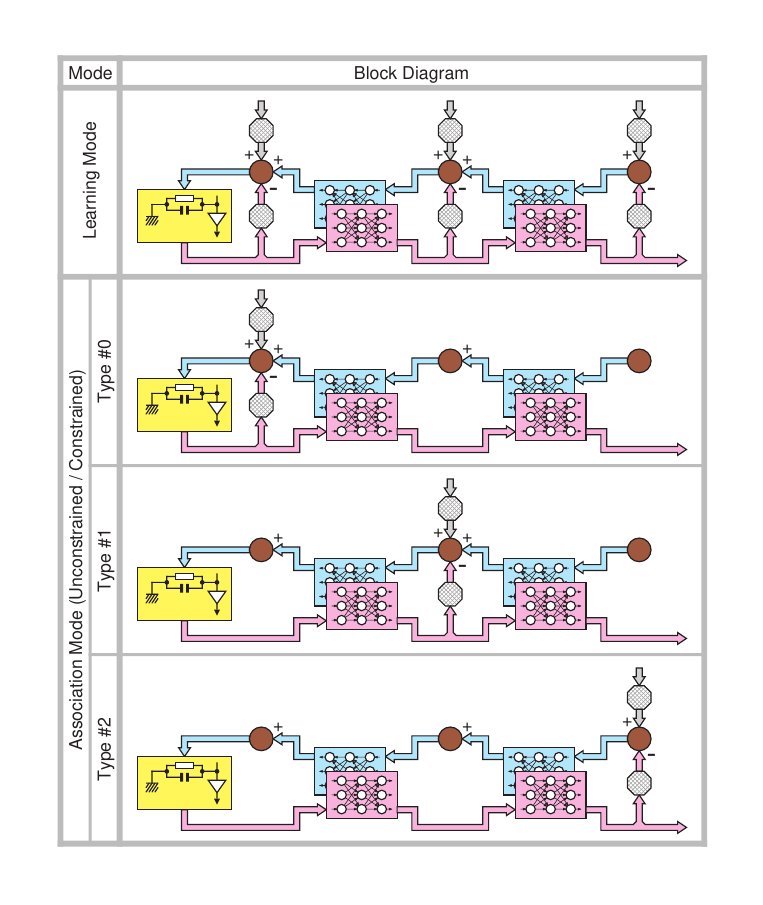}

    \vspace*{-4.00mm}
    \caption{
      Putting \( P \) as the number of {\sl Internetworks}, 
      the model is topologically grouped, 
      basically into two types (two modes), 
      totally into \( P + 2 \) types, 
      depending on the presence or absence 
      of fixed connections in {\sl Subspaces}. 
      This figure illustrates an example with \( P = 2 \). 
      A block diagram in the uppermost row shows the case 
      with all pairs of a feedback path and an input port, 
      and we call it the ``Learning Mode." 
      The next three rows are called the ``Association Mode" 
      in which a pair of a feedback path and an input port 
      exists only in one {\sl Subspace}; 
      each block diagram from top to bottom respectively 
      corresponds to Type \#0, Type \#1, or Type \#2 architecture. 
    } 

\end{figure}

In each {\sl Internetwork} composed of 
a pair of static layered neural networks, 
input signals and teacher (target) ones 
are both needed for its proper training. 
In order to train all of the {\sl Internetworks} in a lump, 
therefore, it is crucial to adequately provide 
a set of input signals and teacher (target) ones for those networks.
\footnote{We will conduct simulation experiments 
for the network's training in the later section.} 
Although we can suppose some methods by which 
each {\sl Internetwork} is trained separately or in stages, 
we will investigate only a lump training here. 
Thus, the architecture for the Learning Mode 
that works based on Eq. (42) (for the activity dynamics) 
and Eq. (43) (for the learning dynamics) is limited to 
only one type with all pairs of a feedback path and an input port 
regardless of the model size (i.e., the number of {\sl Subspaces}). 
On the other hand, the architecture for the Association Mode 
needs to be examined carefully. 
Supposing that the total number of {\sl Subspaces} in the model is \( P + 1 \), 
the combination number of pairs of a feedback path and an input port is 
\( {}_{P + 1} C_{1} + {}_{P + 1} C_{2} + \cdots + {}_{P + 1} C_{P + 1} 
= \sum_{\kappa=1}^{P+1} {}_{P + 1} C_{\kappa} \). 
It is interesting in principle to thoroughly explore dynamical behavior 
of all cases with one, two, three, ... pairs 
of a feedback path and an input port. 
However, we will only analyze, as the Association Mode in this paper, 
the simplest cases in which there exists only one pair 
of a feedback path and an input port in the model.
\footnote{Regarding a model that has 
multiple pairs of a feedback path and an input port, 
its dynamical behavior may be grasped 
as a linear superposition of the simplest cases 
with only one pair of a feedback path and an input port. 
This is a subject for future analysis.} 
Figure 3 depicts types of architectures with \( P = 2 \) as an example; 
the Learning Mode has a unique architecture, and, 
in the Association Mode, there are 3 (\( = P + 1 \)) types of architectures.

When we examine associative dynamics of the model only for a short time, 
the total amount of synaptic modification based on Eq. (43) 
in {\sl Internetworks} must be restricted. 
When signals (as input signals and target ones) 
applied from the outside to input ports 
are not adequate for {\sl Internetwork}'s training, however, 
the synaptic modification will not principally go well. 
And, even at the stage in which 
the amount of such synaptic modification is small, 
the mapping relationships that were acquired 
previously and appropriately in the {\sl Internetworks} 
may be collapsed according to Eq. (43). 
Therefore, in the same way as our basic model, 
it is reasonable, in avoiding this situation, 
that we presume the following condition for Eq. (43) 
in conjunction with topological distinction. 
That is, Eq. (43) is set to be OFF in the Association Mode
when we test how the activity dynamics differ 
depending on such types of architectures 
as Types \( \#0 \), \( \#1 \), and \( \#2 \) in Fig. 3. 
Note that the activity dynamics described by Eq. (42) 
is ON at all times in both the Learning and Association Modes. 
At this stage of modelisation, 
based on the discussion in our previous paper, 
we divide the Association Mode into 
the ``Unconstrained" Association Mode 
and the ``Constrained" Association Mode 
in order to deeply inspect and effectively utilize 
the associative dynamics of the model; 
the former mode is the one in which 
``a set of fixed values" (corresponding to 
a goal point in a coordinate system) 
is given to a unique input port, 
and the latter mode is the one in which 
not a goal point but ``a trajectory" with various speeds 
is applied to a unique input port.
\footnote{In Section 5, 
as a target input in the Constrained Association Mode, 
we will specifically employ 
``a circular periodic trajectory" 
in which a goal point is not explicitly indicated, 
instead of ``a straight trajectory toward a goal point" 
adopted in our previous paper.} 
They are only characterized by different styles of input signals,  
and both are identical from the architectural viewpoint. 
We will discuss these Association Modes 
in detail and separately in later sections.

\section{Learning Mode}

\subsection{Learning Mode in One-Dimensional Model}

\vspace{2.00mm} 
\noindent 
{\large (1) Linear Mapping}

\vspace{2.50mm} 
\noindent 
As stated in the previous section, 
synaptic connections in the derived network are roughly classified into two; 
the ones only inside {\sl Internetworks} are assumed to have plasticity, 
and the ones outside {\sl Internetworks} are all fixed. 
How can those plastic connections be trained? 
Eq. (42) is the activity dynamics of the whole network, 
and Eq. (43) is the learning dynamics for modifiable connections 
in {\sl Internetworks}; 
these two kinds of dynamics always run in the Learning Mode. 
Therefore, any other special learning algorithm is not needed 
even if we discuss training issues for the model. 
All we have to do is to trace time-course behavior of the total network 
when appropriate signals are applied to the model's input ports. 
In this sense, it is the most important subject for the Learning Mode 
to examine what kind of signals should be applied 
to those input ports from the outside 
and then analyze how the training procedure goes.

For training each {\sl Internetwork}, as suggested in Subsection 2.3, 
the information corresponding to both an input signal and a teacher signal 
have to be consistently applied to it. 
In order to let the {\sl Internetworks} acquire 
wide-ranging mapping relationships, 
time-varying signals instead of fixed-value ones 
should be adequate as inputs from the outside to all {\sl Subspaces}. 
In our basic model with a single {\sl Internetwork}, 
when identical regular sinusoidal waves were injected 
to two input ports in front of and behind the {\sl Internetwork} 
of a one-dimensional model, 
the synaptic connections were reasonably modified 
so that the {\sl Internetwork} functioned as a linear mapper. 
In the case of employing deformed sinusoidal waves, 
we were able to make the {\sl Internetwork} 
gain various non-linear mapping relationships.

Simply in line with those previous experiments, 
can we adequately train all plastic connections 
inside the model with two or more {\sl Internetworks}? 
Although we may assume various types of input signals, 
it is one of the options 
to verify the same kind of signals 
as those employed in our basic model. 
When we set \( P \) in general as the number of {\sl Internetworks}, 
there exist \( P + 1 \) {\sl Subspaces} and \( P + 1 \) input ports 
inside the model. 
Because of this structure, 
it is estimated that input signals from the outside 
and back-propagating signals through {\sl Backward Subnets} 
complexly interfere with each other 
even if non-distorted regular sinusoidal waves 
are applied to the input ports. 
Thus, it is quite significant to investigate time-course behavior 
in the Learning Mode of the proposed network 
with multiple {\sl Internetworks}. 
Since the output \( V^{<p>}_{i} \) and the input \( Z^{<p>}_{i} \) 
are supposed to be of short-term average impulse density, 
note again that the sinusoidal or periodic wave 
practically represents repetitive bursting or semi-bursting nerve impulses, 
where \( p = 0, 1, ..., P \).

The numerical value \( P \) in the proposed network can arbitrarily be set. 
Now let us consider a one-dimensional model with \( P = 2 \) 
as the smallest case of \( P \geq 2 \). 
Figure 4(a) illustrates the time-course of network parameters 
when identical regular sinusoidal waves were applied to 
three (\( P + 1 = 3 \)) input ports, 
each of which was located in a {\sl Subspace} one by one. 
Setting the values of \( c_{i} \) and \( \eta_{ij}^{{<p>}{<l>}} \) 
so that the time constants in activity and learning dynamics 
were \( 1 \) {\sf ms} and \( 5 \) {\sf s} respectively, 
we conducted our simulation experiments 
based on the standard Runge-Kutta method 
with a stepsize of \( 0.1 \) {\sf ms}. 
We let \( r_{i} \) be \( \infty \). 
Then, the frequency of all input signals 
applied to three {\sl Subspaces} 
was set to \( 10.0 \) {\sf Hz}. 
As for the fixed connections (conductance), \( T^{<p>} \) and \( S^{<p>} \), 
we put that \( T^{<p>} = 1.0 \) {\sf S} and \( S^{<p>} = 1.0 \) {\sf S}, 
where \( p = 0, 1, 2 \). 
Although the number of layers in an {\sl Internetwork} is arbitrary, 
we assumed here that both the {\sl Forward Subnets} 
and the {\sl Backward Subnets} were three-layered (\(L = 3\)); 
in the input and output layers, 
we employed linear neurons described by Eq. (6) with \( \alpha = 1.0 \), 
and in the hidden layer, we used non-linear ones 
whose characteristics were given by Eq. (7) with \( \alpha = 10.0 \). 
In the one-dimensional model here, 
we employed eight hidden neurons in each {\sl Internetwork}. 
We observed the model's dynamics for \( 10^{6} \) seconds 
from the initial state in our simulation experiments, 
and show the results at five intermediate stages in Fig. 4(a). 
The first, second, and third graphs from the top respectively illustrate 
the input \( Z^{<0>} \) and the output \( V^{<0>} \) in {\sl Subspace \#0}, 
the input \( Z^{<1>} \) and the output \( V^{<1>} \) in {\sl Subspace \#1}, 
and the input \( Z^{<2>} \) and the output \( V^{<2>} \) in {\sl Subspace \#2}. 
The second graph from the bottom and the lowermost one 
respectively indicate 
``the input to {\sl Backward Subnet \#2} \hspace{0.50mm} \( \delta^{<2><3>} \) 
and the output of {\sl Backward Subnet \#2} \hspace{0.50mm} \( \delta^{<2><0>} \)" 
and ``the input to {\sl Backward Subnet \#1} \hspace{0.50mm} \( \delta^{<1><3>} \) 
and the output of {\sl Backward Subnet \#1} \hspace{0.50mm} \( \delta^{<1><0>} \)."

At an early stage of learning, as is clear from these graphs, 
the outputs of two {\sl Backward Subnets}, 
\( \delta^{<2><0>} \) and \( \delta^{<1><0>} \), are nearly zero. 
Therefore, {\sl Subspace \#0} is not influenced 
by signals from the {\sl Backward Subnets}, 
and the output of a dynamical neuron \( V^{<0>} \) 
smoothly converges to the input from the outside \( Z^{<0>} \) 
according to the dynamical neuron's time constant. 
In the other {\sl Subspaces} at an early stage 
when learning in {\sl Internetworks} does not make much progress, 
the output signals from two {\sl Forward Subnets}, 
\( V^{<1>} \) and \( V^{<2>} \), are still small, 
and their values are quite different 
from the input signals from the outside, 
\( Z^{<1>} \) and \( Z^{<2>} \). 
This means that input to each {\sl Backward Subnet} 
has a large amplitude 
which is almost the same as the amplitude of input from the outside, 
and such a situation helps the {\sl Internetworks} 
learn the corresponding input-output relationships appropriately. 
However, synaptic connections in the two {\sl Backward Subnets} 
are kept small values at an early stage of learning, 
so the values of the output signals from the two {\sl Backward Subnets}, 
\( \delta^{<2><0>} \) and \( \delta^{<1><0>} \), 
become almost zero as confirmed above. 
This is the reason why 
the output of a dynamical neuron in {\sl Subspace \#0} 
\hspace{0.50mm} \( V^{<0>} \) 
nearly coincides with the input from the outside 
\hspace{0.50mm} \( Z^{<0>} \) 
without any disturbances at an early period of learning.

\begin{figure}[!t]

    \vspace{1.00mm} 
    \hspace*{-5.85mm} 
    \includegraphics[scale=0.76]{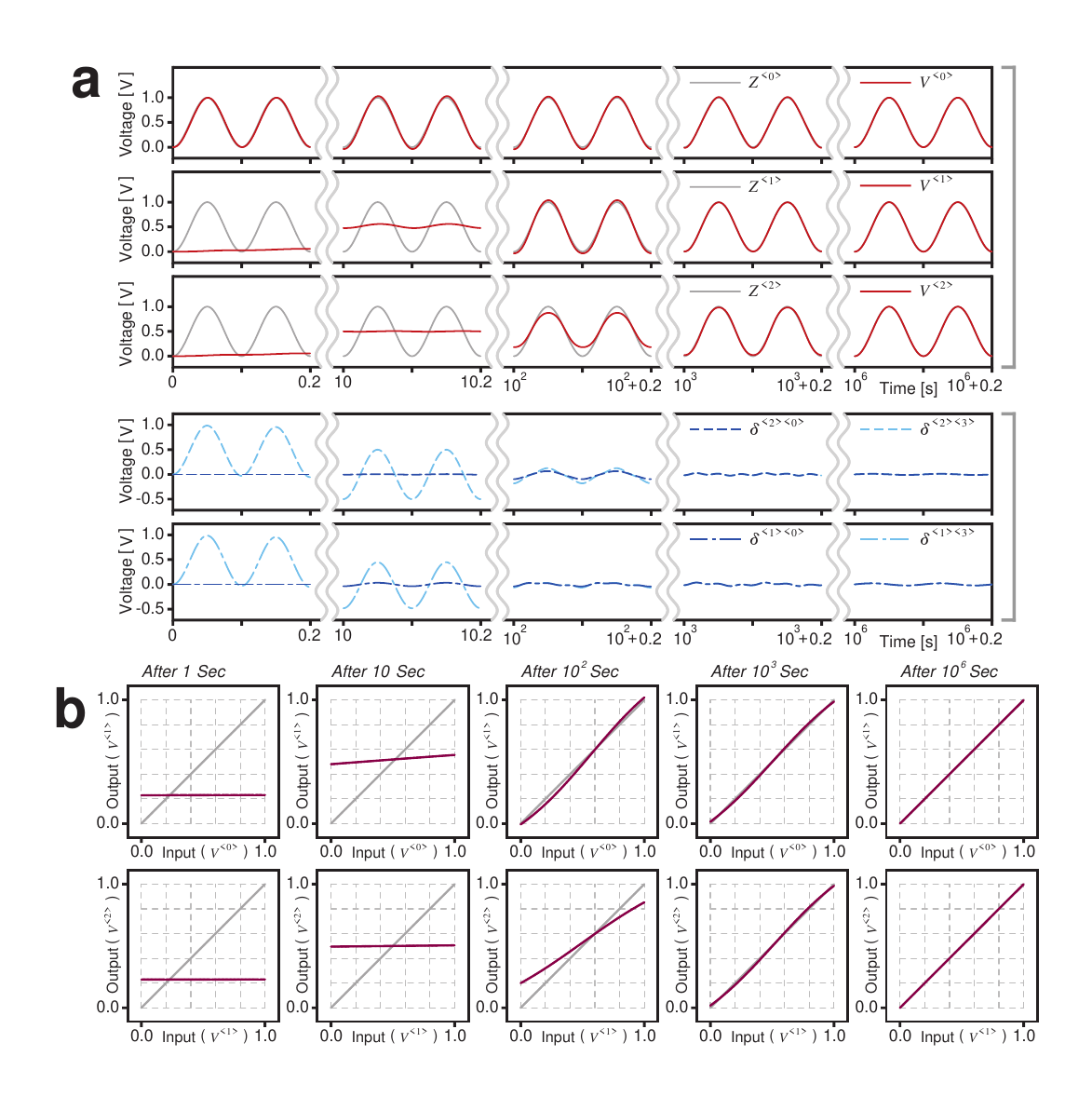}

    \vspace{-4.00mm} 
    \caption{
      Simulation results of the Learning Mode 
      in the one-dimensional model with \( P = 2 \) 
      for the case in which both {\sl Internetworks} {\sl \#1} \& {\sl \#2} 
      obtain linear mapping relationships 
      (Linear \( \rightarrow \) Linear). 
      Frequency of all the periodic input signals is set to 10.0 {\sf Hz}. 
      Model's dynamics were observed for \( 10^{6} \) seconds 
      from the initial state, and the results are 
      separately shown at five intermediate stages. 
      (a) Time-course of network parameters 
      when identical regular sinusoidal waves 
      corresponding to {\sf C0 (= A or E0)} of Fig. 5 
      were applied to all the input ports in {\sl Subspaces} 
      {\sl \#0}, {\sl \#1}, \& {\sl \#2}. 
      (b) Transition of the input-output relationships 
      acquired in the {\sl Internetworks}. 
      The upper and lower graphs respectively illustrate 
      the results for {\sl Forward Subnet} {\sl \#1} 
      and those for {\sl Forward Subnet} {\sl \#2}; 
      both {\sl Forward Subnets} were detached from the model 
      and evaluated. 
    } 

\end{figure}

The modification of plastic connections in the {\sl Internetworks} 
progresses smoothly as time proceeds, 
and an output signal from each {\sl Forward Subnet} 
is finally coincident with an input signal from the outside 
in the corresponding {\sl Subspace}. 
In this regard, some interesting phenomena emerge 
after around \( 10^{2} \) {\sf s} from the beginning. 
Making a comparison between the output of a dynamical neuron \( V^{<0>} \) 
and the input from the outside \( Z^{<0>} \) in {\sl Subspace \#0}, 
the amplitude of \( V^{<0>} \) slightly exceeds that of \( Z^{<0>} \). 
In addition, comparing the output \( V^{<1>} \) with 
the input from the outside \( Z^{<1>} \) in {\sl Subspace \#1}, 
the amplitude of \( V^{<1>} \) is somewhat larger than that of \( Z^{<1>} \). 
Such phenomena were partly confirmed in our basic model with \( P = 1 \); 
it is of great interest that they emerge 
in both \( V^{<0>} \) and \( V^{<1>} \) of the model with \( P = 2 \). 
The outputs of two {\sl Backward Subnets}, 
\( \delta^{<2><0>} \) and \( \delta^{<1><0>} \), 
are applied additively to the corresponding {\sl Subspaces}. 
The slight increases in \( V^{<0>} \) and \( V^{<1>} \) as mentioned above 
practically produce an effect to cancel out those additional signal components 
from the {\sl Backward Subnets} via negative feedback connections.

Figure 4(b) illustrates 
how the {\sl Internetworks} gain 
the input-output relationships over time. 
Specifically, we selected five temporal stages in the same way as Fig. 4(a), 
and evaluated the input-output relationships of {\sl Forward Subnets}, 
each of which was detached from the model. 
The upper graphs show the results for inner {\sl Forward Subnet \#1}, 
and the lower ones for outer {\sl Forward Subnet \#2}.
\footnote{In each of the graphs in Fig. 4(b), 
a one-dimensional goal relationship between input and output 
is depicted with a solid gray line, 
and an evaluation result is illustrated with a solid color line. 
Dotted grid lines with light gray color are reference ones. 
As well as the graphs in Fig. 4(b), 
those in Figs. 6(b) and 7(b) shown later 
are also drawn in this style.} 
Sufficient mapping relationships are not obtained at an early stage. 
However, the learning steadily progresses as time proceeds, 
and linear mapping relationships within the range between \( 0.0 \) and \( 1.0 \) 
are completely attained in the end. 
As is clear from the result after around \( 10^{2} \) {\sf s} 
from the beginning in Fig. 4(a) and Fig. 4(b), 
the learning in inner {\sl Forward Subnet \#1} 
seems to proceed faster than that in outer {\sl Forward Subnet \#2}. 
This does not mean that learning in the former 
is accomplished faster than that in the latter. 
The plastic connections in both of the {\sl Internetworks} 
are continuously tweaked up to the final stage, 
although there are quantitative differences in their modification.

\vspace{3.00mm} 
\noindent 
{\large (2) Non-Linear Mapping}

\vspace{2.50mm} 
\noindent 
In the previous simulation for a one-dimensional model, 
regular sinusoidal waves were equally provided for three input ports, 
each of which was located one by one in a {\sl Subspace}. 
It is quite interesting to study the cases in which 
periodic signals with different waveforms are applied 
to those input ports. 
As shown in Fig. 5, 
we assume the three types of periodic signals 
{\sf (C0 (Ln), C1 (Ct), and C2 (Ep))} 
to which the regular sinusoidal wave {\sf A} is converted 
by employing one of the three types of converting functions 
{\sf (B0 (Ln), B1 (Ct), or B2 (Ep))}; 
the converted waves {\sf (C0 (Ln), C1 (Ct), and C2 (Ep))} 
can further be transformed to the three types of periodic signals 
{\sf (E0 (Ln), E1 (Ct), and E2 (Ep))} 
by using one of the three types of converting functions 
{\sf (D0 (Ln), D1 (Ct), or D2 (Ep))}. 
In the following simulation studies, 
we apply a combined set of these signals 
to the three input ports in the model. 
Among the converting functions, 
both {\sf B0} and {\sf D0} are linear straight lines through the origin; 
therefore, {\sf C0} converted from the non-distorted sinusoidal signal {\sf A} 
by {\sf B0} becomes {\sf A} itself, and 
{\sf E0} transformed from {\sf C0} by {\sf D0} 
is also identical with {\sf C0} (or {\sf A}). 
{\sf B1} (or {\sf D1}) is a converting function 
that is expansive at the edges near \( 0.0 \) and \( 1.0 \) 
but contractive around \( 0.5 \), 
and what the original sine wave {\sf A} (or {\sf C0}) 
is converted to by this {\sf B1} (or {\sf D1}) 
becomes the periodic signal {\sf C1} (or {\sf E1}) 
with an acute sinusoidal waveform. 
In contrast to this, 
the converting function {\sf B2} (or {\sf D2}) is 
contractive at the edges near \( 0.0 \) and \( 1.0 \) 
but expansive around 0.5, 
and what the original sine wave {\sf A} (or {\sf C0}) 
is transformed to by this {\sf B2} (or {\sf D2}) 
becomes the periodic signal {\sf C2} (or {\sf E2}) 
with a sinusoidal waveform rounded at the tip. 
In terms of {\sf B1} (or {\sf D1}) or {\sf B2} (or {\sf D2}), 
we basically employed a non-linear function 
in which a sine wave was superimposed to a straight line y = x; 
specifically, the maximal rising or falling inclination 
of the superimposed sine wave was set to 0.4 or - 0.4 . 
The converted signals {\sf C0}, {\sf C1}, and {\sf C2} 
are the basic elements, 
so we frame them with a dotted line in the figure.

If we define an arbitrary point on {\sf B1} (or {\sf D1}) 
as ( \( u_{1}, v_{1} \) ) and employ \( \sigma \) as a parameter, 
the conversion from the input \( u_{1} \) 
to the output \( v_{1} \) can be expressed 
by the following parametric equation:

\begin{equation} 
\left[ 
\begin{array}{c} 
u_{1} \\ 
\\ 
v_{1} \\
\end{array} 
\right] 
= \left[ 
\begin{array}{c} 
\displaystyle \sigma - \frac{1}{2 \pi} ~\beta ~ \sin{ 2 \pi \sigma} \\ 
\\ 
\displaystyle \sigma + \frac{1}{2 \pi} ~\beta ~ \sin{ 2 \pi \sigma} \\ 
\end{array} 
\right] , 
\end{equation} 

\vspace{1.00mm} 
\hspace{60.00mm} where ~~\( \beta = 0.4 \), ~~\( 0 \leq \sigma \leq 1 \) ~.

\vspace{4.00mm} 
\noindent 
Similarly, if we define an arbitrary point on {\sf B2} (or {\sf D2}) 
as ( \( u_{2}, v_{2} \) ), 
the conversion from the input \( u_{2} \) 
to the output \( v_{2} \) can be written 
as the following parametric equation with \( \sigma \):

\begin{equation} 
\left[ 
\begin{array}{c} 
u_{2} \\ 
\\ 
v_{2} \\
\end{array} 
\right] 
= \left[ 
\begin{array}{c} 
\displaystyle \sigma + \frac{1}{2 \pi} ~\beta ~ \sin{ 2 \pi \sigma} \\ 
\\ 
\displaystyle \sigma - \frac{1}{2 \pi} ~\beta ~ \sin{ 2 \pi \sigma} \\ 
\end{array} 
\right] , 
\end{equation} 

\vspace{1mm} 
\hspace{60mm} where ~~\( \beta = 0.4 \), ~~\( 0 \leq \sigma \leq 1 \) ~.

\vspace{4.00mm} 
\noindent 
Let us provide an additional explanation about the relations 
between the three types of converting functions 
and the three types of converted signals. 
Notice the area from the center to the right edge of Fig. 5. 
For instance, the function {\sf D2 (=B2)} converts {\sf C1} 
(to which {\sf A} is transformed by the function {\sf B1}) 
into {\sf E0 (=C0)} 
that is equivalent to the original sine wave {\sf A}. 
Tracing Eq. (45) and Eq. (46) in this order, 
we can get to know that \( v_{2} =  u_{1} \) 
through the following process:

\begin{equation} 
u_{1} 
\hspace{1mm} \rightarrow \hspace{1mm} 
{\rm Eq. ~(45)}
\hspace{1mm} \rightarrow \hspace{1mm} 
v_{1} 
\hspace{1mm} \rightarrow \hspace{1mm} 
u_{2} ( \stackrel{\bigtriangleup}{=} v_{1} ) 
\hspace{1mm} \rightarrow \hspace{1mm} 
{\rm Eq. ~(46)}
\hspace{1mm} \rightarrow \hspace{1mm} 
v_{2} ( = u_{1} ) ~. 
\end{equation}

\vspace{2.00mm} 
\noindent 
What {\sf C2} 
(to which {\sf A} is transformed by the function {\sf B2}) 
is converted to by the function {\sf D1 (=B1)} 
also becomes {\sf E0 (=C0)}, 
which is equivalent to the original sine wave {\sf A}. 
Tracing Eq. (45) and Eq. (46) in the reverse order, 
we can get to understand that \( v_{1} = u_{2} \) 
through the following process:

\begin{equation} 
u_{2} 
\hspace{1mm} \rightarrow \hspace{1mm} 
{\rm Eq. ~(46)} 
\hspace{1mm} \rightarrow \hspace{1mm} 
v_{2} 
\hspace{1mm} \rightarrow \hspace{1mm} 
u_{1} ( \stackrel{\bigtriangleup}{=} v_{2} ) 
\hspace{1mm} \rightarrow \hspace{1mm} 
{\rm Eq. ~(45)} 
\hspace{1mm} \rightarrow \hspace{1mm} 
v_{1} ( = u_{2} ) ~. 
\end{equation}

\vspace{2.00mm} 
\noindent 
Thus, we can produce a non-distorted sine wave {\sf A} again 
from a deformed sine wave such as 
{\sf C1} (or {\sf E1}) or {\sf C2} (or {\sf E2}) 
by suitably selecting/combining those converting functions. 
We utilize these relations 
between the converting functions and the converted signals 
in the following simulation studies.

\begin{figure}[!t]

    \vspace*{1.0mm} 
    \hspace*{-3.40mm} 
    \includegraphics[scale=0.74]{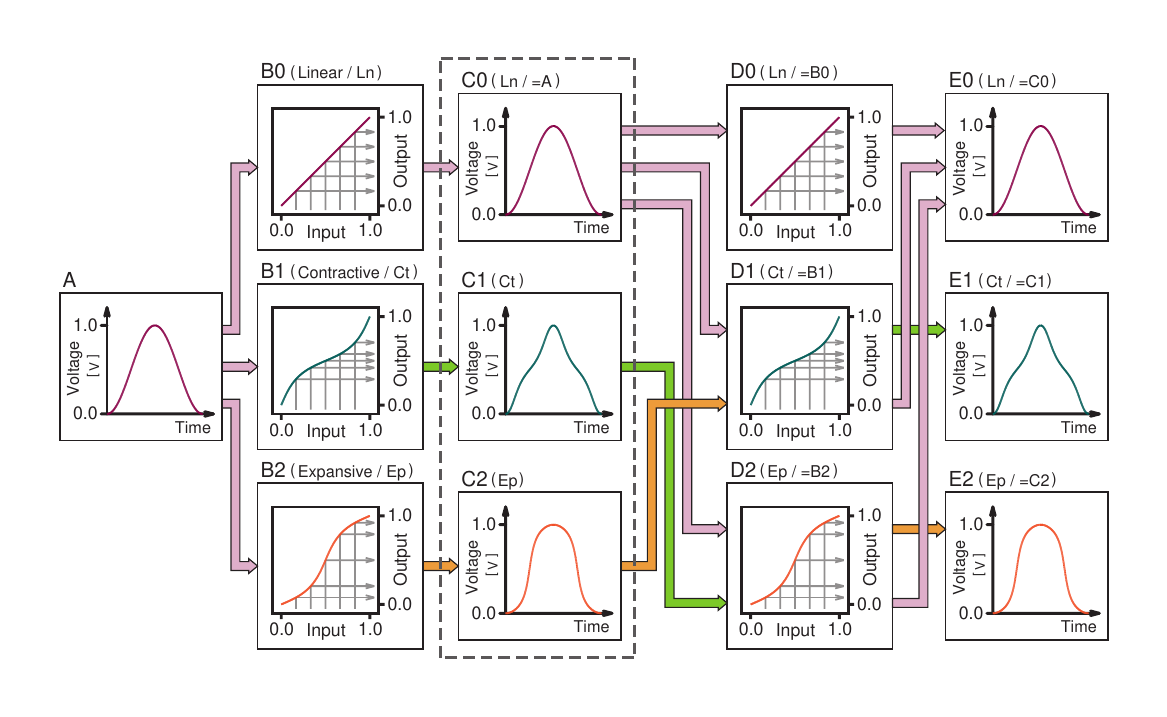}

    \vspace*{-3.00mm} 
    \caption{
      Relation between ``one linear and two non-linear converting functions"
      and ``three sets of sinusoidal and quasi-sinusoidal signals" 
      supposed for evaluating network dynamics in the Learning Mode. 
      The non-distorted regular sinusoidal wave {\sf A} can be converted to 
      the three types of periodic signals 
      {\sf (C0 (Ln), C1 (Ct), and C2 (Ep))} 
      by employing one of the three types of converting functions 
      {\sf (B0 (Ln), B1 (Ct), or B2 (Ep))}. 
      The converted waves {\sf (C0 (Ln), C1 (Ct), and C2 (Ep))} 
      can further be transformed to the three types of periodic signals 
      {\sf (E0 (Ln), E1 (Ct), and E2 (Ep))} 
      by using one of the three types of converting functions 
      {\sf (D0 (Ln), D1 (Ct), or D2 (Ep))}. 
      In the figure, purple, green, and orange colors 
      respectively correspond to 
      Linear (Ln) Case, Contractive (Ct) Case, and Expansive (Ep) Case; 
      for better understanding, we make full use of this coloring 
      also in the other graphs 
      for the mapping relationship of an {\sl Internetwork}. 
      Note that the non-distorted sine wave {\sf A} 
      can be produced again from a deformed sine wave 
      such as {\sf C1} (or {\sf E1}) or {\sf C2} (or {\sf E2}) 
      by fittingly selecting/combining the above-mentioned converting functions; 
      these relations between the converting functions 
      and the converted signals are effectively utilized 
      in simulation studies for the Learning Mode. 
    } 

\end{figure}

Figure 6 illustrates a simulation result in which 
identical non-distorted regular sinusoidal waves 
corresponding to {\sf C0} (= {\sf A}) in Fig. 5 
were applied to the input ports 
in {\sl Subspaces} {\sl \#0} \& {\sl \#2} 
and a quasi-sinusoidal wave 
corresponding to {\sf C1} in Fig. 5 
was put to the input port in {\sl Subspace \#1}. 
Figure 7 is a simulation result in which 
identical quasi-sinusoidal waves 
corresponding to {\sf C1} in Fig. 5 
were given to the input ports 
in {\sl Subspaces} {\#0} \& {\#2} 
and a non-distorted regular sinusoidal wave 
corresponding to {\sf C0} (= {\sf A}) in Fig. 5 
was applied to the input port in {\sl Subspace \#1}. 
The difference from the first simulation experiment shown in Fig. 4 
is only in the waveform of one kind of input signals; 
the frequency of all input signals is also set as \( 10.0 \) {\sf Hz}, 
and the experimental conditions for Figs. 6 and 7 
are common to those in the first experiment 
except for an input signal's waveform. 
Figures 6(a) and 7(a) show the time-courses 
of state variables in the model. 
Figures 6(b) and 7(b) depict the process of growth 
in the input-output relationships of the {\sl Internetworks}; 
in the same way as Fig. 4(b), 
the upper graphs are for inner {\sl Forward Subnet \#1} 
which was set apart from the model and evaluated, 
and the lower ones for detached outer {\sl Forward Subnet \#2}. 
These graphs are illustrated at five temporal stages 
as in the case of Figs. 4(a) and 4(b). 
The varying tendencies of signals in Fig. 6(a) or Fig. 7(a) 
are basically common to those in Fig. 4(a). 
The output signals, 
\( V^{<0>} \), \( V^{<1>} \), and \( V^{<2>} \), 
gradually approach the input ones, 
\( Z^{<0>} \), \( Z^{<1>} \), and \( Z^{<2>} \), 
respectively over time, 
and \( V^{<p>} \) individually correspond to \( Z^{<p>} \) 
in the end, where \( p = 0, 1, 2 \). 
At the same time, these results indicate that 
two {\sl Forward Subnets} in each network 
successfully acquired the expected input-output relationships, 
each of which was coincident with 
one of the converting functions shown in Fig. 5. 
Concretely in Fig. 6(b), 
the mapping relationship of {\sl Forward Subnet \#1} 
corresponds to {\sf B1} in Fig. 5, 
and that of {\sl Forward Subnet \#2} becomes {\sf B2} in Fig. 5. 
In Fig. 7(b), by contrast, 
the mapping relationship of {\sl Forward Subnet \#1} 
overlaps with {\sf B2} in Fig. 5, 
and that of {\sl Forward Subnet \#2} becomes {\sf B1} in Fig. 5.

\begin{figure}[!t]

    \vspace{1.00mm} 
    \hspace*{-5.85mm} 
    \includegraphics[scale=0.76]{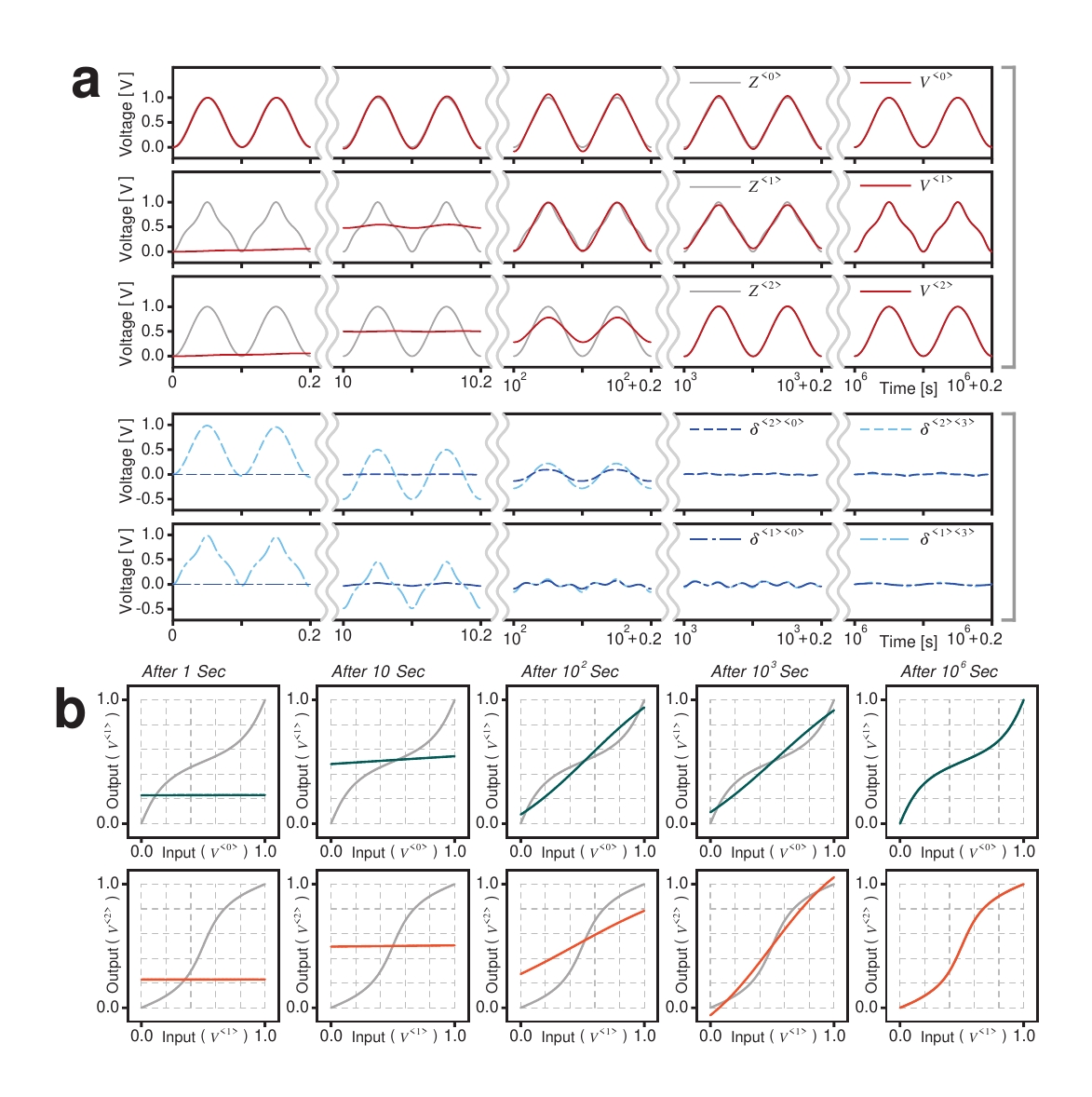}

    \vspace{-4.00mm} 
    \caption{
      Simulation results of the Learning Mode 
      in the one-dimensional model with \( P = 2 \) 
      for the case in which {\sl Internetworks} {\sl \#1} \& {\sl \#2} 
      respectively obtain contractive and expansive mapping relationships 
      (Contractive \( \rightarrow \) Expansive). 
      Frequency of all the periodic input signals is set to 10.0 {\sf Hz}. 
      Model's dynamics were observed for \( 10^{6} \) seconds 
      from the initial state, and the results are 
      separately shown at five intermediate stages. 
      (a) Time-course of network parameters 
      when identical regular sinusoidal waves 
      corresponding to {\sf C0 (= A or E0)} of Fig. 5 were applied 
      to the input ports in {\sl Subspaces} {\sl \#0} \& {\sl \#2} 
      and a quasi-sinusoidal wave 
      corresponding to {\sf C1 (= E1)} of Fig. 5 was put 
      to the input port in {\sl Subspace \#1}. 
      (b) Transition of the input-output relationships 
      acquired in the {\sl Internetworks}. 
      The upper and lower graphs respectively illustrate 
      the results for {\sl Forward Subnet} {\sl \#1} 
      and those for {\sl Forward Subnet} {\sl \#2}; 
      both {\sl Forward Subnets} were detached from the model 
      and evaluated. 
    }

\end{figure}

\begin{figure}[!t]

    \vspace{1.00mm} 
    \hspace*{-5.85mm} 
    \includegraphics[scale=0.76]{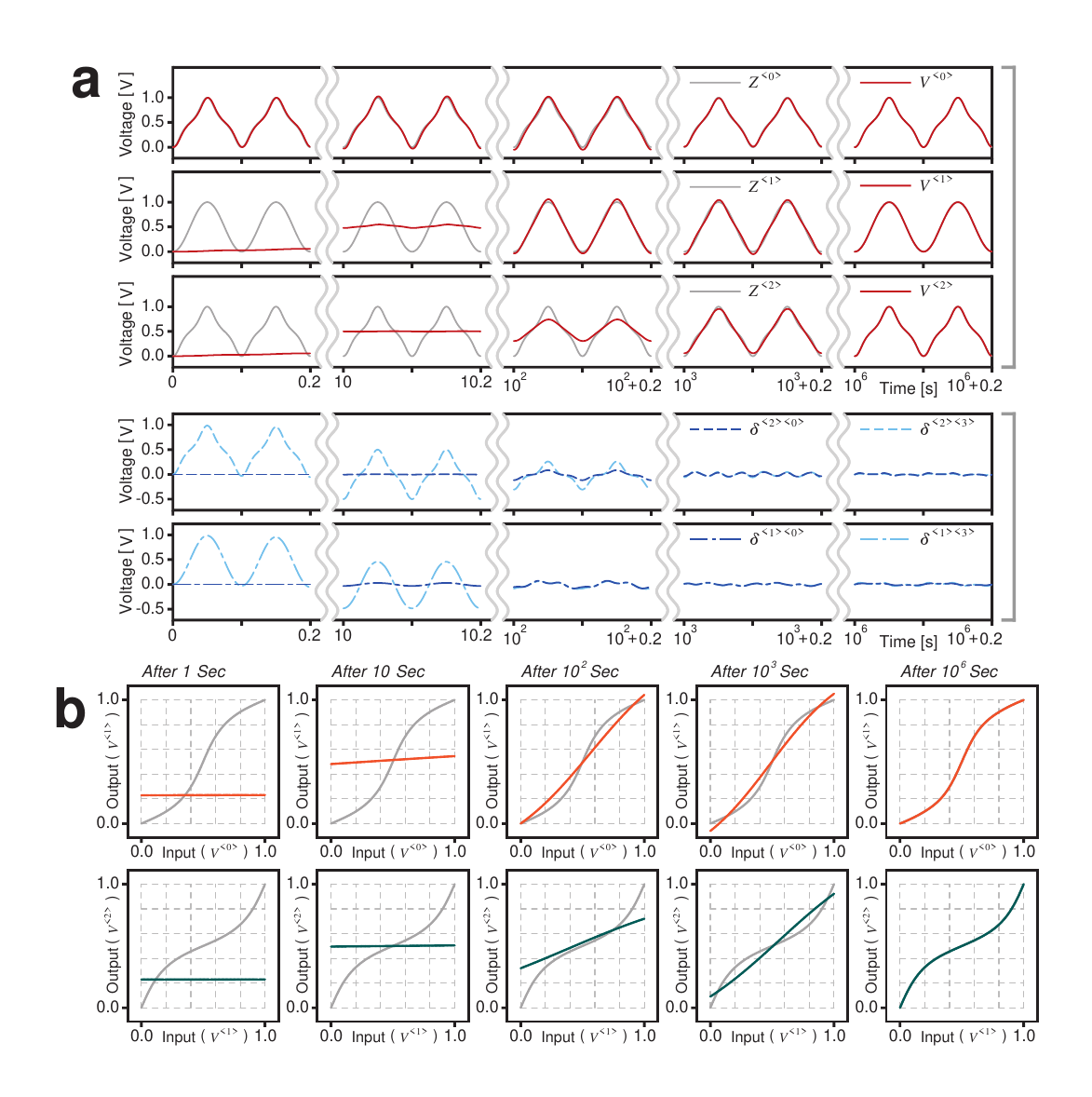}

    \vspace{-4.00mm} 
    \caption{
      Simulation results of the Learning Mode 
      in the one-dimensional model with \( P = 2 \) 
      for the case in which {\sl Internetworks} {\sl \#1} \& {\sl \#2} 
      respectively obtain expansive and contractive mapping relationships 
      (Expansive \( \rightarrow \) Contractive). 
      Frequency of all the periodic input signals is set to 10.0 {\sf Hz}. 
      Model's dynamics were observed for \( 10^{6} \) seconds 
      from the initial state, and the results are 
      separately shown at five intermediate stages. 
      (a) Time-course of network parameters 
      when identical quasi-sinusoidal waves 
      corresponding to {\sf C1 (= E1)} of Fig. 5 were given 
      to the input ports in {\sl Subspaces} {\sl \#0} \& {\sl \#2} 
      and a regular sinusoidal wave 
      corresponding to {\sf C0 (= A or E0)} of Fig. 5 was applied 
      to the input port in {\sl Subspace \#1}. 
      (b) Transition of the input-output relationships 
      acquired in the {\sl Internetworks}. 
      The upper and lower graphs respectively illustrate 
      the results for {\sl Forward Subnet} {\sl \#1} 
      and those for {\sl Forward Subnet} {\sl \#2}; 
      both {\sl Forward Subnets} were detached from the model 
      and evaluated. 
    } 

\end{figure}

When {\sf C2}, {\sf C0}, and {\sf C2} 
(instead of {\sf C0}, {\sf C1}, and {\sf C0}) in Fig. 5 
are given to three input ports of the model, 
the same final results as in Fig. 6 can be obtained; 
if {\sf C0}, {\sf C2}, and {\sf C0} 
(instead of {\sf C1}, {\sf C0}, and {\sf C1}) in Fig. 5 
are applied to three input ports, 
the same final results as in Fig. 7 can be acquired, 
although we omit those figures because of space limitations. 
We can easily understand this situation 
according to the conversion relationship from {\sf C0}, {\sf C1}, and {\sf C2} 
to {\sf E0}, {\sf E1}, and {\sf E2} 
via {\sf D0}, {\sf D1}, and {\sf D2} in Fig. 5.

Carefully looking at the developmental processes 
of the three cases shown in Figs. 4, 6, and 7, 
there are some interesting phenomena. 
For instance, after around \( 10^{2} \) {\sf s} 
from the beginning in Fig. 6(a), 
the amplitude of the output \( V^{<1>} \) 
nearly equals that of the input \( Z^{<1>} \), 
but the amplitude of \( V^{<0>} \) 
is slightly larger than that of \( Z^{<0>} \). 
This means that the range of an input to a {\sl Forward Subnet} 
exceeds the original one covering from \( 0.0 \) to \( 1.0 \); 
it is a phenomenon also observed in Fig. 4(a). 
Then, let us see the state variables after around \( 10^{3} \) {\sf s}. 
In Fig. 6(a), the amplitude of the output \( V^{<1>} \) 
is a little smaller than that of the input \( Z^{<1>} \), 
but the output \( V^{<2>} \) almost overlaps with the input \( Z^{<2>} \). 
At the same second in Fig. 7(a), 
the amplitude of the output \( V^{<0>} \) 
nearly equals that of the input \( Z^{<0>} \), 
but the amplitude of the output \( V^{<1>} \) is a little larger 
than that of the input \( Z^{<1>} \). 
These suggest that each of the input-output relationships 
attained in the corresponding {\sl Forward Subnets} 
must be exceeding the original range from \( 0.0 \) to \( 1.0 \). 
In fact, after around \( 10^{3} \) {\sf s} from the beginning, 
in the lower graph (for {\sl Forward Subnet \#2}) of Fig. 6(b) 
and in the upper graph (for {\sl Forward Subnet \#1}) of Fig. 7(b), 
the output of {\sl Forward Subnet} {\sl \#1} or {\sl \#2} 
exceeds the range between \( 0.0 \) and \( 1.0 \) 
against the input to it in the range between \( 0.0 \) and \( 1.0 \).

In the proposed dynamical model, {\sl Internetworks} are trained 
based on a periodic signal with the ``constant" maximum amplitude 
applied from the outside. 
During that time, however, the maximum amplitudes 
of input signals to the {\sl Internetworks} 
(practically used as both input signals and target ones 
for {\sl Internetwork}'s training) 
are not fixed as shown in Figs. 4(a), 6(a), and 7(a); 
they sometimes go beyond the original range between \( 0.0 \) and \( 1.0 \) 
because of the superimposition of signals from {\sl Backward Subnets}. 
In addition, the maximum amplitudes 
of output signals from {\sl Forward Subnets} 
do not vary monotonously toward the final stage. 
It is worthy of note that they change moment by moment 
as they sometimes increase temporarily and then decrease. 
When we train a ``static" layered neural network, in general, 
we continue to use a set of input and target (teacher) data 
prepared beforehand 
and never exchange it for others during the training. 
That is to say, 
the training of an {\sl Internetwork} in the proposed dynamical model 
is comparable to that of a ``static" layered neural network 
in which a preliminarily determined training set 
(i.e., a set of input and target data) is varied during learning. 
The highly hierarchical and modular architecture 
of the proposed model 
and its dynamics are derived through the process of 
minimizing an energy function. 
Therefore, the above-mentioned distinctive features of the proposed model 
are caused by the search for the steepest descent in an energy field 
due to the cooperation of activity dynamics and learning dynamics; 
to put it the other way around, one can see that 
such a cooperation contributes to 
acceleration and improving efficiency of learning.

Thus, we clarified that, in the generalized model 
with hierarchically-connected multiple {\sl Internetworks} 
as well as in our basic model with only one {\sl Internetwork}, 
various one-dimensional mapping relationships 
were able to be gained in the {\sl Internetworks} 
depending on waveforms of input signals injected from the outside. 
We also confirmed that the variance in their learning speed 
and the synergistic effect of activity dynamics and leaning dynamics 
came out in numerous forms according to circumstances of input signals. 
These might essentially be caused 
by the qualitative and quantitative differences 
between signals propagating through {\sl Backward Subnets}. 
We will specifically discuss this point in later sections.

\subsection{Learning Mode in Two-Dimensional Model}

\vspace{2.00mm} 
\noindent 
{\large (1) Linear Mapping}

\vspace{2.50mm}
\noindent
In this subsection, we consider a two-dimensional model 
based on the results for a one-dimensional model. 
When three identical regular sinusoidal waves 
with the peak-to-peak value of \( 1 \) {\sf V} 
were respectively applied to three input ports 
in a one-dimensional model with \( P = 2 \), 
each of the two {\sl Forward Subnets} 
finally acquired a linear mapping relationship 
that ranged exactly from \( 0.0 \) to \( 1.0 \). 
If six ``identical" sinusoidal waves in total 
are applied to three input ports 
in a two-dimensional model with \( P = 2 \) 
(i.e., two input signals per one input port are necessary 
for the horizontal and vertical axes in each {\sl Subspace}), 
a mapping relationship degenerated merely into a single dimension 
may be obtained in each {\sl Forward Subnet} 
instead of that with a two-dimensional extent. 
Hence, it is a rational decision to first and foremost 
examine the configuration in which 
sinusoidal waves with different frequencies 
are provided to input ports in a two-dimensional model, 
following our previous paper for the basic model.

Figure 8 shows how widely and densely 
two sinusoidal or quasi-sinusoidal waves with different frequencies 
can cover a two-dimensional plane 
when one is employed to express a point on the horizontal axis 
and the other on the vertical axis. 
In terms of gaining a two-dimensional mapping relation, 
evenly and densely covering the entire region of the plane is advisable, 
and this mechanism is similar to the one of a well-known Lissajous curve. 
Figure 8 includes, from the top to the bottom, five examples of 
\( F_{1} = F_{0} \), \( F_{1} = (20/19) \ast F_{0} \), 
\( F_{1} = 2 \ast F_{0} \), \( F_{1} = 5 \ast F_{0} \), 
and \( F_{1} = 20 \ast F_{0} \) as frequency relationships, 
where \( F_{0} \) and \( F_{1} \) are the frequencies 
of signals for the horizontal and vertical axes respectively.
\footnote{We need to take notice of phase relationships 
between signals with \( F_{0} \) and \( F_{1} \), 
but here we only treat the case with no phase shift 
at the initial state for the sake of simplicity. 
For detail about their phase relations, 
refer to Figs. 8, 10, \& 12 and the related descriptions 
in \cite{Tsutsumi2022}.} 
In each row of the leftmost column, 
shown are real periodic signals 
with the three different types of waveforms 
({\sf C0}, {\sf C1}, and {\sf C2} in Fig. 5), 
which are respectively colored in purple, green, and orange. 
The right three columns illustrate Lissajous curves drawn by 
{\sf C0}, {\sf C1}, and {\sf C2} in Fig. 5 
from left to right in turn. 
When two frequencies are slightly different 
like the case of \( F_{1} = (20/19) \ast F_{0} \), 
the entire region of the plane can finely be covered 
as is clearly shown in the figure. 
On the other hand, when \( F_{0} \) and \( F_{1} \) 
have the frequency relation of an integral multiple, 
the coverage area becomes wider and denser 
with the increase of its multiplication ratio. 
It is noted that the influence of frequency relation on the cover area 
depends on how input signals are non-linearly deformed. 
Let us notice the second and bottom rows in Fig. 8.
Thinking of {\sf C0}-based Lissajous curves 
({\sf Ln} / the third column from the right) as the basis, 
denser and sparser regions in {\sf C1}-based Lissajous curves 
({\sf Ct} / the second column from the right) 
are inversely related to those in {\sf C2}-based Lissajous curves 
({\sf Ep} / the rightmost column) by each frequency relation.

\begin{figure}[!t]

    \hspace*{-2.60mm} 
    \includegraphics[scale=0.82]{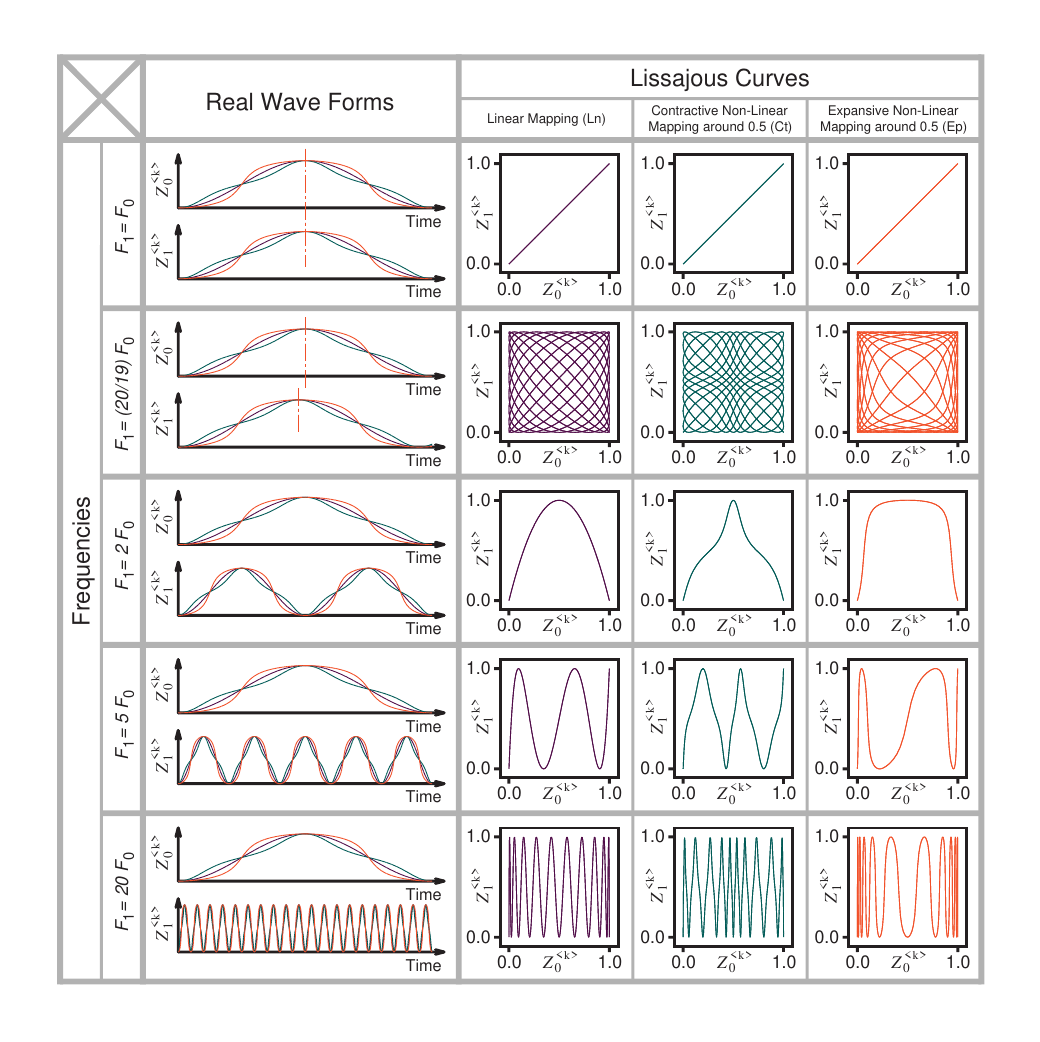}

    \vspace*{-4.00mm} 
    \caption{
      Relationship between two kinds of 
      sinusoidal or quasi-sinusoidal waves with different frequencies. 
      The graphs in the leftmost column indicate real waveforms, 
      and those in the right three columns show 
      how widely and densely two signals for the horizontal and vertical axes 
      can cover a two-dimensional plane 
      regarding Linear (Ln) Case, Contractive (Ct) Case, and Expansive (Ep) Case. 
      From the top row to the bottom one, 
      shown are five examples of 
      \( F_{1} = F_{0} \), \( F_{1} = (20/19) \ast F_{0} \), 
      \( F_{1} = 2 \ast F_{0} \), \( F_{1} = 5 \ast F_{0} \), 
      and \( F_{1} = 20 \ast F_{0} \) 
      as frequency relationships. 
      Also in this figure, 
      purple, green, and orange colors respectively correspond to 
      Linear (Ln) Case, Contractive (Ct) Case, and Expansive (Ep) Case. 
      Note the similarity to a Lissajous curve. 
    } 

\end{figure}

Figure 9 shows an example with \( F_{1} = (20/19) \ast F_{0} \) 
in the Learning Mode; 
a non-distorted regular sinusoidal wave ({\sf C0} in Fig. 5) 
with \( F_{0} = 9.5 \) {\sf Hz} 
is commonly applied to three input ports 
as \( Z^{<0>}_{0} \), \( Z^{<1>}_{0} \), and \( Z^{<2>}_{0} \) 
for the horizontal axes, 
and that with \( F_{1} = 10.0 \) {\sf Hz} 
is also commonly given to three input ports 
as \( Z^{<0>}_{1} \), \( Z^{<1>}_{1} \), and \( Z^{<2>}_{1} \) 
for the vertical axes. 
In a two-dimensional model, 
we put that \( T_{ii} = 1.0 \) {\sf S} 
and \(S_{ii} = 1.0 \) {\sf S}, where \( i = 0, 1 \), 
and set the other \( T \) and \( S \) elements to \( 0.0 \). 
With regard to each {\sl Forward Subnet} or {\sl Backward Subnet}, 
we employed two linear neurons 
respectively in the input and output layers 
and sixteen non-linear neurons in the hidden layer. 
The other conditions are basically common 
to those in the one-dimensional model 
treated in the previous subsection. 
Figure 9(a) depicts the time-course of activity dynamics 
at the signal level. 
Figure 9(b) illustrates the progress of mapping relationships 
developed in the {\sl Internetworks}; 
the upper and lower graphs are the evaluation results 
respectively for {\sl Forward Subnet \#1} 
and {\sl Forward Subnet \#2}, 
each of which was detached from the model and verified.
\footnote{In each of the graphs in Fig. 9(b), 
a two-dimensional goal relationship between input and output 
is drawn with solid gray lines, 
and an evaluation result is depicted with solid color lines. 
As well as the graphs in Figs. 9(b), 
those for the two-dimensional mapping relationship of an {\sl Internetwork} 
in the figures shown later 
are also presented in this style.} 
In Fig. 9(a), the phases of \( Z^{<p>}_{0} \) and \( Z^{<p>}_{1} \) 
from \( 1.0 \) {\sf s} to \( 1.2 \) {\sf s} 
are almost the opposite of 
those from \( 0.0 \) {\sf s} to \( 0.2 \) {\sf s}, 
where \( p = 0, 1, 2 \); 
this is because single periods of the input signals 
with \( 9.5 \) {\sf Hz} and \( 10.0 \) {\sf Hz} 
are \( 2/19 \) {\sf s} and \( 1/10 \) {\sf s} respectively. 
Regarding a two-dimensional model, 
in the same way as a one-dimensional model, 
the activity and learning dynamics smoothly progress 
through a distinctive stage at which 
the amplitudes of \( V^{<0>}_{0} \) and \( V^{<0>}_{1} \) 
exceed those of \( Z^{<0>}_{0} \) and \( Z^{<0>}_{1} \). 
The output signals in all {\sl Subspaces}, 
\( V^{<p>}_{0} \) and \( V^{<p>}_{1} \), 
eventually coincide with the input ones from the outside, 
\( Z^{<p>}_{0} \) and \( Z^{<p>}_{1} \), where \( p = 0, 1, 2 \). 
It is also clear from Fig. 9(b) that 
training goes smoothly over time 
in the two {\sl Internetworks}, 
each of which finally gains 
a two-dimensional linear mapping relationship; 
then, each coordinate component in its mapping relationship 
covers the range exactly from \( 0.0 \) to \( 1.0 \) 
with regard to both input and output.

\begin{figure}[p]

    \vspace*{-4.00mm} 
    \hspace*{11.50mm} 
    \includegraphics[scale=0.58]{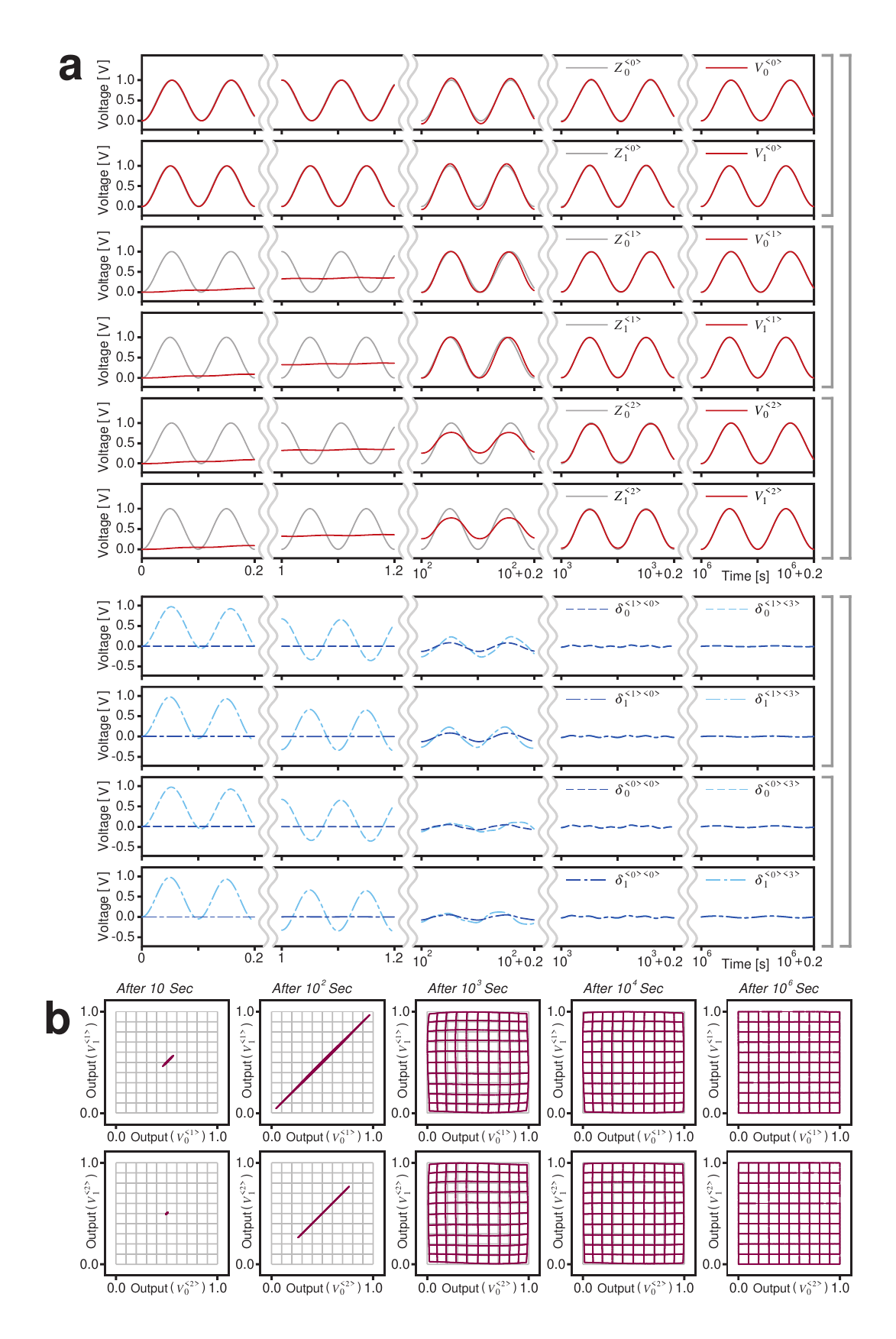}

    \vspace*{-3.00mm} 
    \caption{
      Simulation results of the Learning Mode 
      in the two-dimensional model with \( P = 2 \) 
      for the case in which both {\sl Internetworks} {\sl \#1} \& {\sl \#2} 
      obtain linear mapping relationships 
      (Linear \( \rightarrow \) Linear). 
      The frequency \( F_{0} \) of periodic input signals 
      for the horizontal axes is 9.5 {\sf Hz}, and 
      the frequency \( F_{1} \) of those 
      for the vertical axes is 10.0 {\sf Hz}; 
      the frequency relationship between the two kinds of signals 
      for the horizontal and vertical axes is 
      \( F_{1} = (20/19) * F_{0} \). 
      Model's dynamics were observed for \( 10^{6} \) seconds 
      from the initial state, and the results are 
      separately shown at five intermediate stages. 
      (a) Time-course of network parameters 
      when a set of regular sinusoidal waves 
      (with different frequencies) 
      corresponding to {\sf C0 (= A or E0)} of Fig. 5 was applied 
      to all the input ports in {\sl Subspaces} 
      {\sl \#0}, {\sl \#1}, \& {\sl \#2}. 
      (b) Transition of the input-output relationships 
      acquired in the {\sl Internetworks}. 
      The upper and lower graphs respectively illustrate 
      the results for {\sl Forward Subnet} {\sl \#1} 
      and those for {\sl Forward Subnet} {\sl \#2}; 
      both {\sl Forward Subnets} were detached from the model 
      and evaluated. 
    } 

\end{figure}

Let us take note of network's states after \( 10^{2} \) {\sf s} in Fig. 9(a). 
Judging from the situation in which the amplitudes of 
\( V^{<1>}_{0} \) and \( V^{<1>}_{1} \) become fairly large, 
training of {\sl Internetwork \#1} appears to be already in its final stage. 
However, such comprehension is not correct. 
Watching the waveforms of \( V^{<1>}_{0} \) and \( V^{<1>}_{1} \) prudently, 
\( V^{<1>}_{0} \) goes to the left of \( Z^{<1>}_{0} \) 
and it looks to be moving to a higher frequency; 
in contrast to this signal for the horizontal axis, 
\( V^{<1>}_{1} \) for the vertical axis 
moves to the right of \( Z^{<1>}_{1} \) 
and it seems to be going to a lower frequency. 
Similar phenomena appear in 
the relation between \( V^{<2>}_{0} \) and \( Z^{<2>}_{0} \) 
and that between \( V^{<2>}_{1} \) and \( Z^{<2>}_{1} \), 
although the amplitudes of \( V^{<2>}_{0} \) and \( V^{<2>}_{1} \) 
are not very large at this stage. 
The intervals between zero-crossings remain in original states, 
so their frequencies do not practically change at all. 
However, these behaviors produce an effect 
in which frequencies of two input signals 
for the horizontal and vertical axes are about to mutually get closer. 
Checking the states after \( 10^{2} \) {\sf s} in Fig. 9(b), 
the mapping relationships in both {\sl Internetworks} 
obtained at this moment are not two-dimensional; 
each of them remains almost on a one-dimensional straight line. 
This suggests that only a limited amount of information, 
like a Lissajous curve 
shown in the uppermost row and the third column from the right of Fig. 8 
(i.e., the case of \( F_{1} = F_{0} \) with Linear Mapping), 
is applied to the {\sl Internetworks} at the present stage. 
In general, when a full set of input and target (teacher) data 
necessary for acquirement of a two-dimensional mapping relationship 
is given to a ``static" layered neural network, 
a one-dimensional mapping relationship is first formed in the network 
and then the mapping begins to have a two-dimensional spread 
with the progress of training. 
In the proposed dynamical model with multiple {\sl Internetworks}, 
periodic signals applied to input ports intricately interfere with each other 
while they propagate directly and indirectly through {\sl Backward Subnets}. 
Even in such a complex situation, 
a one-dimensional mapping relationship is temporarily shaped 
and then a two-dimensional one is attained 
in all of the {\sl Internetworks}. 
Although a set of input signals 
with two different but constant frequencies 
were applied from the outside to input ports in all {\sl Subspaces}, 
the waveforms of output signals 
particularly in {\sl Subspaces} {\sl \#1} \& {\sl \#2} 
changed at an early stage, 
as if input signals with a single frequency 
had been apparently provided for {\sl Internetwork}'s training. 
It is of great interest that 
what were observed in our basic model with only one {\sl Internetwork} 
appear also in this generalized model with two (or more) {\sl Internetworks}. 
The fact that such an interesting way of learning is chosen 
reflects the mutual interaction between activity dynamics and learning dynamics 
derived from an energy function through its minimizing procedure.

\vspace{3.00mm} 
\noindent 
{\large (2) Non-Linear Mapping}

\vspace{2.50mm} 
\noindent 
Figure 10 shows a simulation result with \( F_{1} = (20/19) \ast F_{0} \) 
in the Learning Mode under no phase shift at the initial state 
when a combination of non-distorted and distorted 
sinusoidal waves was applied to three input ports, 
each of which was located in a {\sl Subspace} on a one-by-one basis. 
Concretely, {\sf C2}, {\sf C0}, and {\sf C2} in Fig. 5 
with \( 9.5 \) {\sf Hz} 
were respectively applied to the corresponding input ports 
as \( Z^{<0>}_{0} \), \( Z^{<1>}_{0} \), and \( Z^{<2>}_{0} \) 
for the horizontal axes, 
and {\sf C2}, {\sf C0}, and {\sf C2} in Fig. 5 
with \( 10.0 \) {\sf Hz} 
were put correspondingly to the input ports 
as \( Z^{<0>}_{1} \), \( Z^{<1>}_{1} \), and \( Z^{<2>}_{1} \) 
for the vertical axes. 
Figure 10(a) illustrates the time-course 
of activity dynamics at the signal level. 
The upper and lower graphs in Fig. 10(b) respectively 
show the transition of mapping relationships obtained 
in inner {\sl Forward Subnet \#1} and outer {\sl Forward Subnet \#2}, 
each of which was detached from the model and evaluated. 
In both figures, the results at five temporal stages are depicted 
in the same format as Fig. 9. 
As is clear from the periodic signals after \( 10^{2} \) {\sf s} 
from the beginning in Fig. 10(a), 
activity dynamics go through the following stages 
in the same manner as in Fig. 9(a):

\vspace{1.00mm}
\begin{itemize}

  \item[\( \diamondsuit \)] 
  The amplitudes of \( V^{<0>}_{0} \) and \( V^{<0>}_{1} \) 
  become a little greater than those of \( Z^{<0>}_{0} \) and \( Z^{<0>}_{1} \). 

  \item[\( \diamondsuit \)] 
  \( V^{<1>}_{0} \) is about to move to the left side of \( Z^{<1>}_{0} \), 
  that is, to have a higher frequency, 
  whereas \( V^{<1>}_{1} \) is going to the right side of \( Z^{<1>}_{1} \), 
  that is, to take a lower frequency. 

\end{itemize}

\vspace{1.00mm}
\noindent
The outputs in each {\sl Subspace}, \( V^{<p>}_{0} \) and \( V^{<p>}_{1} \), 
finally coincide with the corresponding inputs from the outside, 
\( Z^{<p>}_{0} \) and \( Z^{<p>}_{1} \), where \( p = 0, 1, 2 \). 
Then, {\sl Forward Subnets} {\sl \#1} \& {\sl \#2} 
respectively obtain contractive and expansive 
two-dimensional mapping relationships as expected.

\begin{figure}[p]

    \vspace*{-4.00mm} 
    \hspace*{11.50mm} 
    \includegraphics[scale=0.58]{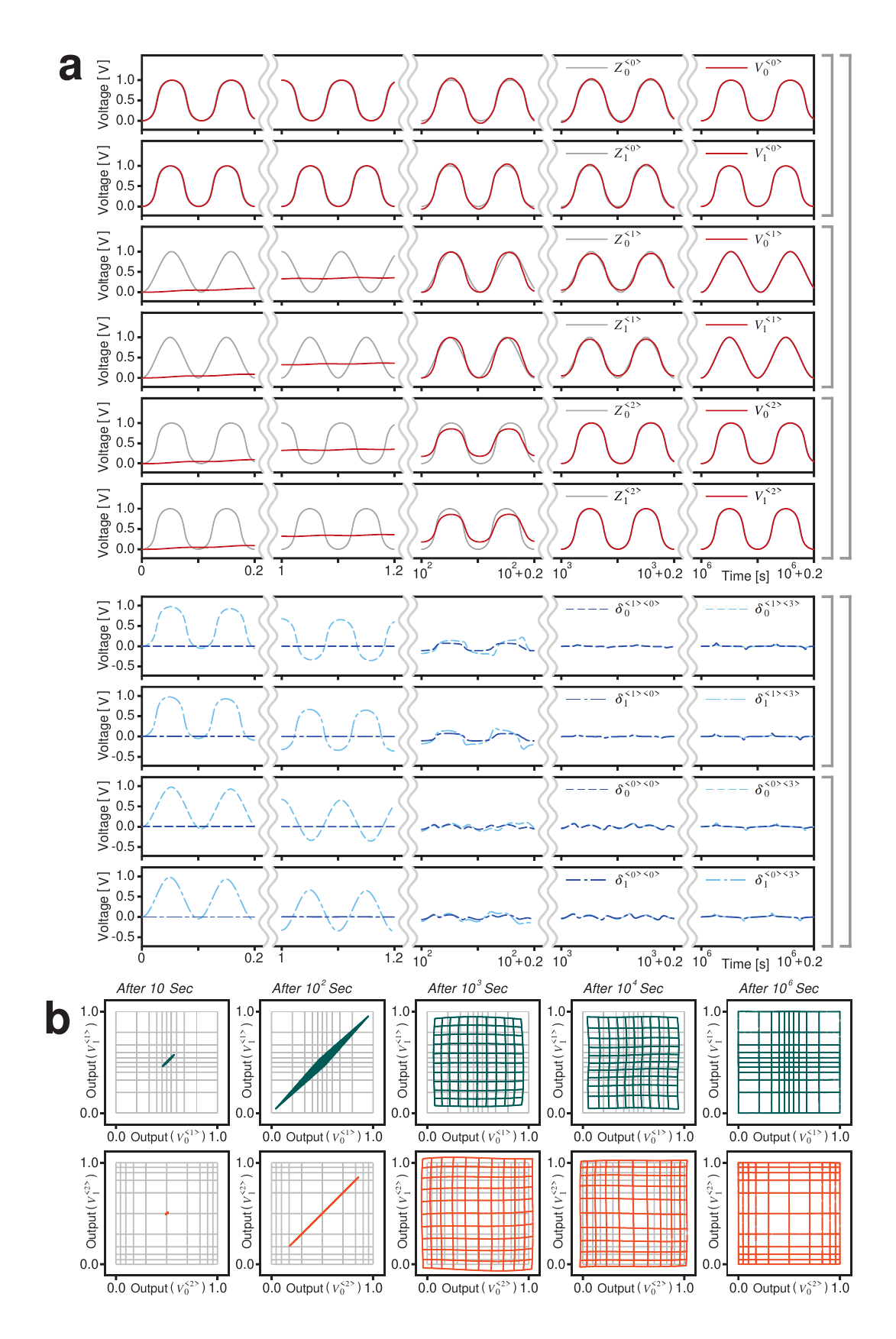}

    \vspace*{-3.00mm} 
    \caption{
      Simulation results of the Learning Mode 
      in the two-dimensional model with \( P = 2 \) 
      for the case in which {\sl Internetworks} {\sl \#1} \& {\sl \#2} 
      respectively obtain contractive and expansive mapping relationships 
      (Contractive \( \rightarrow \) Expansive). 
      The frequency \( F_{0} \) of periodic input signals 
      for the horizontal axes is 9.5 {\sf Hz}, and 
      the frequency \( F_{1} \) of those 
      for the vertical axes is 10.0 {\sf Hz}; 
      the frequency relationship between the two kinds of signals 
      for the horizontal and vertical axes is 
      \( F_{1} = (20/19) * F_{0} \). 
      Model's dynamics were observed for \( 10^{6} \) seconds 
      from the initial state, and the results are 
      separately shown at five intermediate stages. 
      (a) Time-course of network parameters 
      when a set of quasi-sinusoidal waves 
      (with different frequencies) 
      corresponding to {\sf C2 (= E2)} of Fig. 5 was applied 
      to the input ports in {\sl Subspaces} {\sl \#0} \& {\sl \#2}, 
      and a set of regular sinusoidal waves 
      (with different frequencies) 
      corresponding to {\sf C0 (= A or E0)} of Fig. 5 was given 
      to the input port in {\sl Subspaces} {\sl \#1}. 
      (b) Transition of the input-output relationships 
      acquired in the {\sl Internetworks}. 
      The upper and lower graphs respectively illustrate 
      the results for {\sl Forward Subnet} {\sl \#1} 
      and those for {\sl Forward Subnet} {\sl \#2}; 
      both {\sl Forward Subnets} were detached from the model 
      and evaluated. 
    } 

\end{figure}

\begin{figure}[p]

    \vspace*{-4.00mm} 
    \hspace*{11.50mm} 
    \includegraphics[scale=0.58]{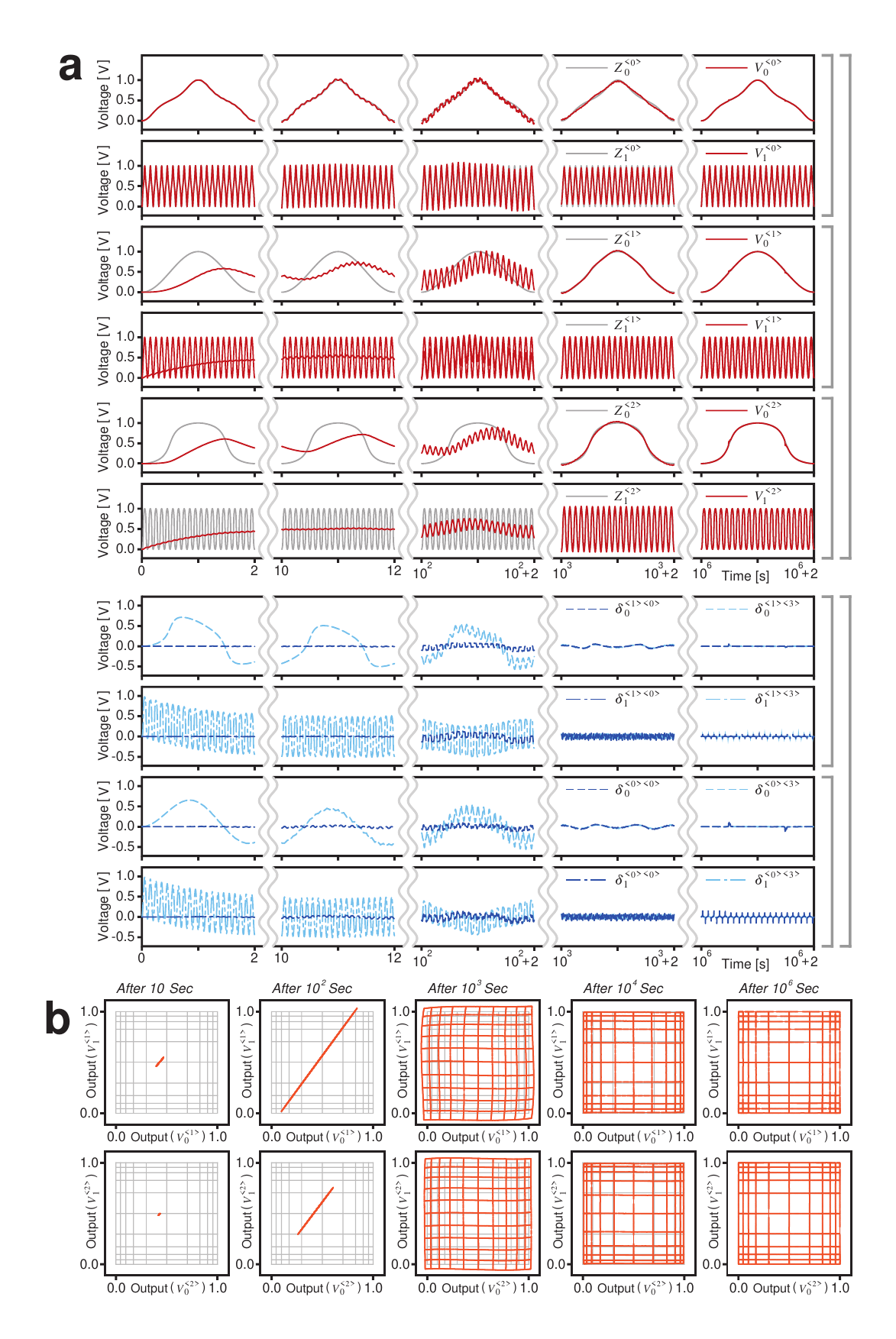}

    \vspace*{-3.00mm} 
    \caption{
      Simulation results of the Learning Mode 
      in the two-dimensional model with \( P = 2 \) 
      for the case in which both {\sl Internetworks} {\sl \#1} \& {\sl \#2} 
      obtain expansive mapping relationships 
      (Expansive \( \rightarrow \) Expansive). 
      The frequency \( F_{0} \) of periodic input signals 
      for the horizontal axes is 0.5 {\sf Hz}, and 
      the frequency \( F_{1} \) of those 
      for the vertical axes is 10.0 {\sf Hz}; 
      the frequency relationship between two kinds of signals 
      for the horizontal and vertical axes is \( F_{1} = 20 * F_{0} \). 
      Model's dynamics were observed for \( 10^{6} \) seconds 
      from the initial state, and the results are 
      separately shown at five intermediate stages. 
      (a) Time-course of network parameters 
      when a set of quasi-sinusoidal waves 
      (with different frequencies) 
      corresponding to {\sf C1 (= E1)} of Fig. 5, 
      a set of regular sinusoidal waves 
      (with different frequencies) 
      corresponding to {\sf C0 (= A or E0)} of Fig. 5, 
      and a set of quasi-sinusoidal waves 
      (with different frequencies) 
      corresponding to {\sf C2 (= E2)} of Fig. 5 
      were respectively applied to the input ports 
      in {\sl Subspaces} {\sl \#0}, {\sl \#1}, \& {\sl \#2}. 
      (b) Transition of the input-output relationships 
      acquired in the {\sl Internetworks}. 
      The upper and lower graphs respectively illustrate 
      the results for {\sl Forward Subnet} {\sl \#1} 
      and those for {\sl Forward Subnet} {\sl \#2}; 
      both {\sl Forward Subnets} were detached from the model 
      and evaluated. 
    } 

\end{figure}

Figure 11 shows the result of a simulation experiment 
in the Learning Mode under the condition that \( F_{1} = 20 * F_{0} \), 
which is vastly different from the frequency relationship 
in the previous experiments for a two-dimensional model; 
we also presume no phase shift between the two kinds of input signals 
at the initial state. 
Specifically, {\sf C1}, {\sf C0}, and {\sf C2} in Fig. 5 
with \( 0.5 \) {\sf Hz} 
were used correspondingly 
as \( Z^{<0>}_{0} \), \( Z^{<1>}_{0} \), and \( Z^{<2>}_{0} \) 
for the horizontal axes, 
and {\sf C1}, {\sf C0}, and {\sf C2} in Fig. 5 
with \( 10.0 \) {\sf Hz} 
were employed respectively 
as \( Z^{<0>}_{1} \), \( Z^{<1>}_{1} \), and \( Z^{<2>}_{1} \) 
for the vertical axes. 
Figure 11(a) shows the time-course of activity dynamics 
at the signal level. 
Although \( F_{1} \) may look very high at first glance of Fig. 11(a) 
compared with Fig. 9(a) and Fig. 10(a), 
note that \( F_{1} \) is commonly set to \( 10.0 \) {\sf Hz} 
and the value of \( F_{0} \) is very small in this example. 
Figure 11(b) depicts how the obtainment of mapping relationships 
in the {\sl Internetworks} progresses; 
the upper and lower graphs are respectively 
for {\sl Forward Subnet \#1} and {\sl Forward Subnet \#2}, 
each of which was detached from the model and evaluated. 
As is clear from the results, 
also when we employ the input signals 
having the relation of \( F_{1} = 20 * F_{0} \), 
the outputs of each {\sl Subspace}, 
\( V^{<p>}_{0} \) and \( V^{<p>}_{1} \), 
ultimately accord with the corresponding inputs, 
\( Z^{<p>}_{0} \) and \( Z^{<p>}_{1} \), 
where \( p = 0, 1, 2 \). 
Then, the expected mapping relationships, 
i.e., the mapping relationships 
that are both two-dimensionally expansive in this case, 
are smoothly gained in the {\sl Internetworks} 
through the process similar to Fig. 9(b) and Fig. 10(b).

Let us notice the lower graph (for {\sl Forward Subnet \#2}) 
in Fig. 10(b) after \( 10^{3} \) {\sf s} from the beginning. 
Each coordinate component 
of the acquired two-dimensional mapping relationship 
is beyond the range from \( 0.0 \) to \( 1.0 \) 
against an input in the range between \( 0.0 \) and \( 1.0 \). 
Looking at the mapping relationships obtained 
after \( 10^{3} \) {\sf s} from the beginning in Fig. 11(b), 
they exceed the range from \( 0.0 \) to \( 1.0 \) 
in both the upper graph (for {\sl Forward Subnet \#1}) 
and the lower one (for {\sl Forward Subnet \#2}). 
These phenomena equally occur 
when a mapping relationship to be trained is expansive. 
They are caused by a large signal propagating through a {\sl Backward Subnet}, 
and also appear commonly in the one-dimensional cases such as 
{\sl Forward Subnet \#2} after  \( 10^{3} \) {\sf s} in Fig. 6(b) 
and {\sl Forward Subnet \#1} after \( 10^{3} \) {\sf s} in Fig. 7(b). 
Even though those interesting phenomena occur on the way of learning, 
two-dimensionally non-linear mapping relationships 
can successfully be obtained between adjacent {\sl Subspaces} 
in the generalized model with multiple {\sl Internetworks} 
depending on how periodic signals applied to input ports are deformed, 
when \( F_{0} \) and \( F_{1} \) have a relation such that 
they are either slightly different (Fig. 9 and Fig. 10) 
or very different (Fig. 11).

If the frequency relationship between \( F_{0} \) and \( F_{1} \) 
takes a low integral multiple 
such as \( F_{1} = 2 * F_{0} \) or \( F_{1} = 5 * F_{0} \) in Fig. 8, 
how do the overall network dynamics develop 
and what sort of final states do they converge on? 
The following is a summary of the results, 
although we omit their specific illustrations due to limitations of space:
\footnote{For more detailed information, 
refer to Section 3 in \cite{Tsutsumi2022}.}

\vspace{1.00mm}
\begin{itemize}

  \item[\( \diamondsuit \)] 
  The value of the energy function 
  defined by Eq. (10) decreases over time, 
  and training of {\sl Internetworks} is finally completed. 
  Then, the outputs of {\sl Subspace \#\(p\)}, 
  \( V^{<p>}_{0} \) and \( V^{<p>}_{1} \), 
  are respectively coincident with the inputs from the outside, 
  \( Z^{<p>}_{0} \) and \( Z^{<p>}_{1} \), 
  where \( p = 0, 1, 2 \). 

  \item[\( \diamondsuit \)] 
  However, perfect two-dimensional mapping relations 
  are not attained in {\sl Forward Subnets} unlike the cases 
  with \( F_{1} = (20/19) * F_{0} \) and \( F_{1} = 20 * F_{0} \). 
  This is because only limited information ineffective 
  for complete learning is provided 
  within the square two-dimensional plane. 

\end{itemize}

\vspace{1.00mm}
\noindent
In any cases when \( F_{0} \) and \( F_{1} \) 
are related with a small integral multiple, 
both the activity dynamics and the learning dynamics 
in and of themselves progress smoothly to a final state, 
but the mapping relationships gained in {\sl Internetworks} 
do not become rich 
even with all the excellent generalization capability 
of a static layered neural network. 
Thus, with respect to training of the proposed dynamical network, 
we can conclude that, 
as suggested by the mechanism of a Lissajous curve, 
preferable is ``only" the situation in which 
the frequencies of two kinds of periodic input signals 
for the horizontal and vertical axes on a coordinate plane 
are either slightly different or very different.

\section{Unconstrained Association Mode}

\subsection{Overview}

\vspace{2.00mm} 
\noindent 
Regarding the proposed model, 
we supposed two modes, Learning and Association, 
in Subsection 2.3; 
they can topologically be distinguished
depending on whether a pair of ``a feedback path" 
and ``an input port connecting with the outside" 
exists in a {\sl Subspace} or not. 
More specifically, 
the Learning Mode has a unique type of architecture 
with full pairs of a feedback path and an input port. 
In contrast to this, 
putting the number of {\sl Internetworks} as \( P \), 
the Association Mode is assumed 
to have \( P + 1 \) types of architectures, 
each of which owns a pair of a feedback path and an input port 
in one of the \( P + 1 \) {\sl Subspaces}.

Within the range of {\sl Subspaces} {\sl \#0} \& {\sl \#1}, 
``the structure and the dynamics of Type \#0 architecture" 
and ``those of Type \#1 architecture" 
are respectively common 
throughout all the models with \( P \geq 1 \), 
as can easily be inferred from an example with \( P = 2 \) 
depicted in Fig. 3.
\footnote{Our basic model presented in the previous paper 
is a special case of the current general model, 
obtained by setting \( P = 1 \). 
For an example with \( P = 1 \), 
refer to Fig. 3 in \cite{Tsutsumi2022}.} 
Taking account of this nature, 
we can reasonably estimate dynamical features 
for Type \#0 and Type \#1 architectures 
in the Association Mode of this generalized model 
with three or more {\sl Subspaces}, 
based on the results of our basic model 
proposed in the previous paper. 
Type \#0 architecture has a pair of 
a feedback path and an input port 
only in {\sl Subspace \#0}. 
In this case, there is no signal 
flowing through {\sl Backward Subnets} in principle, 
and output of {\sl Subspace \#0} 
is established by dynamical neurons 
with a direct feedback loop. 
The output of {\sl Subspace \#0} is 
simply and statically mapped 
in turn to the outer {\sl Subspaces} 
by the corresponding {\sl Forward Subnets}. 
In Type \#1 architecture, on the other hand, 
a pair of a feedback path and an input port 
exists only in {\sl Subspace \#1}. 
The associative dynamics in {\sl Subspaces} {\sl \#0} \& {\sl \#1} 
are determined by the circuitry with a detoured feedback loop 
in which output of dynamical neurons in {\sl Subspace \#0} 
is returned to {\sl Subspace \#0} 
through {\sl Forward Subnet \#1} and {\sl Backward Subnet \#1} in this order. 
Therefore, it is needed to pay attention to 
signals flowing through {\sl Backward Subnet \#1} 
in order to grasp dynamics of Type \#1 architecture 
in the Association Mode. 
In {\sl Backward Subnet \#1}, 
as is clear from Eqs. (15) and (16), 
the products of ``input to the {\sl Backward Subnet}" 
and ``\( \partial V^{<1>}_{i_{1}} / \partial V^{<0>}_{i} \)" 
are summed up for \( i_{1} \). 
In a one-dimensional model, 
what ``input to {\sl Backward Subnet \#1}" is multiplied by 
is simply \( \partial V^{<1>} / \partial V^{<0>} \), 
which means how the mapping relationship 
acquired in the corresponding {\sl Forward Subnet \#1} warps. 
Thus, this computation affects 
the activity dynamics in both {\sl Subspaces} 
in front of and behind the {\sl Forward Subnet}, 
and also influences the convergence speed of an output trajectory 
in each {\sl Subspace}. 
In Type \#1 architecture of the model 
with \( P \geq 2 \) in general, 
the output of {\sl Subspace \#1} is 
simply and statically mapped 
in turn to the outer {\sl Subspaces} 
by the corresponding {\sl Forward Subnets}.

\vspace{2.00mm} 
\subsection{Dynamics}

\vspace{3.00mm} 
\noindent 
{\large (1) Common Points for Simulation Results}

\vspace{3.00mm} 
\noindent 
It is essential to analyze dynamics of the Association Mode 
in the generalized model with \( P \geq 2 \) 
in order to understand the difference with the basic model 
and further grasp the tendency of features 
produced by generalizing the model. 
As stated in the previous subsection, 
we can predict the fundamental dynamics 
of Type \#0 and Type \#1 architectures 
in the model with \( P \geq 2 \) 
from the simulation results in our basic model. 
However, it is absolutely necessary to examine 
associative dynamics of the other architecture such as Type \#2 
with new simulation studies, 
and, even in such a case, it is quite important 
to comprehensively investigate what differences 
with Type \#0 and Type \#1 architectures emerge.
In this subsection, 
we will scrutinize how the generalized model dynamically behaves 
particularly in the Unconstrained Association Mode 
in which a set of fixed values is simply applied 
to a unique input port, 
focusing on the two-dimensional model with \( P = 2 \) 
that is the simplest case of \( P \geq 2 \).

Figures 12, 13, and 14 respectively illustrate experimental results 
for Type \#0, Type \#1, and Type \#2 architectures with \( P = 2 \). 
With regard to each architecture, 
commonly choosing five kinds of combinations 
for the mapping relationships of two {\sl Internetworks}, 
studied for their five networks 
was network behavior starting from various initial states 
ending at the fixed point \( (0.5, 0.5) \) 
given from the outside to a unique input port; 
we show these simulation results as a table with five rows. 
Each set of mapping relationships 
in those five networks of Fig. 12, 13, or 14 
was obtained through the Learning Mode 
in which six (i.e., \( 3 \hspace{0.5mm} {\rm locations} 
\times 2 \hspace{0.5mm} {\rm frequencies} \)) 
input signals with the frequency relation of \( F_{1} = (20/19) * F_{0} \) 
were applied to three two-dimensional input ports.
\footnote{\( F_{0} \) and \( F_{1} \) are respectively 
for the horizontal axis and the vertical axis 
in the same way as stated in Section 3.} 
Specifically,  
the mapping relationships for each network 
depicted from the top row to the bottom row of each table 
respectively correspond to the ones 
obtained after a set of input signals such as 
({\sf C0}, {\sf C0}, {\sf C0}), 
({\sf C2}, {\sf C0}, {\sf C1}),
({\sf C2}, {\sf C0}, {\sf C2}),
({\sf C1}, {\sf C0}, {\sf C1}), or 
({\sf C1}, {\sf C0}, {\sf C2}) 
was applied in sufficient time.
\footnote{Regarding the correspondence 
between a symbol and a waveform, see Fig. 5.} 
For the horizontal direction in each table, 
the leftmost graphs show the outputs of {\sl Subspace \#0}, 
and their right ones are the mapping relationships 
of {\sl Forward Subnet \#1}. 
The next right graphs depict the outputs of {\sl Subspace \#1}, 
and their right ones are the mapping relationships 
of {\sl Forward Subnet \#2}. 
The rightmost graphs illustrate the outputs of {\sl Subspace \#2}. 
In these graphs, small black circles indicate the initial points, 
16 of which are selected for each case; 
all the 16 initial points were arranged in such a manner 
that their positions in {\sl Subspace \#1} 
are common throughout the five networks in each table. 
We choose five trajectories in each graph 
and draw ten arrows at most on each of the trajectories every 0.2 ms; 
a longer interval between the arrows 
denotes a faster change with time.
We appended an index to the upper-right part of each graph 
in order to make referring easier. 
In addition, we drew a block diagram of the whole network 
along with wide gray arrow lines 
in the upper part of each figure 
for ease of understanding the meaning of each graph.

The following properties are also common to Figs. 12, 13, and 14:

\vspace{2.00mm}
\begin{itemize}

  \setlength{\leftskip}{4mm}
  \setlength{\labelsep}{2mm}

  \item[(CM-a)] When an initial point is 
                on the diagonal line \( V^{<p>}_{1} = V^{<p>}_{0} \) or 
                \( V^{<p>}_{1} = - V^{<p>}_{0} + 1.0 \), 
                the network state converges straightforwardly 
                to the center \( (0.5, 0.5) \) 
                since \( V^{<p>}_{0} \) and \( V^{<p>}_{1} \) 
                vary in the same way regardless of 
                mapping relationship's warping in each {\sl Internetwork}, 
                where \( p = 0, 1, 2 \). 

  \item[(CM-b)] When an initial point is on the straight line 
                \( V^{<p>}_{1} = 0.5 \) or \( V^{<p>}_{0} = 0.5 \), 
                the varying tendencies of \( V^{<p>}_{0} \) and \( V^{<p>}_{1} \) 
                are different from each other, where \( p = 0, 1, 2 \). 
                However, since one parameter already stays in a converging state 
                and only the other parameter varies, 
                the network state converges straightforwardly 
                to the center \( (0.5, 0.5) \) also in this case. 

\end{itemize}

\vspace{2.00mm}
\noindent 
Note that, even if a network's output trajectory is straight, 
the speed of the convergence depends on the case. 
This point is important when seeing the graphs shown in this section.

\vspace{4.00mm} 
\noindent 
{\large (2) Type \#0 Architecture}

\vspace{3.00mm} 
\noindent 
Let us look at Fig. 12 in detail. 
In this Association Mode in which a pair of a feedback path 
and an input port exists only in {\sl Subspace \#0}, 
output of dynamical neurons 
with a local feedback loop inside {\sl Subspace \#0} 
directly follows input applied from the outside 
since there is no signal 
flowing through {\sl Backward Subnets} in principle. 
Eventually, the output of {\sl Subspace \#0} 
\hspace{0.5mm} \( V^{<0>}_{i} \) 
draws a straight trajectory; 
it is statically converted in turn 
to \( V^{<1>}_{i} \) and \( V^{<2>}_{i} \) 
by the corresponding {\sl Forward Subnets}, 
where \( i = 0, 1 \). 
As explained in (1) of this subsection, 
we arrange the initial points in {\sl Subspace \#1} 
so that they are common in each of Figs. 12, 13, and 14. 
Because of that, output responses in {\sl Subspace \#0} 
may look different case by case. 
However, note that they are all the same as each other 
throughout the five cases 
in this Fig. 12 for Type \#0 architecture. 
\footnote{In fact, if the initial positions in ``{\sl Subspace \#0}" 
are arranged to be identical in Fig. 12, 
all the five graphs for {\sl Subspace \#0} 
become exactly the same as each other.} 
On the basis of these features, 
we summarize the convergence nature for the five cases in the following.

\vspace{3.00mm}
\begin{itemize}

  \setlength{\leftskip}{4mm}
  \setlength{\labelsep}{2mm}

  \item[(12-S0)] Both of the {\sl Forward Subnets} have 
                 non-warped linear mapping relationships 
%
%
                 ({\tt 12-S0 -MP1} and {\tt 12-S0-MP2}). 
                 For that reason, 
                 outputs of {\sl Subspaces} {\sl \#1} \& {\sl \#2} 
                 ({\tt 12-S0-V<1>} and {\tt 12-S0-V<2>}) are both identical 
                 with an output of {\sl Subspace \#0} ({\tt 12-S0-V<0>}) 
                 in terms of initial positions, straight trajectories, 
                 and convergence behaviors (i.e., intervals between arrows). 

  \item[(12-S1)] Since both of the {\sl Forward Subnets} have 
                 contractive mapping relationships 
%
%
                 ({\tt 12-S1 -MP1} and {\tt 12-S1-MP2}), 
                 the outputs of {\sl Subspace \#1}, 
                 \( V^{<1>}_{0} \) and \( V^{<1>}_{1} \), 
                 converge to the goal point 
                 curving toward the diagonal line 
                 \( V^{<1>}_{1} = V^{<1>}_{0} \) 
                 or \( V^{<1>}_{1} = - V^{<1>}_{0} + 1.0 \) 
                 in all the situations except for (CM-a) and (CM-b). 
                 An output trajectory in the outer {\sl Subspace} 
                 becomes more contractive 
                 based on the mapping relationship of {\sl Forward Subnet \#2}, 
                 and it bends more at the same time, 
                 although the difference in curving between 
                 {\tt 12-S1-V<1>} and {\tt 12-S1-V<2>} 
                 may be a little difficult to distinguish 
                 on account of non-identical initial positions. 

  \item[(12-S2)] Because the mapping relationship of {\sl Forward Subnet \#1} 
                 ({\tt 12-S2-MP1}) is the same 
                 as that in the case right above ({\tt 12-S1-MP1}), 
                 the trajectories of \( V^{<1>}_{0} \) and \( V^{<1>}_{1} \) here 
                 ({\tt 12-S2-V<1>}) 
                 become identical with those in the upper case 
                 ({\tt 12-S1-V<1>}). 
                 On the contrary, outer {\sl Forward Subnet \#2} 
                 has an expansive mapping relationship ({\tt 12-S2-MP2}), 
                 and curving of an output trajectory 
                 generated by this {\sl Forward Subnet \#2} 
                 is opposite to that by inner {\sl Forward Subnet \#1}. 
                 In Type \#0 architecture, 
                 curvings of output trajectories 
                 produced by two {\sl Forward Subnets} 
                 with mutually inverse mapping relationships 
                 seem to be completely canceled out. 
                 As a result, an output trajectory in {\sl Subspace \#2} 
                 ({\tt 12-S2-V<2>}) becomes straight 
                 and it is identical with that in {\sl Subspace \#0} 
                 ({\tt 12-S2-V<0>}). 

  \item[(12-S3)] Inner {\sl Forward Subnet \#1} is expansive ({\tt 12-S3-MP1}) 
                 and outer {\sl Forward Subnet \#2} is contractive ({\tt 12-S3-MP2}); 
                 the mutual relation in the mapping relationships 
                 of the two {\sl Forward Subnets} here 
                 is exactly opposite to that in the case right above 
                 ({\tt 12-S2-MP1} and {\tt 12-S2-MP2}). 
                 Then, \( V^{<1>}_{0} \) and \( V^{<1>}_{1} \) 
                 in all the situations except for (CM-a) and (CM-b) 
                 converge to the goal point with curves 
                 in such a manner that the network state goes away from 
                 the neighbor diagonal line \( V^{<1>}_{1} = V^{<1>}_{0} \) or 
                 \( V^{<1>}_{1} = - V^{<1>}_{0} + 1.0 \). 
                 On the contrary, curving of an output trajectory 
                 generated by outer {\sl Forward Subnet \#2} 
                 with a contractive mapping is opposite 
                 to that by inner {\sl Forward Subnet \#1}. 
                 Judging from the graphs for (12-S3) here, 
                 curvatures of output trajectories 
                 yielded by these two {\sl Forward Subnets} 
                 with mutually inverse mapping relationships 
                 seem to be fully canceled out 
                 in the same way as in the graphs for (12-S2) right above. 
                 Consequently, an output trajectory in {\sl Subspace \#2} 
                 ({\tt 12-S3-V<2>}) is straight 
                 and it becomes the same as that in {\sl Subspace \#0}
                 ({\tt 12-S3-V<0>}). 

  \item[(12-S4)] Since the mapping relationship of {\sl Forward Subnet \#1} 
                 is expansive ({\tt 12-S4-MP1}), 
                 output trajectories in {\sl Subspace \#1} here 
                 ({\tt 12-S4-V<1>}) 
                 are identical with those in the upper case 
                 ({\tt 12-S3-V<1>}). 
                 {\sl Forward Subnet \#2} also has 
                 an expansive mapping relationship ({\tt 12-S4-MP2}), 
                 so output trajectories in {\sl Subspace \#2} 
                 become more expansive 
                 and they curve more at the same time ({\tt 12-S4-V<2>}). 

\end{itemize}

\vspace{2.50mm}
\noindent 
In Type \#0 architecture, the convergence speed with which 
output of {\sl Subspace \#0} goes toward a goal point 
depends simply on the time constant of a dynamical neuron. 
The convergence speeds of output trajectories in the outer {\sl Subspaces} 
are about the same as 
the convergence speed of an output trajectory in {\sl Subspace \#0}, 
while there are some differences in speed 
depending on how each of the relevant {\sl Forward Subnets} 
warps statically.

\begin{figure}[p]

    \vspace*{-4.00mm} 
    \hspace*{-1.50mm} 
    \includegraphics[scale=0.78]{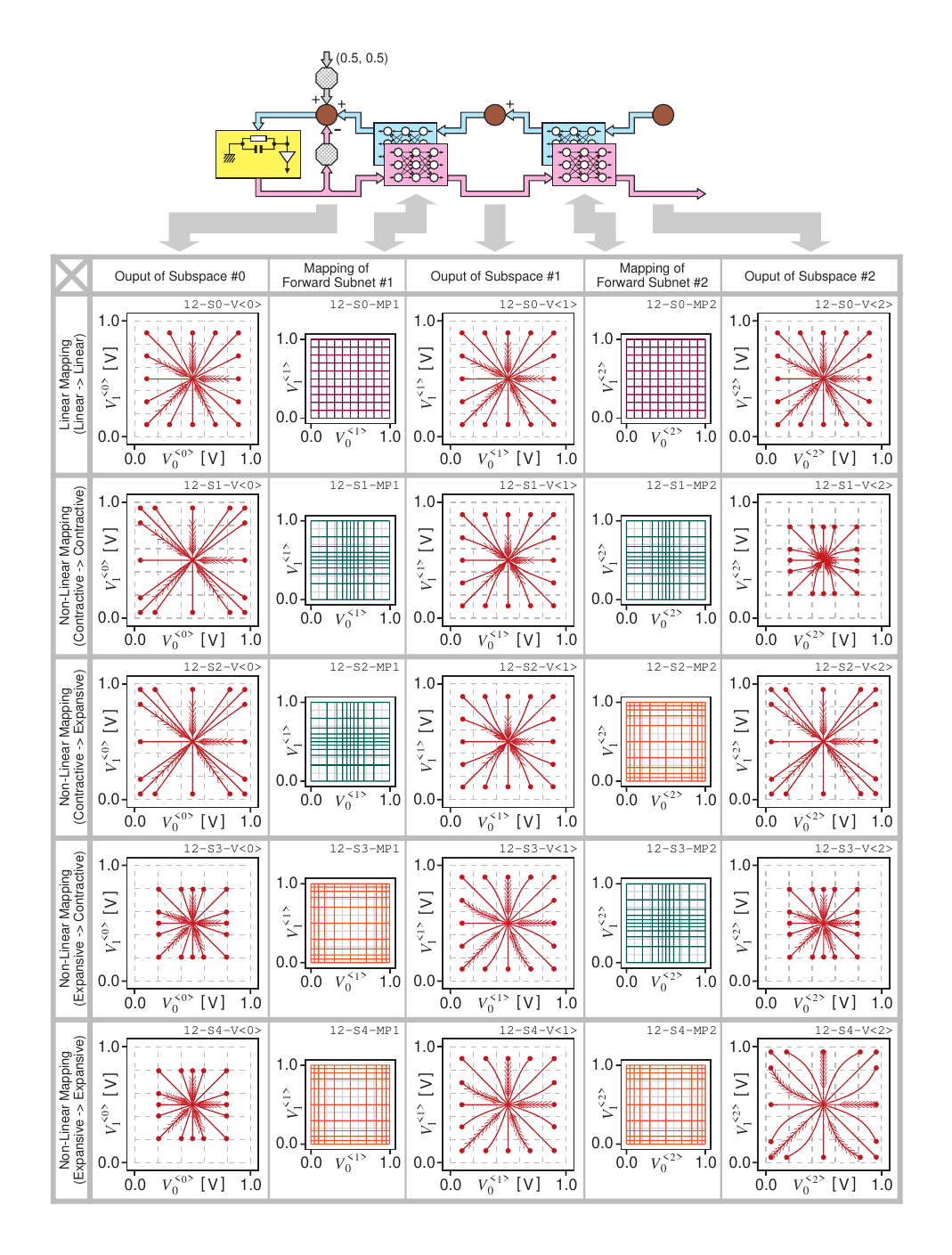}

    \vspace*{-4.00mm} 
    \caption{ 
      Simulation results for Type \#0 architecture 
      in the Unconstrained Association Mode 
      of the two-dimensional model with \( P = 2 \). 
      In this case, the position 
      of a pair of a feedback path and an input port 
      is limited to {\sl Subspace \#0}. 
      Choosing five kinds of combinations 
      for mapping relationships of {\sl Internetworks} {\#1} \& {\#2}, 
      dynamical behavior starting from various initial states 
      is depicted for each network 
      when a set of the fixed values \( (0.5, 0.5) \) 
      was put from the outside to the unique input port. 
      For ease of understanding the meaning of each graph, 
      a block diagram of the whole network 
      with gray wide arrow lines is drawn in the upper part. 
    }

\end{figure}

\begin{figure}[p]

    \vspace*{-4.00mm} 
    \hspace*{-1.50mm} 
    \includegraphics[scale=0.78]{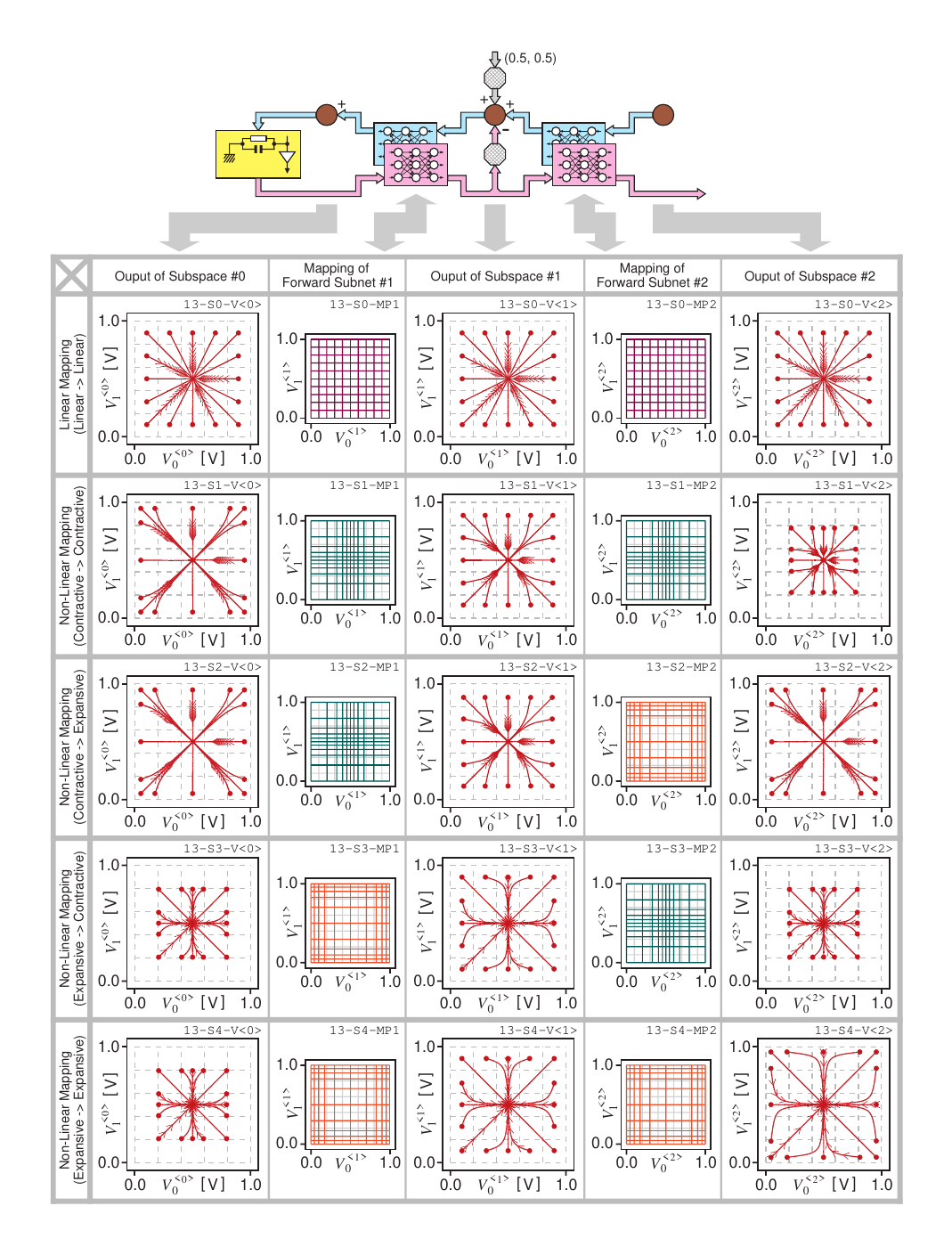}

    \vspace*{-4.00mm} 
    \caption{ 
      Simulation results for Type \#1 architecture 
      in the Unconstrained Association Mode 
      of the two-dimensional model with \( P = 2 \). 
      In this case, the position 
      of a pair of a feedback path and an input port 
      is limited to {\sl Subspace \#1}. 
      Choosing five kinds of combinations 
      for mapping relationships of {\sl Internetworks} {\#1} \& {\#2}, 
      dynamical behavior starting from various initial states 
      is depicted for each network 
      when a set of the fixed values \( (0.5, 0.5) \) 
      was put from the outside to the unique input port. 
      For ease of understanding the meaning of each graph, 
      a block diagram of the whole network 
      with gray wide arrow lines is drawn in the upper part. 
    }

\end{figure}

\begin{figure}[p]

    \vspace*{-4.00mm} 
    \hspace*{-1.50mm} 
    \includegraphics[scale=0.78]{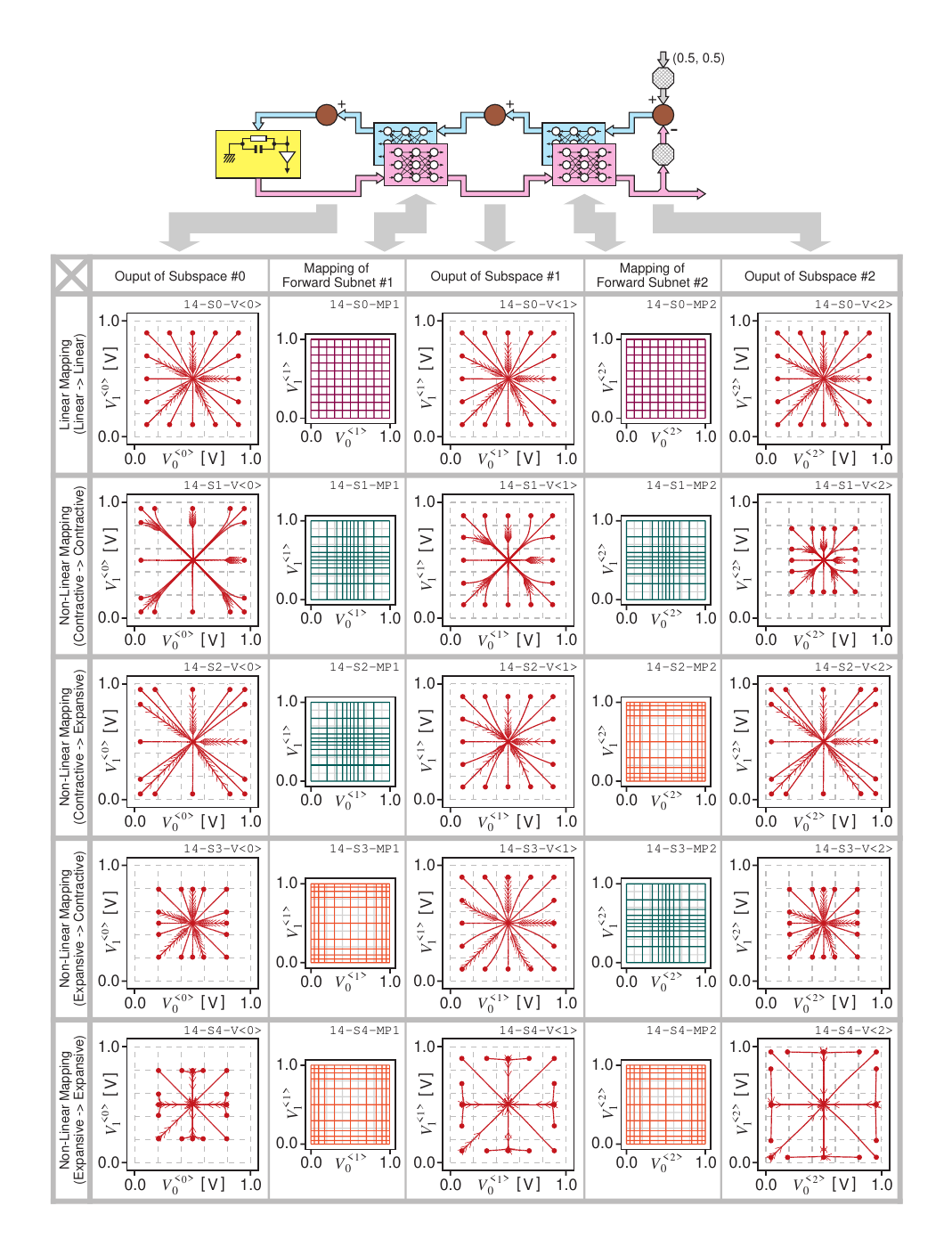}

    \vspace*{-4.00mm} 
    \caption{ 
      Simulation results for Type \#2 architecture 
      in the Unconstrained Association Mode 
      of the two-dimensional model with \( P = 2 \). 
      In this case, the position 
      of a pair of a feedback path and an input port 
      is limited to {\sl Subspace \#2}. 
      Choosing five kinds of combinations 
      for mapping relationships of {\sl Internetworks} {\#1} \& {\#2}, 
      dynamical behavior starting from various initial states 
      is depicted for each network 
      when a set of the fixed values \( (0.5, 0.5) \) 
      was put from the outside to the unique input port. 
      For ease of understanding the meaning of each graph, 
      a block diagram of the whole network 
      with gray wide arrow lines is drawn in the upper part. 
    }

\end{figure}

\vspace{4.00mm} 
\noindent 
{\large (3) Type \#1 Architecture}

\vspace{3.00mm} 
\noindent 
Let us move on to Fig. 13. 
In this Association Mode 
in which a pair of a feedback path and an input port 
exists only in {\sl Subspace \#1}, 
output of {\sl Subspace \#0} is sent 
to {\sl Subspace \#1} by {\sl Forward Subnet \#1}, 
and the variation occurring there is basically fed back 
to {\sl Subspace \#0} through {\sl Backward Subnet \#1}; 
the output of {\sl Subspace \#0} is again transmitted 
to {\sl Subspace \#1} by {\sl Forward Subnet \#1}. 
Due to such a detoured loop, 
there may be a certain level of signals 
flowing through {\sl Backward Subnet \#1} 
depending on the mapping relationship 
of the corresponding {\sl Forward Subnet \#1}, 
and then the outputs of {\sl Subspace \#0}, 
\( V^{<0>}_{0} \) and \( V^{<0>}_{1} \), 
are affected by the outputs of {\sl Subspace \#1}, 
\( V^{<1>}_{0} \) and \( V^{<1>}_{1} \). 
In contrast to the circumstances 
in {\sl Subspaces} {\sl \#0} \& {\sl \#1}, 
what the outputs of {\sl Subspace \#1} 
were statically converted to by {\sl Forward Subnet \#2} 
appear in {\sl Subspace \#2} 
as \( V^{<2>}_{0} \) and \( V^{<2>}_{1}\). 
Confirming these points, 
we will show the convergence properties for the five cases below.

\vspace{3.00mm}
\begin{itemize}

  \setlength{\leftskip}{4mm}
  \setlength{\labelsep}{2mm}

  \item[(13-S0)] In the uppermost row, the difference between 
                 ``output of non-warped linear {\sl Forward Subnet \#1}" 
                 and ``input from the outside" 
                 is computationally processed in {\sl Subspace \#1}, 
                 and it flows through the corresponding {\sl Backward Subnet \#1}
                 into {\sl Subspace \#0}. 
                 The effect that this detoured signal brings about 
                 in {\sl Subspace \#0} is identical with 
                 that by the direct feedback loop inside {\sl Subspace \#0}. 
                 Therefore, the output of {\sl Subspace \#1} here 
                 ({\tt 13-S0-V<1>}) is the same as 
                 that of {\sl Subspace \#0} in Fig. 12 
                 ({\tt 12-S0-V<0>}). 
                 At the same time, the output of {\sl Subspace \#1} 
                 becomes identical with 
                 that of {\sl Subspace \#0} ({\tt 13-S0-V<0>}) 
                 before passing through {\sl Forward Subnet \#1}, 
                 and it is further sent to {\sl Subspace \#2} as it is 
                 by non-warped linear {\sl Forward Subnet \#2} 
                 ({\tt 13-S0-V<2>}). 
                 As with the uppermost case in Fig. 12, 
                 the outputs of all {\sl Subspaces} here 
                 eventually become the same as each other 
                 in terms of initial positions, straight trajectories, 
                 and convergence behaviors (i.e., intervals between arrows). 

  \item[(13-S1)] Since the mapping relationships of both {\sl Forward Subnets} 
                 are contractive, an output of {\sl Subspace \#0} 
                 is basically reduced in size and conveyed to {\sl Subspace \#1}, 
                 whose output is further sent to {\sl Subspace \#2} 
                 in a static manner as an even smaller one. 
                 This situation is similar to 
                 that in the graphs for (12-S1) of Type \#0 architecture.  
                 However, there are clear differences 
                 in {\tt 13-S1-V<0>} here and {\tt 12-S1-V<0>} in Fig. 12. 
                 The output in the latter draws a straight trajectory, 
                 while the output in the former 
                 converges to the goal point with curves 
                 approaching the diagonal line \( V^{<0>}_{1} = V^{<0>}_{0} \) or 
                 \( V^{<0>}_{1} = - V^{<0>}_{0} + 1.0 \) in all the situations 
                 except for the cases of (CM-a) and (CM-b), 
                 owing to the detoured feedback loop 
                 via contractive {\sl Internetwork \#1}. 
                 With regard to {\sl Subspace \#1}, 
                 comparing {\tt 13-S1-V<1>} here 
                 with {\tt 12-S1-V<1>} in Fig. 12, 
                 we can confirm that curvature of an output trajectory 
                 in the former is greater than that in the latter; 
                 this is also because of the detoured feedback loop 
                 with non-linear {\sl Internetwork \#1} 
                 in the former case. 
                 Output trajectories in {\sl Subspace \#2} ({\tt 13-S1-V<2>}) 
                 curve more than those in {\sl Subspace \#1} 
                 due to contractive {\sl Forward Subnet \#2}, 
                 although these are slightly indistinguishable 
                 because of different initial positions in each {\sl Subspace}. 
                 Since the mapping relationship of {\sl Forward Subnet \#1} 
                 is contractive, the strength of signals flowing 
                 through the corresponding {\sl Backward Subnet \#1} 
                 becomes smaller and smaller 
                 as output of {\sl Forward Subnet \#1} approaches the center; 
                 then not fully produced is an effect 
                 that brings the difference between 
                 ``output of the {\sl Forward Subnet}" 
                 and ``input from the outside" in {\sl Subspace \#1} 
                 to the inner {\sl Subspace} 
                 and reduces it there. 
                 As a result, 
                 the outputs of {\sl Subspaces} {\sl \#0} \& {\sl \#1} 
                 both approach their respective goal points very slowly, 
                 and the time necessary for reaching the center 
                 becomes very large. 
                 Comparing the graphs for this (13-S1) 
                 with those for previous (12-S1) on the whole, 
                 we can see, in the former, 
                 that the intervals between small subsidiary arrows 
                 on output trajectories in every {\sl Subspace} 
                 become much narrower at a stage 
                 when the network's state is still far from the center. 

  \item[(13-S2)] The mapping relationship 
                 between {\sl Subspaces} {\sl \#0} \& {\sl \#1} 
                 is the same as that in the upper case, 
                 whereas {\sl Forward Subnet \#2} here is expansive 
                 and curving of an output trajectory by this mapping 
                 is opposite to that by inner {\sl Forward Subnet \#1}. 
                 In Type \#1 architecture, as already stated, 
                 output trajectories in {\sl Subspace \#0 \& \#1} 
                 ({\tt 13-S2-V<0>} and {\tt 13-S2-V<1>}) 
                 are determined based on the detoured feedback loop 
                 via {\sl Internetwork \#1}. 
                 Differently, an output trajectory in {\sl Subspace \#2} 
                 ({\tt 13-S2-V<2>}) becomes what was only statically converted to 
                 from that in {\sl Subspace \#1} 
                 based on the mapping relationship of {\sl Forward Subnet \#2}. 
                 In this fashion, the generating processes 
                 of output trajectories in the three {\sl Subspaces} 
                 are different from each other, 
                 but trajectory's curvatures caused by signals 
                 passing through the two {\sl Forward Subnets} 
                 with mutually inverse mapping relationships 
                 seem to be completely canceled out. 
                 As a matter of fact, 
                 an output trajectory in {\sl Subspace \#2} 
                 is identical with that in {\sl Subspace \#0}. 
                 Judging from the intervals between small subsidiary arrows 
                 on output trajectories in every {\sl Subspace}, 
                 the convergence speeds in the graphs here are slow overall 
                 in the same way as those in the upper row graphs. 
                 This suggests that dynamics based on the detoured feedback loop 
                 via contractive {\sl Internetwork \#1} are dominant 
                 in this case. 
                 It is also interesting that 
                 the convergence speed in {\sl Subspace \#2} 
                 is identical with that in {\sl Subspace \#0}. 

  \item[(13-S3)] The mapping relationship of {\sl Forward Subnet \#1} is expansive, 
                 and that of {\sl Forward Subnet \#2} is contractive; 
                 their mutual relation is opposite to 
                 that in the upper row. 
                 Output trajectories in {\sl Subspace \#1} ({\tt 13-S3-V<1>}) 
                 have a tendency to move in a direction away from 
                 the diagonal line \( V^{<1>}_{1} = V^{<1>}_{0} \) or 
                 \( V^{<1>}_{1} = - V^{<1>}_{0} + 1.0 \) 
                 in all the situations except for (CM-a) and (CM-b). 
                 This property is similar to 
                 that of the graph for {\sl Subspace \#1} 
                 in the fourth row from the top of Fig. 12, 
                 even though curvature in {\tt 13-S3-V<1>} 
                 is a little larger than that in {\tt 12-S3-V<1>}. 
                 By contrast, for {\sl Subspace \#0}, 
                 there is a big difference between 
                 {\tt 13-S3-V<0>} here and {\tt 12-S3-V<0>} in Fig. 12; 
                 the latter's trajectories are straight, 
                 and the former's ones curve in the same direction 
                 as those in {\sl Subspace \#1} 
                 in all the situations except for (CM-a) and (CM-b). 
                 Since {\sl Forward Subnet \#1} is expansive, 
                 a signal back-propagating through 
                 the corresponding {\sl Backward Subnet \#1} 
                 keeps a certain level even near the center, 
                 and it produces a fairly strong effect that makes 
                 the difference between ``an output of {\sl Forward Subnet \#1}"
                 and ``an input from the outside" in {\sl Subspace \#1} 
                 decrease in {\sl Subspace \#0}. 
                 Eventually, the network states 
                 in {\sl Subspaces} {\sl \#0} \& {\sl \#1} 
                 converge very quickly toward the center. 
                 Intervals between small subsidiary arrows 
                 in {\tt 13-S3-V<0>} and {\tt 13-S3-V<1>} here 
                 are larger than those 
                 in {\tt 12-S3-V<0>} and {\tt 12-S3-V<1>} in Fig. 12 
                 by each {\sl Subspace}, 
                 so we can see, also from this comparison, 
                 that dynamics based on the detoured feedback loop 
                 via expansive {\sl Internetwork \#1} 
                 contribute to the convergence speed's raising  
                 within {\sl Subspaces} {\sl \#0} \& {\sl \#1}. 
                 On the other hand, outer {\sl Forward Subnet \#2} is contractive, 
                 and curving of an output trajectory by this mapping 
                 is opposite to that by inner {\sl Forward Subnet \#1}. 
                 As with the graphs for (13-S2) right above, 
                 trajectory's curvings caused by signals 
                 flowing through two {\sl Forward Subnets} 
                 with mutually inverse mapping relationships 
                 seem to be entirely canceled out, 
                 although the generating processes of output trajectories 
                 in the three {\sl Subspaces} are different from each other 
                 in this case too. 
                 In fact, an output trajectory in {\sl Subspace \#2} 
                 is identical with that in {\sl Subspace \#0}. 
                 The convergence speed is high in every {\sl Subspace}. 
                 It is quite interesting that 
                 the convergence speed in {\sl Subspace \#2} 
                 is exactly the same as that in {\sl Subspace \#0} 
                 also in this case. 

  \item[(13-S4)] Within the range between {\sl Subspaces} {\sl \#0} \& {\sl \#1}, 
                 {\tt 13-S4-V<0>} and {\tt 13-S4-V<1>} here 
                 are identical with 
                 {\tt 13-S3-V<0>} and {\tt 13-S3-V<1>} right above, 
                 since, in both cases, 
                 the dynamics in their {\sl Subspaces} 
                 are based on Type \#1 architecture 
                 with expansive {\sl Internetwork \#1}. 
                 From another perspective, 
                 if we contrast {\tt 13-S4-V<0>} and {\tt 13-S4-V<1>} here 
                 (Type \#1 architecture) 
                 with {\tt 12-S4-V<0>} and {\tt 12-S4-V<1>} in Fig. 12 
                 (Type \#0 architecture), 
                 they are very dissimilar from each other, 
                 since {\sl Forward Subnet \#1}'s mapping relationships 
                 are common in both cases 
                 but the architectures differ from each other. 
                 Focusing on {\sl Subspace \#0}, 
                 output trajectories here ({\tt 13-S4-V<0>}) 
                 curve in such a manner that 
                 the network state keeps away from 
                 the diagonal line \( V^{<0>}_{1} = V^{<0>}_{0} \) 
                 or \( V^{<0>}_{1} = - V^{<0>}_{0} + 1.0 \) 
                 in all the situations except for (CM-a) and (CM-b), 
                 whereas those in Fig. 12 ({\tt 12-S4-V<0>}) 
                 are all straight. 
                 As for {\sl Subspace \#1}, 
                 curvings of output trajectories in {\tt 13-S4-V<1>} here 
                 (based on the detoured loop via {\sl Internetwork \#1})
                 are greater than those in {\tt 12-S4-V<1>} in Fig. 12 
                 (based on ``the direct feedback loop inside {\sl Subspace \#0}" 
                 and ``the static mapping by {\sl Forward Subnet \#1}"). 
                 Output trajectories in {\sl Subspace \#2} ({\tt 13-S4-V<2>}) 
                 curve more than those in {\sl Subspace \#1} ({\tt 13-S4-V<1>}) 
                 owing to expansive {\sl Forward Subnet \#2}; 
                 at the same time, they bend more than 
                 those in {\sl Subspace \#2} of Fig. 12 ({\tt 12-S4-V<2>}). 
                 When {\sl Forward Subnet \#1} has an expansive mapping relationship 
                 in Type \#1 architecture, 
                 the convergence speeds of outputs 
                 of {\sl Subspaces} {\sl \#0} \& {\sl \#1} going toward the center 
                 are fairly high 
                 in the same way as those in the graphs for (13-S3) right above. 
                 Making a comparison between 
                 the lowest graphs in this Fig. 13 and those in Fig. 12 
                 on the whole, 
                 intervals between small subsidiary arrows 
                 in all {\sl Subspaces} of the former 
                 are wider than those of the latter 
                 already at an early stage, 
                 and this aspect also stands for the overall fast convergence 
                 in Type \#1 architecture 
                 with expansive {\sl Internetworks} {\sl \#1} \& {\sl \#2}. 

\end{itemize}

\vspace{4.00mm} 
\noindent 
{\large (4) Type \#2 Architecture}

\vspace{3.00mm} 
\noindent 
At the end of the Unconstrained Association Mode, 
let us look at Fig. 14. 
In this architecture 
in which a pair of a feedback path and an input port 
exists only in {\sl Subspace \#2}, 
there may be a certain level of signals 
flowing through {\sl Backward Subnets} 
{\sl \#2} \& {\sl \#1} in this order 
depending on how the mapping relationships of 
the corresponding {\sl Forward Subnets} are, 
and then the outputs of {\sl Subspace \#0},
\( V^{<0>}_{0} \) and \( V^{<0>}_{1} \), 
and those of {\sl Subspace \#1}, 
\( V^{<1>}_{0} \) and \( V^{<1>}_{1} \), 
are affected by those of {\sl Subspace \#2},
\( V^{<2>}_{0} \) and \( V^{<2>}_{1} \). 
Figuring out this specific situation for Type \#2 architecture, 
we will describe convergence properties for the five cases below.

\vspace{3.00mm}
\begin{itemize}

  \setlength{\leftskip}{4mm}
  \setlength{\labelsep}{2mm}

  \item[(14-S0)] When both {\sl Forward Subnets} 
                 have non-warped linear mapping relationships, 
                 the difference between ``output of {\sl Forward Subnet \#2}" 
                 and ``input from the outside" in {\sl Subspace \#2} 
                 flows through two {\sl Backward Subnets} 
                 into {\sl Subspace \#0} as it is. 
                 So output trajectories in every {\sl Subspace} 
                 are fully the same as each other. 
                 Moreover, when all {\sl Internetworks} have 
                 non-warped linear mapping relationships, 
                 as is clear from the comparison of 
                 the top row graphs in Figs. 12, 13, and 14, 
                 the results for every architecture 
                 in the Unconstrained Association Mode 
                 become identical with each other from any point of view 
                 such as initial positions, straight trajectories, 
                 and convergence behaviors (i.e., intervals between arrows). 

  \item[(14-S1)] Both {\sl Forward Subnets} have 
                 contractive mapping relationships, 
                 so an output of {\sl Subspace \#0} 
                 is basically reduced in size and sent to {\sl Subspace \#1}, 
                 whose output is further transferred to {\sl Subspace \#2} 
                 as an even smaller one. 
                 This tendency applies to the second rows from the top 
                 of Figs. 12 and 13. 
                 However, there are some clear differences between them 
                 because of the distinction of architectures. 
                 Comparing {\tt 14-S1-V<0>} here 
                 with {\tt 13-S1-V<0>} in Fig. 13 for instance, 
                 degree of trajectory's warping in the former (Type \#2 architecture) 
                 is greater than that in the latter (Type \#1 architecture) 
                 in all the situations except for (CM-a) and (CM-b). 
                 Looking overall at the output trajectories 
                 in the second rows from the top of Figs. 12, 13, and 14, 
                 output trajectory's curvings by each {\sl Subspace} increase 
                 in order of Type \#0, Type \#1, and Type \#2 architectures. 
                 In addition, intervals between small subsidiary arrows 
                 by each {\sl Subspace} become narrower also in this order; 
                 this means that the convergence speeds decrease overall 
                 in order of Type \#0, Type \#1, and Type \#2 architectures. 
                 It can be said that such phenomena 
                 for trajectory's warping and convergence speed's lowering 
                 are the unique features of Type \#2 architecture
                 with serially connected two contractive {\sl Internetworks}. 

  \item[(14-S2)] {\sl Forward Subnet \#1} is contractive 
                 as with the previous case, 
                 whereas the output trajectories 
                 in {\sl Subspaces} {\sl \#0} \& {\sl \#1} here, 
                 {\tt 14-S2-V<0>} and {\tt 14-S2-V<1>}, are quite different 
                 from {\tt 14-S1-V<0>} and {\tt 14-S1-V<1>} right above. 
                 This is because dynamics in Type \#2 architecture 
                 are determined based on a widely detoured feedback loop 
                 via two {\sl Internetworks} 
                 and so signals flowing through 
                 {\sl Backward Subnets} {\sl \#2} \& {\sl \#1} in this order 
                 greatly differ depending on the mapping relationship 
                 of ``{\sl Forward Subnet \#2}." 
                 Curvature of an output trajectory 
                 caused by expansive {\sl Forward Subnet \#2}
                 is opposite to that by contractive {\sl Forward Subnet \#1}.  
                 As with the graphs in the third rows from the top of Figs. 12 and 13,  
                 it is of considerable interest that 
                 warpings of output trajectories 
                 yielded by serially connected two {\sl Forward Subnets} 
                 with mutually inverse mapping relationships are canceled out; 
                 some canceling effect seems to occur 
                 in signals propagating through the corresponding {\sl Backward Subnets}. 
                 In fact, output trajectories in {\sl Subspace \#2} 
                 ({\tt 14-S2-V<2>}) are identical with 
                 those in {\sl Subspace \#0} ({\tt 14-S2-V<0>}). 
                 The convergence speeds in {\sl Subspaces} {\sl \#0} \& {\sl \#2} 
                 are also the same as each other. 
                 Totally comparing the graphs in the third row from the top in Fig. 12 
                 (the graphs for (12-S2)) 
                 with those in Fig. 14 
                 (the graphs for (14-S2)), 
                 we can see that output trajectories in all {\sl Subspaces} 
                 are identical with each other 
                 in Type \#0 and Type \#2 architectures 
                 when two {\sl Internetworks} have 
                 this set of mapping relationships. 

  \item[(14-S3)] {\sl Forward Subnet \#1} is expansive and 
                 {\sl Forward Subnet \#2} is contractive; 
                 the relation between {\sl Forward Subnets} {\sl \#1} \& {\sl \#2} here 
                 is quite contrary to that in the upper case. 
                 Regarding {\sl Subspace \#1}, its output 
                 approaches the center with a curved trajectory 
                 in such a manner that the network state goes away from 
                 the neighbor diagonal line \( V^{<1>}_{1} = V^{<1>}_{0} \) 
                 or \( V^{<1>}_{1} = - V^{<1>}_{0} + 1.0 \) 
                 in all the situations except for (CM-a) and (CM-b); 
                 in fact, curving in {\tt 14-S3-V<1>} here 
                 is exactly the opposite of that in {\tt 14-S2-V<1>} above it. 
                 On the other hand, 
                 as with the graphs in the third row from the top 
                 (the graphs for (14-S2) right above) 
                 and those in the third and fourth rows from the top 
                 in Figs. 12 and 13 
                 (the graphs for (12-S2), (12-S3), (13-S2), and (13-S3)), 
                 trajectory's curvatures generated 
                 by serially connected two {\sl Forward Subnets} 
                 with mutually inverse mapping relationships 
                 seem to be entirely canceled out. 
                 As a result, an output of {\sl Subspace \#2} 
                 draws a locus identical with that of {\sl Subspace \#0}. 
                 We can also see that the convergence speeds 
                 in {\sl Subspaces} {\sl \#0} \& {\sl \#2} 
                 are the same as each other. 
                 Moreover, it is important that the graphs for (14-S3) here, 
                 {\tt 14-S3-V<0>}, {\tt 14-S3-V<1>}, and {\tt 14-S3-V<2>}, 
                 are completely identical with 
                 those in the fourth row from the top in Fig. 12, 
                 {\tt 12-S3-V<0>}, {\tt 12-S3-V<1>}, and {\tt 12-S3-V<2>}; 
                 this means that dynamics in Type \#2 architecture 
                 are completely the same as those in Type \#0 one 
                 when two {\sl Internetworks} have 
                 this set of mapping relationships. 
                 Judging from ``the relation between 
                 the graphs for (14-S3) here and those for (12-S3)" 
                 as well as ``that between 
                 the graphs for (14-S2) right above and those for (12-S2)," 
                 when serially connected two {\sl Forward Subnets} 
                 have mutually inverse mapping relationships, 
                 some canceling effect seems to be generated in signals 
                 flowing through the corresponding {\sl Backward Subnets} 
                 in Type \#2 architecture. 

  \item[(14-S4)] Since both {\sl Forward Subnets} are expansive, 
                 output trajectories in {\sl Subspace \#0} 
                 ({\tt 14-S4-V<0>}) 
                 are sent to the outer {\sl Subspaces} in turn 
                 basically as more expanded trajectories 
                 ({\tt 14-S4-V<1>} and {\tt 14-S4-V<2>}). 
                 This point is common to the graphs 
                 in the bottom rows of Figs. 12 and 13. 
                 However, comparing the graphs 
                 in the bottom rows of Figs. 12, 13, and 14 with each other, 
                 there are intriguing contrasts between them 
                 due to the distinction of architectures; 
                 degrees of output trajectory's curvings by each {\sl Subspace} 
                 increase in order of Type \#0, Type \#1, and Type \#2 
                 architectures in all the situations except for (CM-a) and (CM-b). 
                 Additionally, intervals between small subsidiary arrows 
                 by each {\sl Subspace} become wider 
                 already at an early stage also in this order; 
                 this suggests that the convergence speeds become higher overall 
                 in order of Type \#0, Type \#1, and Type \#2 architectures. 
                 Such phenomena for 
                 output trajectory's curving and convergence speed's raising 
                 are thought to be essential features of Type \#2 architecture 
                 with serially connected two expansive {\sl Internetworks}. 

\end{itemize}

\vspace{1.00mm}
\subsection{On relation between adjacent Internetworks}

\vspace{2.50mm}
\noindent 
In the previous subsections for the Unconstrained Association Mode, 
we found that an output trajectory sometimes curved more 
every time it went through a {\sl Forward Subnet}. 
In some cases, output trajectories 
in {\sl Subspaces} {\sl \#0} \& {\sl \#2} 
became identical with each other 
for whatever reason that was related to the cancellation 
of signal components passing through {\sl Backward Subnets}.

Now we suppose three different networks, each of which 
has any of Type \#0, Type \#1, or Type \#2 architecture 
and commonly owns two {\sl Forward Subnets} 
with a certain set of mapping relationships. 
The location of a pair of a feedback path and an input port 
is architecture specific 
and so output trajectories 
in the same {\sl Subspace} 
differ from each other depending on architectures, 
although two {\sl Forward Subnets} inside each network 
have the common set of mapping relationships. 
Even under such conditions, 
regardless of architectures
(i.e., no matter what the location 
of a pair of a feedback path and an input port is), 
an output trajectory in a certain {\sl Subspace} 
and that in its adjacent {\sl Subspace} 
seem to be associated with each other 
simply based on the ``static" mapping relationship 
of the {\sl Forward Subnet} located between those {\sl Subspaces}. 
In the following, we will confirm this feature 
concretely looking at the experimental results again.

Let us focus on the graphs in the second rows from the top 
(Contractive \( \rightarrow \) Contractive) 
and the ones in the bottom rows 
(Expansive \( \rightarrow \) Expansive) in Figs. 12, 13, and 14. 
As already mentioned, 
degrees of output trajectory's curvings differ from each other 
depending on architectures, 
even if mapping relationships of {\sl Forward Subnets} 
inside a network are common throughout the three architectures. 
However, carefully looking at their graphs, 
we can understand that, in each architecture, 
the mutual relation between 
output trajectories in a certain {\sl Subspace} 
and those in its adjacent {\sl Subspace} 
is only governed by the static mapping relationship 
of the corresponding {\sl Forward Subnet}.

With regard to the graphs in the third rows from the top 
(Contractive \( \rightarrow \) Expansive) 
and those in the fourth rows from the top 
(Expansive \( \rightarrow \) Contractive) 
in Figs. 12, 13, and 14, 
we can confirm that, in each architecture, 
the interrelation between output trajectories 
in the three adjacent {\sl Subspaces} is just decided statically 
by {\sl Forward Subnets} {\sl \#1} \& {\sl \#2} 
whose mapping relationships are mutually inverse. 
On these examples, it is easy to understand their relations 
because, in each case, an output trajectory in {\sl Subspace \#2} 
eventually becomes identical with that in {\sl Subspace \#0}. 
Specifically comparing 
{\tt 12-S2-V<0>} and {\tt 12-S2-V<2>}, 
and {\tt 12-S3-V<0>} and {\tt 12-S3-V<2>} in Fig. 12, 
{\tt 13-S2-V<0>} and {\tt 13-S2-V<2>}, 
and {\tt 13-S3-V<0>} and {\tt 13-S3-V<2>} in Fig. 13, 
and 
{\tt 14-S2-V<0>} and {\tt 14-S2-V<2>}, 
and {\tt 14-S3-V<0>} and {\tt 14-S3-V<2>} in Fig. 14 
respectively with each other, 
it is obvious at a glance that they are 
individually the same.

Here let us look at Fig. 13 once again. 
Output trajectories in {\sl Subspaces \#0 \& \#1} 
are dynamically determined 
based on the detoured feedback loop including {\sl Internetwork \#1}. 
Output trajectories in {\sl Subspace \#2} become 
what those in {\sl Subspace \#1} were statically mapped to 
by {\sl Forward Subnet \#2}. 
In this way, concerning Type \#1 architecture, 
an output trajectory in each {\sl Subspace} 
is generated based on different processes. 
Regardless of such a situation, 
it is quite interesting that, as confirmed above, 
output trajectories in the three adjacent {\sl Subspaces} 
are associated with each other 
based only on the static mapping relationships 
of the two {\sl Forward Subnets}. 
In particular, with respect to the graphs 
in the third and fourth rows from the top of this figure 
(i.e., the graphs in which {\sl Internetworks \#1 \& \#2} 
have mutually inverse mapping relationships in Type \#1 architecture), 
it is worthy of remark again that 
output trajectories in {\sl Subspaces \#0 \& \#2} 
are fully the same as each other.

As suggested in the previous subsection, 
qualitatively conspicuous phenomena appear 
regarding the graphs in the third and fourth rows of Fig. 14, 
each of which is for Type \#2 architecture
with {\sl Forward Subnets} 
having mutually inverse mapping relationships. 
Interestingly, 
{\tt 14-S2-V<0>}, {\tt 14-S2-V<1>}, and {\tt 14-S2-V<2>} in Fig. 14 
are identical respectively with 
{\tt 12-S2-V<0>}, {\tt 12-S2-V<1>}, and {\tt 12-S2-V<2>} in Fig. 12. 
Moreover, 
{\tt 14-S3-V<0>}, {\tt 14-S3-V<1>}, and {\tt 14-S3-V<2>} in Fig. 14 
are respectively the same as 
{\tt 12-S3-V<0>}, {\tt 12-S3-V<1>}, and {\tt 12-S3-V<2>} in Fig. 12. 
That is, when serially connected two {\sl Forward Subnets} 
have mutually inverse mapping relationships, 
output trajectories in Type \#2 architecture 
in which there exists a pair of a feedback path and an input port 
only in {\sl Subspace \#2} 
are identical with those in Type \#0 architecture 
in which there is a pair of a feedback path and an input port 
only in {\sl Subspace \#0}. 
In Type \#2 architecture, 
the difference between ``output of {\sl Forward Subnet \#2}" 
and ``input from the outside" in {\sl Subspace \#2} 
propagates through {\sl Backward Subnets} {\sl \#2} \& {\sl \#1} 
in said order. 
In spite of this situation, 
for the graphs in the third and fourth rows from the top of Fig. 14, 
output trajectories in {\sl Subspace \#2} 
are completely the same as those in {\sl Subspace \#0}. 
This suggests that 
additional or deducted components of signals 
after passing through {\sl Backward Subnet \#2} 
are completely canceled out by {\sl Backward Subnet \#1}.

Taking notice of a {\sl Subspace} 
without a pair of a feedback path and an input port, 
let us consider a signal 
flowing through serially connected two {\sl Backward Subnets} 
that are located inside the very {\sl Subspace} 
and on the outer side of it. 
This means, for instance, that we pay attention to 
{\sl Subspace \#1} of Type \#2 architecture 
in the Association Mode shown in Fig. 3 with \( P = 2 \). 
Putting \( p = 1, 2, ..., P \) and 
letting \( N^{<p>} \) be  
the input-output relation of {\sl Backward Subnet \#\(p\)} 
for the multi-dimensional model in general, 
\( N^{<p>} \) can be written as the following equation 
based on Eqs. (15) and (16):

\vspace{3.00mm} 
\begin{equation} 
N^{<p>} \stackrel{\triangle}{=} \left[ 
\begin{array}{cccc} 
\displaystyle \frac{\partial V^{<p>}_{0}}{\partial V^{<p-1>}_{0}} & 
\displaystyle \frac{\partial V^{<p>}_{1}}{\partial V^{<p-1>}_{0}} & 
\displaystyle \frac{\partial V^{<p>}_{2}}{\partial V^{<p-1>}_{0}} & 
\cdots \\ 
\\ 
\displaystyle \frac{\partial V^{<p>}_{0}}{\partial V^{<p-1>}_{1}} & 
\displaystyle \frac{\partial V^{<p>}_{1}}{\partial V^{<p-1>}_{1}} & 
\displaystyle \frac{\partial V^{<p>}_{2}}{\partial V^{<p-1>}_{1}} & 
\cdots \\ 
\\ 
\displaystyle \frac{\partial V^{<p>}_{0}}{\partial V^{<p-1>}_{2}} & 
\displaystyle \frac{\partial V^{<p>}_{1}}{\partial V^{<p-1>}_{2}} & 
\displaystyle \frac{\partial V^{<p>}_{2}}{\partial V^{<p-1>}_{2}} & 
\cdots \\ 
\\ 
\vdots & 
\vdots & 
\vdots & 
\ddots \\ 
\end{array} 
\right] ~. \\ 
\end{equation}

\[
\hspace{41.10mm}
(~ p = 1, 2, ..., P ~) 
\]

\vspace{2.00mm} 
\noindent 
In our simulation experiments 
for the Unconstrained Association Mode 
shown in the previous subsection, 
the model is limited to be two-dimensional and  
the relational expression given by Eq. (9) 
is restricted as follows:

\vspace{2.00mm} 
\begin{equation} 
V^{<p>}_{i} \stackrel{\triangle}{=} 
f^{<p>}_{i}(~ V^{<p-1>}_{i} ~) ~. 
\hspace{15.00mm} 
(~ p = 1, 2, ..., P, ~~i = 0, ~1 ~) 
\end{equation}

\vspace{2.00mm} 
\noindent 
Anew putting \( p = 1, 2, ..., P - 1 \) under such assumptions, 
\( N^{<p>} N^{<p+1>} \), which is the operation 
by serially connected two {\sl Backward Subnets} 
from {\sl Subspace \#\( (p + 1) \)} 
through {\sl Subspace \#\( (p - 1) \)}, 
is represented as follows:

\vspace{2.00mm} 
\begin{equation} 
N^{<p>} N^{<p+1>} = \left[ 
\begin{array}{cc} 
\displaystyle \frac{\partial V^{<p+1>}_{0}}{\partial V^{<p-1>}_{0}} & 
0 \\ 

0                                                                   &
\displaystyle \frac{\partial V^{<p+1>}_{1}}{\partial V^{<p-1>}_{1}} \\ 
\end{array} 
\right] ~. \\ 
\end{equation}

\[
\hspace{47.00mm}
(~ p = 1, 2, ..., P - 1 ~) 
\]

\vspace{4.00mm} 
\noindent 
This equation means that 
a signal applied to \( N^{<p + 1>} \) is amplified or attenuated 
and then sent to {\sl Subspace \#\( (p - 1) \)} 
in accordance with how warped the output of {\sl Subspace \#\( (p + 1) \)} is 
based on that of {\sl Subspace \#\( (p - 1) \)}. 
If \( N^{<p + 1>} \) and \( N^{<p>} \) are conversion matrices 
with either both expansive mapping relationships or both contractive ones, 
the difference between 
``output via a feedback path" and ``input from the outside" 
in {\sl Subspace \#\( (p + 1) \)} 
is greatly amplified or attenuated 
and then transmitted to {\sl Subspace \#\( (p - 1) \)}; 
large curvature of an output trajectory 
and big variance in convergence speed are also produced. 
On the contrary, if \( N^{<p + 1>} \) and \( N^{<p>} \) 
are conversion matrices with mutually inverse mapping relationships, 
Eq. (51) becomes an identity matrix, 
and therefore, the difference between 
``output via a feedback path" and ``input from the outside" 
in {\sl Subspace \#\( (p + 1) \)} 
is sent to {\sl Subspace \#\( (p - 1) \)} as it is. 
As a matter of fact, 
with regard to the graphs 
in the third and fourth rows from the top 
of Fig. 14 for Type \#2 architecture, 
the error information in {\sl Subspace \#2} 
is conveyed to {\sl Subspace \#0} as it is, 
owing to cancellation in additional or deducted components 
of signals passing through {\sl Backward Subnets}. 
It is due to this 
that the graphs in the third and fourth rows from the top of Fig. 14 
are respectively identical with those in the same rows of Fig. 12 
as previously confirmed. 
\footnote{As long as \( N^{<p>} N^{<p + 1>} = I \) holds, 
the above-mentioned cancellation can occur
even in the general case 
without any specific conditions such as Eq. (50), 
where \( I \) means an identity matrix. 
Other cases than those discussed here 
are subjects for future analysis.}

\section{Constrained Association Mode}

\subsection{Unconstrained vs Constrained}

\vspace{2.00mm} 
\noindent
In the previous section, we investigated 
dynamical behavior of the proposed network 
in the Unconstrained Association Mode 
where fixed values were applied to a unique input port. 
We then confirmed that signals going through {\sl Backward Subnets} 
qualitatively and quantitatively differed 
depending on which {\sl Subspace} had 
a pair of ``a feedback path" and ``an input port," 
and that according to these variations, 
convergence trajectories sometimes curved largely. 
If what is applied to an input port from the outside is 
not ``a fixed goal point" but ``a trajectory," 
how does the total network behave in the Association Mode?
It is quite interesting to further study 
how the trajectory's ``speed" affects 
dynamical characteristics of the network. 
Now let us pick up, as an example, Type \#2 architecture 
in which a pair of a feedback path and an input port
exists only in {\sl Subspace \#2}. 
As the speed of a target trajectory 
applied from the outside becomes higher, 
the difference between feedback signals and input ones 
in {\sl Subspace \#2} increases, 
and therefore, signals going through {\sl Backward Subnets} 
should get larger depending on 
how the mapping relationships 
of the corresponding {\sl Forward Subnets} are. 
This situation is similar to the Unconstrained Association Mode 
discussed in the previous section, and 
a large change is estimated to occur 
in curvature and/or speed 
of an output trajectory in each {\sl Subspace} 
when the input trajectory is a straight one moving toward the center. 
On the contrary, if the speed of a target trajectory 
applied from the outside is very low, 
the difference between feedback signals and input ones 
in {\sl Subspace \#2} becomes extremely small 
due to the function of dynamical neurons in {\sl Subspace \#0}; 
therefore, the input/output signals to/from both {\sl Backward Subnets} 
considerably lessen regardless of how the mapping relationships 
of the corresponding {\sl Forward Subnets} are. 
Then, we can expect that an output trajectory in each {\sl Subspace} 
does not curve beyond the curvatures 
based on ``static" mapping relationships of the {\sl Forward Subnets}, 
even if those {\sl Forward Subnets} have 
strongly non-linear mapping relationships.

In our previous paper for the basic model, 
we conducted some simulation experiments 
for the ``Constrained" Association Mode 
and analyzed the associative dynamics 
in which a ``straight" target trajectory toward a fixed goal point 
was applied to a unique input port from the outside. 
As suggested in Subsection 2.3 of this paper, 
we here bring up the Constrained Association Mode; 
in the present section, we will investigate in detail 
the network behavior of a two-dimensional model 
particularly when applying a set of ``periodic" signals 
with different phases, 
which is able to draw a full-orbed target trajectory 
in a two-dimensional plane. 
We specifically show those input signals in Fig. 15. 
The two upper left graphs illustrate examples of 
\( Z^{<p>}_{0} \) and \( Z^{<p>}_{1} \) 
whose phases are mutually \( 90^{\circ} \) shifted, 
where \( p = 0, 1, 2 \); 
they can actually generate a circular trajectory 
with the radius \( 0.375 \) in a two-dimensional plane 
as depicted on the right side. 
In the following simulation experiments, 
we prepare signals, 
as well as with different amplitudes, 
with three periods, 
\( 1 {\sf ms} \), ~\( 5 {\sf ms} \), and ~\( 500 {\sf ms} \), 
and designate them as ``Fast (High Speed) Input," 
``Medium Speed Input," and ``Slow (Low Speed) Input" respectively.

\subsection{Dynamics}

\vspace{2.00mm} 
\noindent
{\large (1) Simulation I (Contractive \( \rightarrow \) Expansive)}

\vspace{3.00mm} 
\noindent
(1-1) Overview

\vspace{3.00mm} 
\noindent
Figures 16, 17, and 18 show simulation results 
when mapping relationships of {\sl Internetworks} {\sl \#1} \& {\sl \#2} 
were set to be contractive and expansive in the order given. 
From an architectural point of view, 
these figures respectively correspond to 
Type \#0, Type \#1, and Type \#2 architectures 
in the Association Mode of Fig. 3. 
The main part of each figure is composed of three rows; 
the frequency of input signals applied from the outside 
gets lower from the top to the bottom. 
For the horizontal direction in each row, 
the three graphs on the left correspond to output responses 
in {\sl Subspace \#0}, {\sl Subspace \#1}, and {\sl Subspace \#2}
respectively from left to right, 
and the rightmost graph shows an input signal 
applied from the outside to a unique input port 
in the corresponding {\sl Subspace}. 
In the simulation studies for each row in these tables, 
periodic target signals with five kinds of amplitude were employed. 
To aid understanding of the meaning of each graph, 
we drew a block diagram of the whole network 
and mapping relationships of {\sl Internetworks} 
along with wide gray arrow lines in the upper part of each figure. 
In the graphs, small black circles indicate the initial points. 
Arrows on each trajectory are drawn at regular intervals; 
the specific value of an interval on each row 
is shown in the leftmost part. 
We also appended an index to each graph 
in order to quote it in the following explanation as needed.

\begin{figure}[!t]

    \hspace*{7.65mm} 
    \includegraphics[scale=0.70]{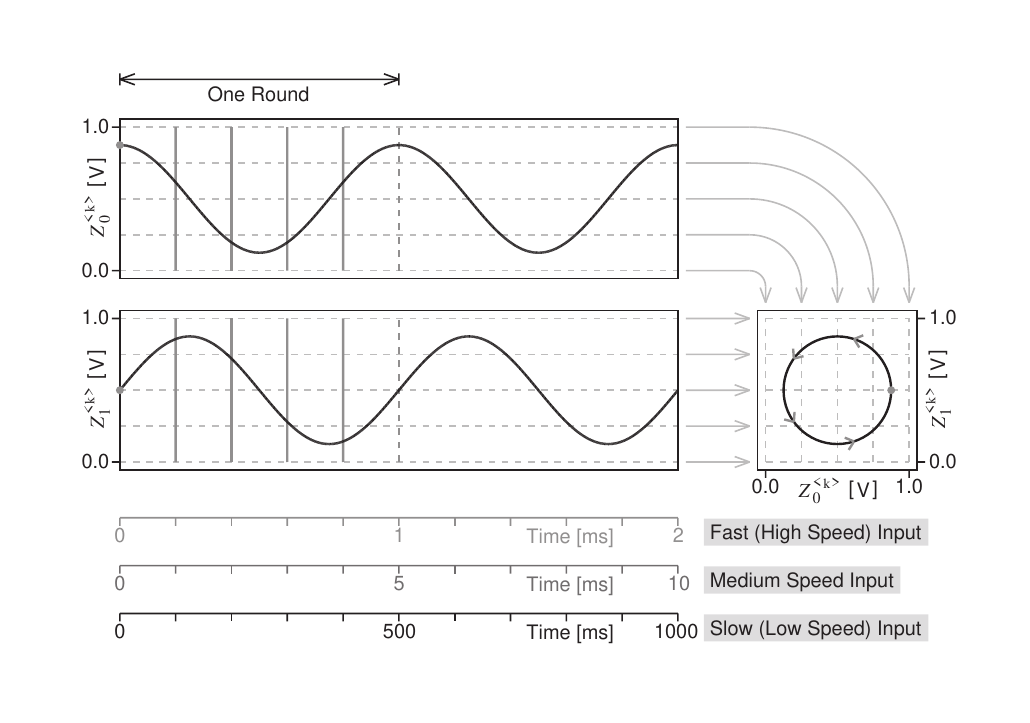}

    \vspace*{-8.00mm} 
    \caption{ 
      A set of ``periodic (sinusoidal)" target signals 
      with different phases 
      that is applied to a unique input port 
      for testing the network behavior 
      in the Constrained Association Mode 
      of the two-dimensional model. 
      The two upper left graphs illustrate examples of 
      \( Z^{<p>}_{0} \) and \( Z^{<p>}_{1} \) 
      whose phases are mutually \( 90^{\circ} \) shifted, 
      where \( p = 0, 1, 2 \); 
      their signals can actually generate a full-orbed trajectory 
      with the radius \( 0.375 \) in a two-dimensional plane 
      as depicted on the right side. 
      In the simulation experiments, 
      we prepare signals, as well as with different amplitudes, 
      with three periods, 
      \( 1 {\sf ms} \), ~\( 5 {\sf ms} \), and ~\( 500 {\sf ms} \), 
      and designate them as ``Fast (High Speed) Input," 
      ``Medium Speed Input," and ``Slow (Low Speed) Input" 
      respectively, as indicated at the bottom. 
    } 

\end{figure}

\vspace{3.00mm} 
\noindent
(1-2) Type \#0 Architecture

\vspace{3.00mm} 
\noindent
First of all, let us look at Fig. 16 precisely. 
In Type \#0 architecture 
in which there exists a pair of a feedback path 
and an input port only in {\sl Subspace \#0}, 
no signal passes through two {\sl Backward Subnets} in principle 
as already stated in the previous section. 
Output of {\sl Subspace \#0} is emitted 
by dynamical neurons with direct feedback connections, 
and it is only statically mapped in turn
to {\sl Subspaces} {\sl \#1} \& {\sl \#2} 
through two {\sl Forward Subnets}.

The top row graphs show output responses 
when sinusoidal waves with the period \( 1 {\rm ms} \) 
(Fast Input) were applied 
to the input port in {\sl Subspace \#0}. 
Since the amplitudes of these input signals change rapidly 
compared with the time constant of a dynamical neuron, 
outputs of {\sl Subspace \#0} 
can not sufficiently follow the target signals 
and become very small circular trajectories near the center
({\tt 16-S2-Fast-V<0>}). 
Those poor outputs of {\sl Subspace \#0} are sent 
to {\sl Subspace \#1} by {\sl Forward Subnet \#1} 
as more shrunk signals ({\tt 16-S2-Fast-V<1>}), 
and the outputs there are expanded and transmitted 
to {\sl Subspace \#2} through {\sl Forward Subnet \#2} 
({\tt 16-S2-Fast-V<2>}). 
The output trajectories 
in {\sl Subspaces} {\sl \#0} \& {\sl \#2} 
consequently become the same as each other.

The mid row graphs show output responses 
when applying sinusoidal waves with the period \( 5 {\rm ms} \) 
(Medium Speed Input). 
These input signals' change is slower 
in comparison with that in the upper row, 
so the tracking feature improves and 
output trajectories in {\sl Subspace \#0} become larger circles 
({\tt 16-S2-Medium-V<0>}). 
In the same way as the top row graphs, 
an output of {\sl Subspace \#1} becomes 
a statically shrunk version of that of {\sl Subspace \#0} 
({\tt 16-S2-Medium-V<1>}) 
because of contractive {\sl Forward Subnet \#1}, 
and an output of {\sl Subspace \#2} 
represents a statically expanded version of that of {\sl Subspace \#1} 
({\tt 16-S2-Medium-V<2>}) 
due to expansive {\sl Forward Subnet \#2}. 
Eventually, the output trajectories in {\sl Subspaces \#2} 
are identical with those in {\sl Subspaces \#0} also in this case.

The bottom row graphs correspond to output responses 
when sinusoidal waves with the period \( 500 {\rm ms} \) 
(Slow Input) were applied. 
Since the amplitudes of these input signals vary very slowly 
compared with the time constant of a dynamical neuron 
in {\sl Subspace \#0}, 
output trajectories in {\sl Subspace \#0} 
almost perfectly follow the full-orbed target signals 
({\tt 16-S2-Slow-V<0>}). 
Here we should take notice of 
the trajectories shown in {\tt 16-S2-Slow-V<1>}. 
Let us compare the following two points 
across which an output trajectory in {\sl Subspace \#0} goes: 
One is a point on the diagonal line 
\( V^{<0>}_{1} = V^{<0>}_{0} \) 
or \( V^{<0>}_{1} = - V^{<0>}_{0} + 1.0 \), 
and the other is a point on the straight line 
\( V^{<0>}_{1} = 0.5 \) or \( V^{<0>}_{0} = 0.5 \). 
Since the mapping relationship of {\sl Forward Subnet \#1} 
is two-dimensionally contractive in a square on a coordinate plane, 
an output trajectory in {\sl Subspace \#1} 
corresponding to the former point 
must pass through the area closer to the center, 
compared with that corresponding to the latter one. 
Consequently, output trajectories in {\sl Subspace \#1} 
are slightly small rounded squares rotated by \( 45^{\circ} \), 
as drawn in {\tt 16-S2-Slow-V<1>}. 
In the same manner as the top row and mid row graphs, 
what the output trajectories in {\sl Subspace \#1} 
were statically expanded to by {\sl Forward Subnet \#2} 
come to be those in {\sl Subspace \#2}  
as shown in {\tt 16-S2-Slow-V<2>}. 
Also in this case, the output trajectories in {\sl Subspace \#2} 
are identical with those in {\sl Subspace \#0}; 
both of them coincide with the full-orbed input signals.

\begin{figure}[p]

    \vspace*{-2.00mm}
    \hspace*{-2.25mm}
    \includegraphics[scale=0.88]{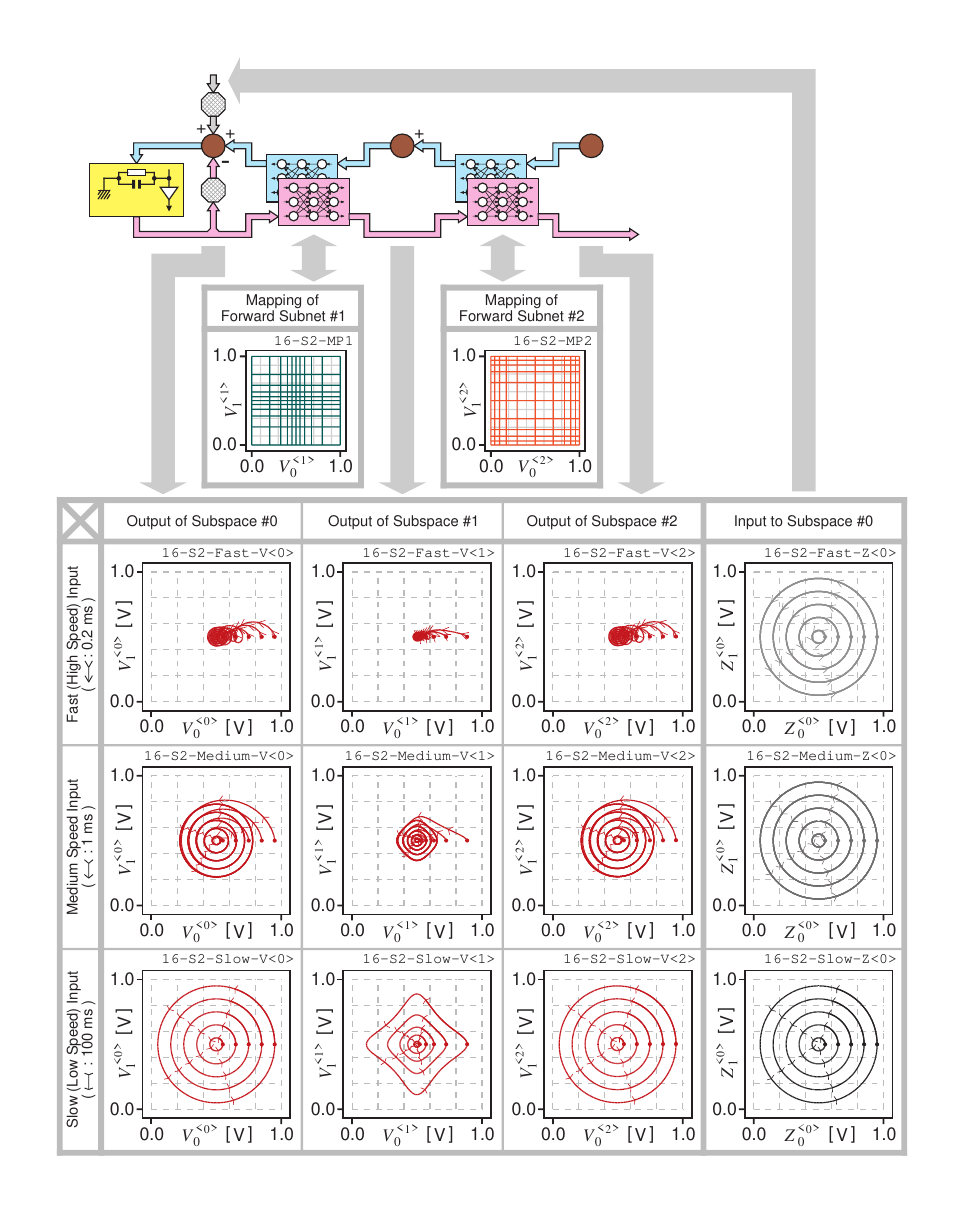}

    \vspace*{-4.00mm}
    \caption{ 
      Simulation results for Type \#0 architecture 
      in the Constrained Associative Mode 
      of the two-dimensional model with \( P = 2 \) 
      when mapping relationships 
      of {\sl Internetworks} {\sl \#1} \& {\sl \#2} 
      are set to be contractive and expansive in this order. 
      The main part (the bottom half) of the figure 
      is composed of three rows; 
      the frequency of the periodic input signals 
      applied from the outside 
      gets lower from the top to the bottom. 
      For the horizontal direction in each row, 
      the three graphs corresponding to the output responses 
      in {\sl Subspaces} {\sl \#0}, {\sl \#1}, \& {\sl \#2} 
      and the one graph for the circular target trajectories 
      are illustrated respectively from the left to the right. 
      To help understanding of the meaning of this figure, 
      a block diagram of the whole network 
      and mapping relationships of the two {\sl Internetworks} 
      are drawn along with wide gray arrow lines in the upper part. 
    } 

\end{figure}

\begin{figure}[p]

    \vspace*{-2.00mm}
    \hspace*{-2.25mm}
    \includegraphics[scale=0.88]{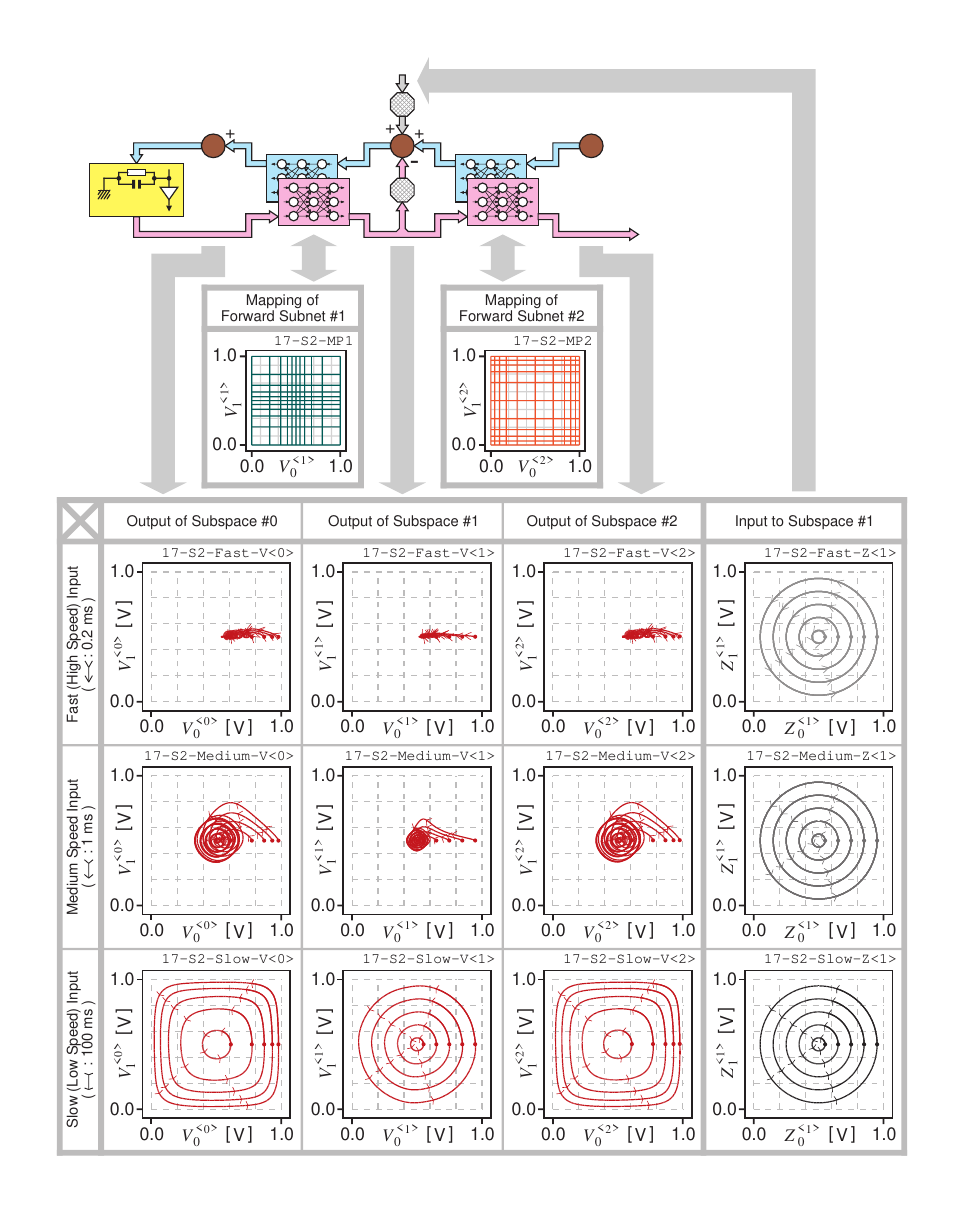}

    \vspace*{-4.00mm}
    \caption{ 
      Simulation results for Type \#1 architecture 
      in the Constrained Associative Mode 
      of the two-dimensional model with \( P = 2 \) 
      when mapping relationships 
      of {\sl Internetworks} {\sl \#1} \& {\sl \#2} 
      are set to be contractive and expansive in this order. 
      The main part (the bottom half) of the figure 
      is composed of three rows; 
      the frequency of the periodic input signals 
      applied from the outside 
      gets lower from the top to the bottom. 
      For the horizontal direction in each row, 
      the three graphs corresponding to the output responses 
      in {\sl Subspaces} {\sl \#0}, {\sl \#1}, \& {\sl \#2} 
      and the one graph for the circular target trajectories 
      are illustrated respectively from the left to the right. 
      To help understanding of the meaning of this figure, 
      a block diagram of the whole network 
      and mapping relationships of the two {\sl Internetworks} 
      are drawn along with wide gray arrow lines in the upper part. 
    } 

\end{figure}

\begin{figure}[p]

    \vspace*{-2.00mm}
    \hspace*{-2.25mm}
    \includegraphics[scale=0.88]{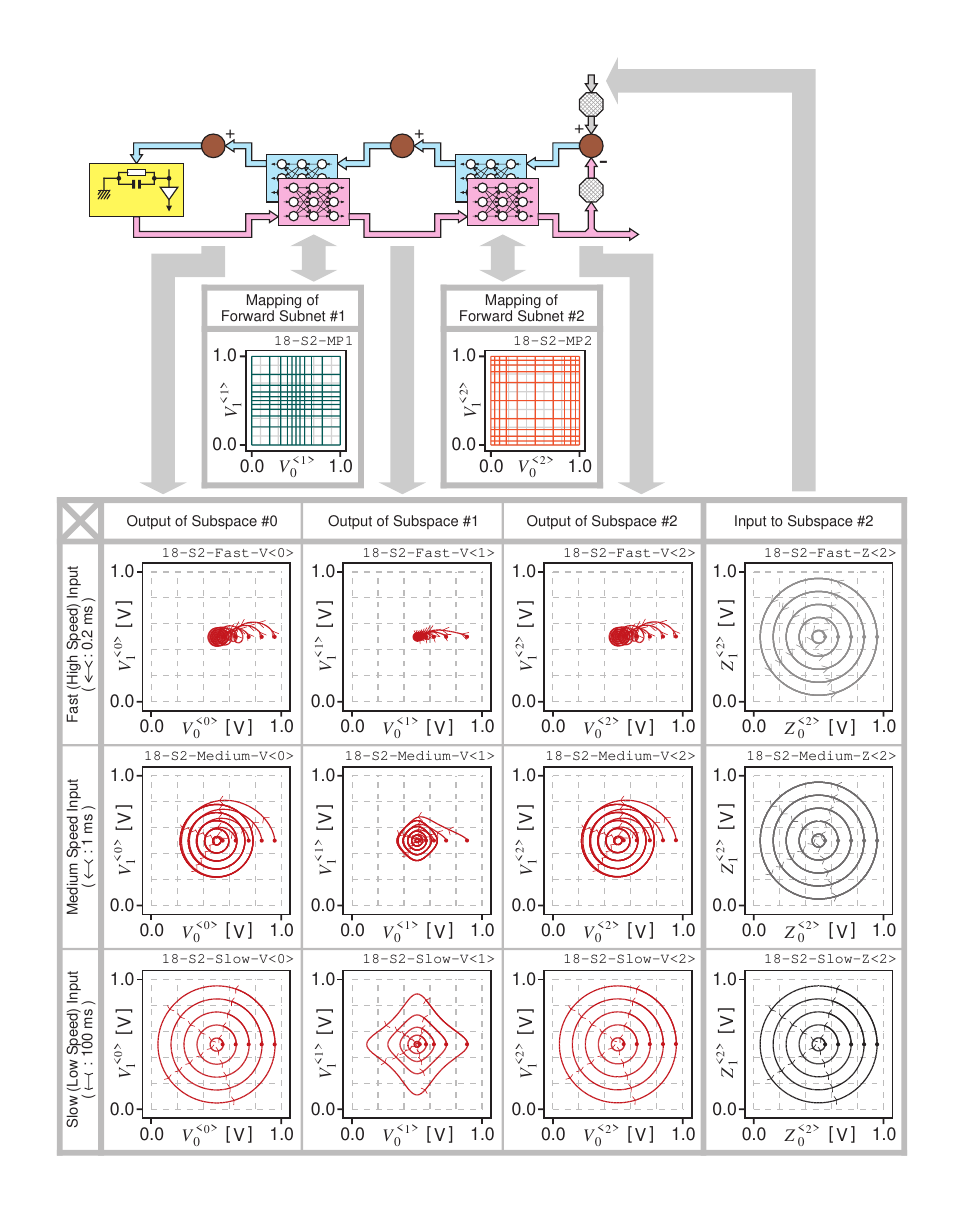}

    \vspace*{-4.00mm}
    \caption{ 
      Simulation results for Type \#2 architecture 
      in the Constrained Associative Mode 
      of the two-dimensional model with \( P = 2 \) 
      when mapping relationships 
      of {\sl Internetworks} {\sl \#1} \& {\sl \#2} 
      are set to be contractive and expansive in this order. 
      The main part (the bottom half) of the figure 
      is composed of three rows; 
      the frequency of the periodic input signals 
      applied from the outside 
      gets lower from the top to the bottom. 
      For the horizontal direction in each row, 
      the three graphs corresponding to the output responses 
      in {\sl Subspaces} {\sl \#0}, {\sl \#1}, \& {\sl \#2} 
      and the one graph for the circular target trajectories 
      are illustrated respectively from the left to the right. 
      To help understanding of the meaning of this figure, 
      a block diagram of the whole network 
      and mapping relationships of the two {\sl Internetworks} 
      are drawn along with wide gray arrow lines in the upper part. 
    } 

\end{figure}

\vspace{3.00mm} 
\noindent
(1-3) Type \#1 Architecture

\vspace{3.00mm} 
\noindent
Secondly, let us check Fig. 17. 
With regard to Type \#1 architecture, 
activity dynamics in {\sl Subspaces} {\sl \#0} \& {\sl \#1} 
are determined based on the detoured feedback loop 
through {\sl Internetwork \#1}, 
and the output of {\sl Subspace \#1} is mapped in a static manner 
to {\sl Subspace \#2} by {\sl Forward Subnet \#2}.

Output of {\sl Backward Subnet \#1} becomes very small 
because of the nature of the corresponding 
contractive {\sl Forward Subnet \#1}, 
even if input to {\sl Backward Subnet \#1} is comparatively large; 
such a situation can be predicted 
from the experiments for Type \#1 architecture 
of the Unconstrained Association Mode in Section 4. 
This property does not produce an effect that makes 
the difference between feedback signals and input ones 
in {\sl Subspace \#1} decrease in {\sl Subspace \#0}, 
and the processing is repeatedly executed without any improvement 
based on the detoured feedback loop through {\sl Internetwork \#1}. 
In the top row graphs, what were applied to the input port 
are sinusoidal waves with a high frequency (Fast Input). 
Hence, the difference between feedback signals and input ones 
in {\sl Subspace \#1} remains large. 
According to the mechanism mentioned above, 
however, output of {\sl Subspace \#0} eventually 
draws very small circular trajectories near the center, 
and that of {\sl Subspace \#1} 
comes to be even smaller trajectories 
due to contractive {\sl Forward Subnet \#1}. 
In terms of {\sl Subspace \#0}, 
let us make a comparison 
between {\tt 17-S2-Fast-V<0>} here 
and previous {\tt 16-S2-Fast-V<0>}. 
As for {\sl Subspace \#1}, 
compare {\tt 17-S2-Fast-V<1>} here 
with previous {\tt 16-S2-Fast-V<1>}. 
It is obvious that, 
by {\sl Subspace \#0} or {\sl Subspace \#1}, 
the radii of the periodic output trajectories 
in Fig. 17 (Type \#1 architecture) 
are smaller than those in Fig. 16 (Type \#0 architecture). 
Next let us check the difference between 
``{\tt 17-S2-Fast-V<1>} here for {\sl Subspace \#1}" and 
``previous {\tt 16-S2-Fast-V<0>} for {\sl Subspace \#0}." 
``The radii of the periodic output trajectories 
in {\sl Subspace \#1} of Fig. 17 (Type \#1 architecture) 
based on the detoured feedback loop 
through contractive {\sl Internetwork \#1}" 
are clearly smaller than 
``those in {\sl Subspace \#0} of Fig. 16 (Type \#0 architecture) 
according to the direct feedback loop inside {\sl Subspace \#0}." 
It can be said that these verifications 
(on the difference as to whether a non-linear {\sl Internetwork} 
is located inside a feedback loop or outside it) 
fully reveal typical features of Type \#0 \& Type \#1 architectures. 
That is to say, 
when an {\sl Internetwork} with a non-linear mapping relationship 
is located inside a feedback loop, 
a {\sl Subspace} on the inner side of the {\sl Internetwork} 
comes to be affected by non-linearity of the mapping, 
and then a {\sl Subspace} on the outer side of the {\sl Internetwork} 
is more strongly influenced by non-linearity of the mapping. 
It is important that output trajectories 
in {\sl Subspaces} {\sl \#0} \& {\sl \#2} 
are identical with each other also in this case, 
although trackability is fairly bad overall.

In the mid row graphs with Medium Speed Input, 
tracking properties improve compared with the top row graphs, 
and periodic output trajectories with a little larger radii 
are drawn in {\sl Subspaces} {\sl \#0} \& {\sl \#1}. 
In fact, with regard to {\sl Subspace \#0}, 
the radii of periodic output trajectories 
in {\tt 17-S2-Medium-V<0>} here 
are larger than those in upper {\tt 17-S2-Fast-V<0>}; 
similarly for {\sl Subspace \#1}, 
the radii of periodic output trajectories 
in {\tt 17-S2-Medium-V<1>} here 
are bigger than those in upper {\tt 17-S2-Fast-V<1>}. 
Next checking the difference of Type \#1 architecture 
against Type \#0 one, 
in terms of {\sl Subspace \#0}, 
the radii of periodic output trajectories 
in {\tt 17-S2-Medium-V<0>} here 
are smaller than those 
in {\tt 16-S2-Medium-V<0>} of Fig. 16; 
likewise for the output responses in {\sl Subspace \#1}, 
the radii of periodic output trajectories 
in {\tt 17-S2-Medium-V<1>} here 
are smaller than those 
in {\tt 16-S2-Medium-V<1>} of Fig. 16. 
The distinction in these characteristics is generated 
depending on whether contractive {\sl Internetwork \#1} 
is located inside a feedback loop or outside it. 
The output of {\sl Subspace \#1} here 
is simply converted in a static manner to that of {\sl Subspace \#2}, 
which becomes identical with output of {\sl Subspace \#0} 
in the same way as the relation found in the previous examples.

The bottom row graphs illustrate the case with Slow Input. 
As can be seen in 
%
%
{\tt 17-S2-Slow -V<1>}, 
output trajectories in {\sl Subspace \#1} 
nearly completely track the input signals from the outside
that are full-orbed ones; 
then, there are almost no signals 
propagating through {\sl Backward Subnet \#1}. 
Noteworthy here is {\tt 17-S2-Slow-V<0>}. 
Due to the functions of ``dynamical neurons in {\sl Subspace \#0}" and 
``{\sl Internetwork \#1} between {\sl Subspaces} {\sl \#0} \& {\sl \#1}," 
the output trajectories in {\sl Subspace \#0} are required to bend 
so that those in {\sl Subspace \#1} become full-orbed. 
Since the mapping relationship of {\sl Forward Subnet \#1} 
is two-dimensionally contractive in a square on a coordinate plane, 
the contraction rate in the area near 
the diagonal line \( V^{<1>}_{1} = V^{<1>}_{0} \) 
or \( V^{<1>}_{1} = - V^{<1>}_{0} + 1.0 \) 
is greater than that in the area near 
the straight line \( V^{<1>}_{1} = 0.5 \) or \( V^{<1>}_{0} = 0.5 \) 
in {\sl Subspace \#1}. 
Conversely, this means that 
output of {\sl Subspace \#0} is required to pass further from the center 
when it is near the diagonal line \( V^{<0>}_{1} = V^{<0>}_{0} \) 
or \( V^{<0>}_{1} = - V^{<0>}_{0} + 1.0 \), 
compared with the case when it is near the straight line 
\( V^{<0>}_{1} = 0.5 \) or \( V^{<0>}_{0} = 0.5 \). 
As shown in {\tt 17-S2-Slow-V<0>}, 
output trajectories in {\sl Subspace \#0} 
eventually become largish rounded squares 
to which full-orbed target trajectories are non-linearly expanded. 
Note that output trajectories in {\sl Subspace \#2} 
are identical with those in {\sl Subspace \#0} 
as shown in {\tt 17-S2-Slow-V<2>} and {\tt 17-S2-Slow-V<0>}; 
this accordance is common to that in the top or mid row graphs.

\vspace{3.00mm} 
\noindent
(1-4) Type \#2 Architecture

\vspace{3.00mm} 
\noindent
As the last case for this combination of mapping relationships, 
let us take a look at Fig. 18. 
In Type \#2 architecture, 
activity dynamics of the whole network are determined 
based on the widely detoured feedback loop 
through {\sl Internetworks} {\sl \#1} \& {\sl \#2}.

In this example where {\sl Internetwork \#2} 
has an expansive mapping relationship, 
output emitted from {\sl Backward Subnet \#2} 
becomes a certain level, 
even if the difference between feedback signals and input ones 
in {\sl Subspace \#2} is not very large. 
In spite of this situation, 
output of {\sl Backward Subnet \#1} comes to be fairly small 
due to the nature of contractive {\sl Internetwork \#1}. 
These characteristics do not generate an effect that makes 
the difference between feedback signals and input ones 
in {\sl Subspace \#2} diminish in {\sl Subspace \#0}; 
therefore, such difference in {\sl Subspace \#2} is kept large. 
As for the case with Fast Input, 
output of {\sl Subspace \#0} 
eventually draws small periodic trajectories 
shown in the leftmost graph of the top row 
({\tt 18-S2-Fast-V<0>}). 
Because of contractive {\sl Forward Subnet \#1}, 
output of {\sl Subspace \#1} becomes 
even smaller trajectories ({\tt 18-S2-Fast-V<1>}),  
which are expanded and sent to {\sl Subspace \#2} 
({\tt 18-S2-Fast-V<2>}) by {\sl Forward Subnet \#2} 
with the inverse mapping relationship to {\sl Forward Subnet \#1}'s; 
then, in the same way as before, 
the output of {\sl Subspace \#2} 
comes to be identical with that of {\sl Subspace \#0}. 
In addition, it is not to be missed that 
these processes are repeated without any improvement 
in accordance with the widely detoured feedback loop 
through {\sl Internetworks} {\sl \#1} \& {\sl \#2}. 
From a different perspective, 
note that {\tt 18-S2-Fast-V<2>} here
is completely the same as {\tt 16-S2-Fast-V<0>} in Fig. 16. 
This congruence means that 
``the output of {\sl Subspace \#2} in Type \#2 architecture 
based on the widely detoured feedback loop 
through two {\sl Internetworks} 
with mutually inverse mapping relationships" 
and 
``the output of {\sl Subspace \#0} in Type \#0 architecture 
according to the direct feedback loop" 
are identical with each other.

The mid row graphs show the case with Medium Speed Input. 
Output responses at this input speed quite improve 
as shown in {\tt 18-S2-Medium-V<0>}, 
{\tt 18-S2-Medium-V<1>}, and {\tt 18-S2-Medium-V<2>}, 
and the radii of the periodic output trajectories get larger overall 
compared with the top row graphs.

The bottom row graphs are simulation results 
for the case with Slow Input. 
Output trajectories in {\sl Subspace \#2} 
become full-orbed ones ({\tt 18-S2-Slow-V<2>}) 
that are the target signals themselves. 
In contrast to the situation in {\sl Subspace \#2}, 
output trajectories in {\sl Subspace \#1} must be deformed 
so that those in {\sl Subspace \#2} are full-orbed 
after going through expansive {\sl Forward Subnet \#2}. 
Output trajectories in {\sl Subspace \#0} have to be distorted 
so that those in {\sl Subspace \#1} represent {\tt 18-S2-Slow-V<1>} 
after passing through contractive {\sl Forward Subnet \#1}; 
the output trajectories in {\sl Subspace \#0} 
eventually become full-orbed as shown in {\tt 18-S2-Slow-V<0>}, 
which is the same as the output of {\sl Subspace \#2}.

As partly confirmed in the above explanation for Fig. 18, 
activity dynamics in Type \#2 architecture 
are identical with those in Type \#0 one 
regardless of how input signals from the outside are, 
when serially connected two {\sl Internetworks} have 
mutually inverse mapping relationships. 
It is actually intriguing that 
Fig. 18 for Type \#2 architecture 
and Fig. 16 for Type \#0 one 
are congruent for every frequency of periodic input signals.

\vspace{4.00mm} 
\noindent
{\large (2) Simulation I\hspace{-.1em}I (Expansive \( \rightarrow \) Expansive)}

\vspace{3.00mm} 
\noindent
(2-1) Overview

\vspace{3.00mm} 
\noindent
Figures 19, 20, and 21 illustrate another set of simulation results 
when both {\sl Internetworks} {\sl \#1} \& {\sl \#2} 
had expansive mapping relationships. 
These figures respectively correspond to 
Type \#0, Type \#1, and Type \#2 architectures 
of the Association Mode shown in Fig. 3. 
The construction of the graphs here 
is basically the same as that in Figs. 16, 17, and 18.

\vspace{3.00mm} 
\noindent
(2-2) Type \#0 Architecture

\vspace{3.00mm} 
\noindent
First, let us take a closer look at Fig. 19. 
The experimental conditions are common to those for Fig. 16 
except that {\sl Internetwork \#1} here is expansive. 
Since dynamical behavior of Type \#0 architecture 
is fundamentally determined only in {\sl Subspace \#0}, 
the output responses in {\sl Subspace \#0} here 
are identical with those in the same {\sl Subspace} of Fig. 16 
by each frequency of periodic input trajectories. 
What the output of {\sl Subspace \#0} was statically and expansively 
mapped to by {\sl Forward Subnet \#1} becomes output of {\sl Subspace \#1}, 
which is further expansively mapped in a static manner 
to {\sl Subspace \#2} by {\sl Forward Subnet \#2}.

In the top row graphs with Fast Input, output of {\sl Subspace \#0} 
becomes very small circular trajectories ({\tt 19-S4-Fast-V<0>}), 
which is identical with the leftmost graph 
in the top row of Fig. 16 ({\tt 16-S2-Fast-V<0>}) 
as mentioned above. 
And, the output of {\sl Subspace \#0} is statically and expansively 
mapped to {\sl Subspace \#1} as shown in {\tt 19-S4-Fast-V<1>}. 
Here, {\tt 19-S4-Fast-V<2>} is worthy of note. 
The output of {\sl Subspace \#2} is required to 
pass through the more peripheral area 
when it is near the diagonal line \( V^{<2>}_{1} = V^{<2>}_{0} \) 
or \( V^{<2>}_{1} = - V^{<2>}_{0} + 1.0 \), 
in comparison with the case when it is around the straight line 
\( V^{<2>}_{1} = 0.5 \) or \( V^{<2>}_{0} = 0.5 \). 
Eventually, the output trajectories in {\sl Subspace \#2} 
become rounded squares that are larger than 
those in {\sl Subspace \#1}.

In the mid row graphs with Medium Speed Input, 
the tracking characteristics of output of {\sl Subspace \#0} 
fairly improve in comparison with the top row graphs. 
In the same way as the case with Fast Input, 
what the output of {\sl Subspace \#0} 
was statically mapped to 
based on expansive {\sl Forward Subnet \#1} 
becomes output of {\sl Subspace \#1}, 
which is further mapped expansively to {\sl Subspace \#2} 
in a static manner by {\sl Forward Subnet \#2}. 
As a result, 
as shown in {\tt 19-S4-Medium-V<1>} and {\tt 19-S4-Medium-V<2>}, 
output trajectories in {\sl Subspaces} {\sl \#1} \& {\sl \#2} 
become larger periodic ones overall 
compared with those in the upper row; 
at the same time, they are larger than output trajectories 
in the mid row of Fig. 16 
({\tt 16-S2-Medium-V<1>} and {\tt 16-S2-Medium-V<2>}) 
by each {\sl Subspace} 
due to different mapping relationships of {\sl Forward Subnet \#1}.

In the bottom row graphs with Slow Input, 
an output trajectory in {\sl Subspace \#0} 
well follows a set of the target signals 
by means of dynamical neurons with the direct feedback loop, 
and it shows a full-orbed one 
that is identical with the input trajectory 
given from the outside. 
Note that the relation between 
{\tt 19-S4-Slow-V<0>} and {\tt 19-S4-Slow-V<1>} 
generated based on {\sl Forward Subnet \#1} here 
is completely the same as that between 
{\tt 17-S2-Slow-V<1>} and {\tt 17-S2-Slow-V<2>} 
produced according to {\sl Forward Subnet \#2} in Fig. 17. 
This is because there are no signals or little signals 
propagating through the corresponding {\sl Backward Subnets} 
in both cases 
and the output trajectories in these {\sl Subspaces} 
are determined only by static mapping relationships 
of the relevant {\sl Forward Subnets}. 
The output trajectories in {\sl Subspace \#1} are statically expanded 
toward the perimeter of a square on a coordinate plane 
by {\sl Forward Subnet \#2}, 
and output trajectories in {\sl Subspace \#2} become much larger ones 
in the shape of a rounded square ({\tt 19-S4-Slow-V<2>}).

\begin{figure}[p]

    \vspace*{-2.00mm}
    \hspace*{-2.25mm}
    \includegraphics[scale=0.88]{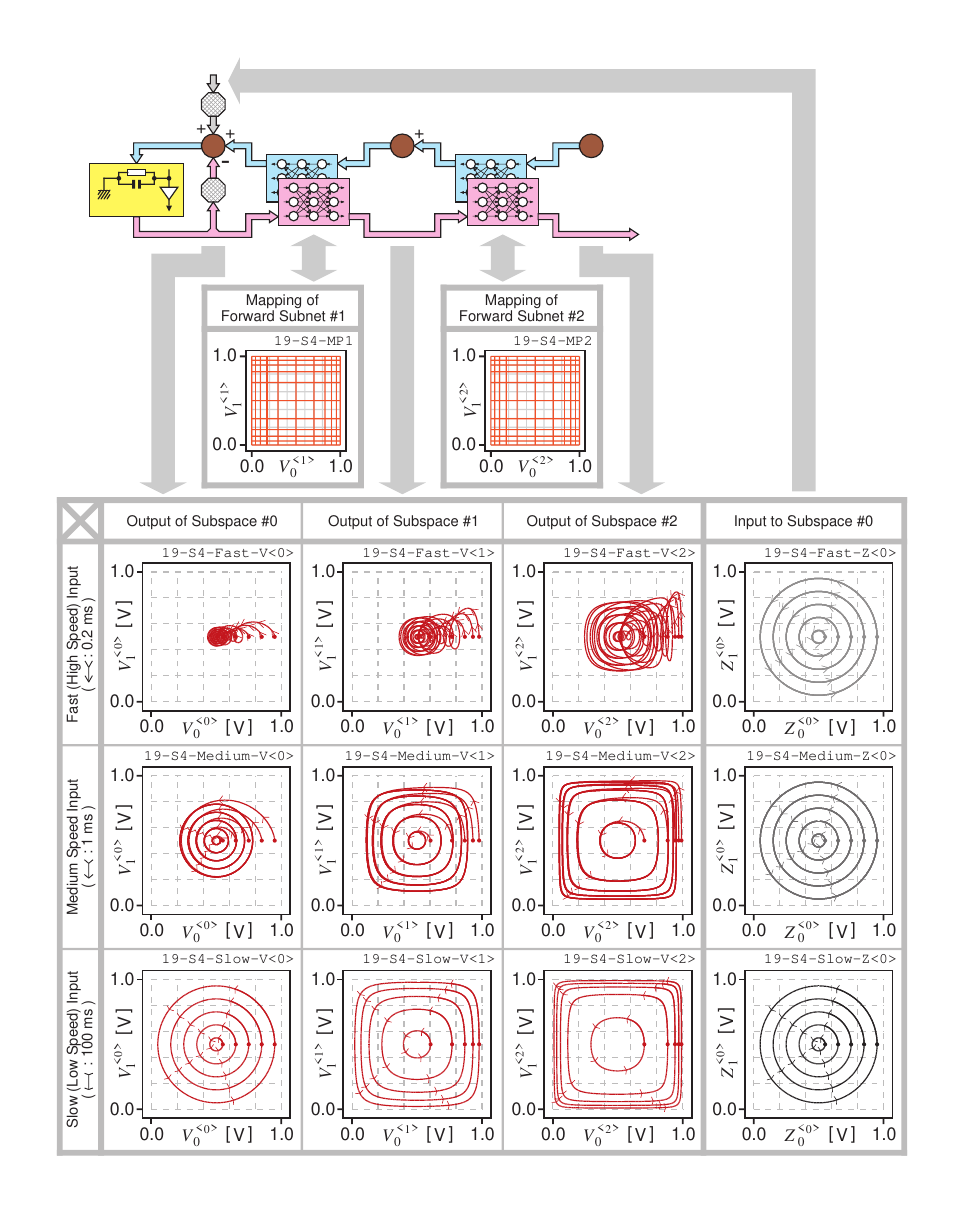}

    \vspace*{-4.00mm}
    \caption{ 
      Simulation results for Type \#0 architecture 
      in the Constrained Associative Mode 
      of the two-dimensional model with \( P = 2 \) 
      when mapping relationships 
      of {\sl Internetworks} {\sl \#1} \& {\sl \#2} 
      are both set to be expansive. 
      The main part (the bottom half) of the figure 
      is composed of three rows; 
      the frequency of the periodic input signals 
      applied from the outside 
      gets lower from the top to the bottom. 
      For the horizontal direction in each row, 
      the three graphs corresponding to the output responses 
      in {\sl Subspaces} {\sl \#0}, {\sl \#1}, \& {\sl \#2} 
      and the one graph for the circular target trajectories 
      are illustrated respectively from the left to the right. 
      To help understanding of the meaning of this figure, 
      a block diagram of the whole network 
      and mapping relationships of the two {\sl Internetworks} 
      are drawn along with wide gray arrow lines in the upper part. 
    } 

\end{figure}

\begin{figure}[p]

    \vspace*{-2.00mm}
    \hspace*{-2.25mm}
    \includegraphics[scale=0.88]{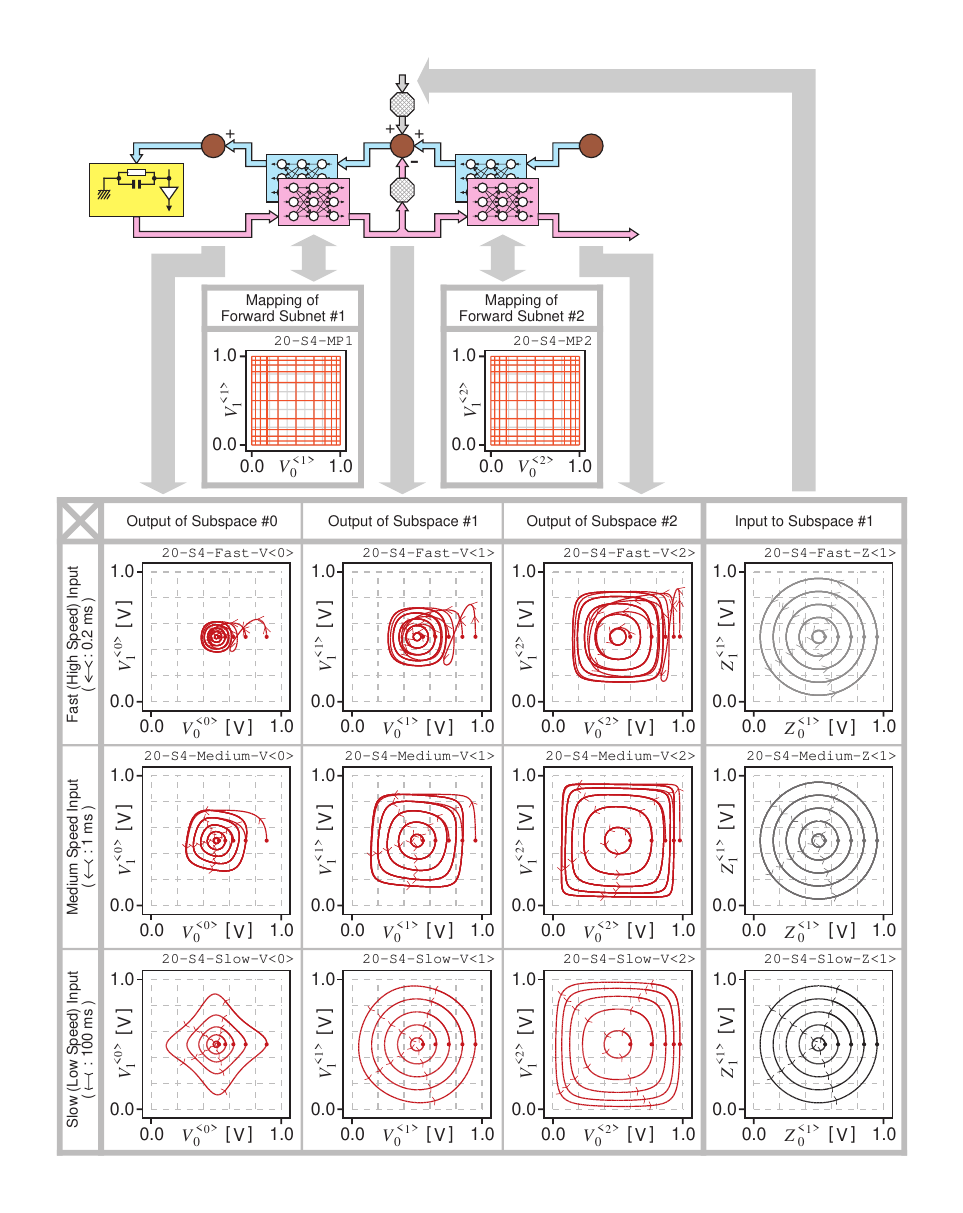}

    \vspace*{-4.00mm}
    \caption{ 
      Simulation results for Type \#1 architecture 
      in the Constrained Associative Mode 
      of the two-dimensional model with \( P = 2 \) 
      when mapping relationships 
      of {\sl Internetworks} {\sl \#1} \& {\sl \#2} 
      are both set to be expansive. 
      The main part (the bottom half) of the figure 
      is composed of three rows; 
      the frequency of the periodic input signals 
      applied from the outside 
      gets lower from the top to the bottom. 
      For the horizontal direction in each row, 
      the three graphs corresponding to the output responses 
      in {\sl Subspaces} {\sl \#0}, {\sl \#1}, \& {\sl \#2} 
      and the one graph for the circular target trajectories 
      are illustrated respectively from the left to the right. 
      To help understanding of the meaning of this figure, 
      a block diagram of the whole network 
      and mapping relationships of the two {\sl Internetworks} 
      are drawn along with wide gray arrow lines in the upper part. 
    } 

\end{figure}

\begin{figure}[p]

    \vspace*{-2.00mm}
    \hspace*{-2.25mm}
    \includegraphics[scale=0.88]{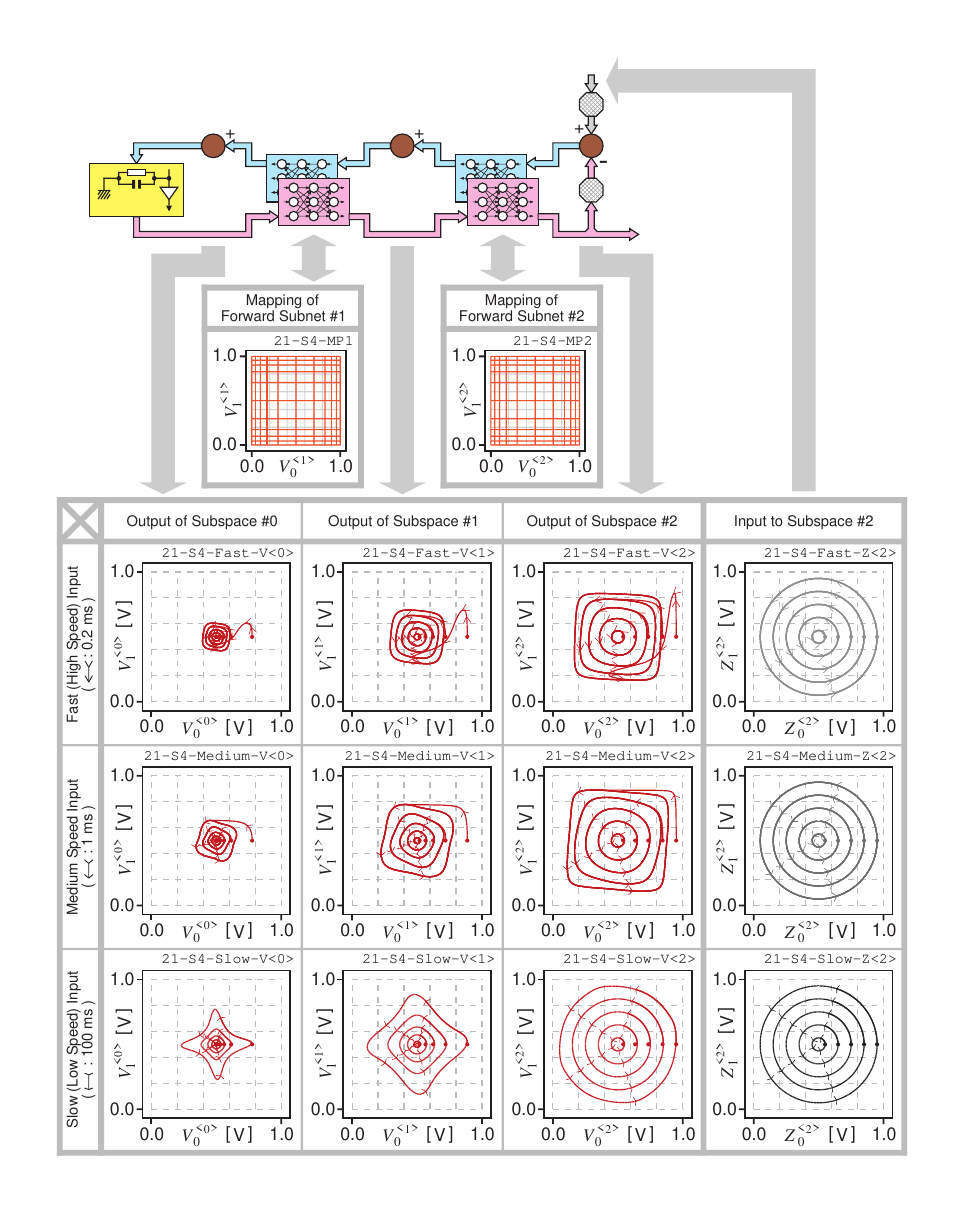}

    \vspace*{-4.00mm}
    \caption{ 
      Simulation results for Type \#2 architecture 
      in the Constrained Associative Mode 
      of the two-dimensional model with \( P = 2 \) 
      when mapping relationships 
      of {\sl Internetworks} {\sl \#1} \& {\sl \#2} 
      are both set to be expansive. 
      The main part (the bottom half) of the figure 
      is composed of three rows; 
      the frequency of the periodic input signals 
      applied from the outside 
      gets lower from the top to the bottom. 
      For the horizontal direction in each row, 
      the three graphs corresponding to the output responses 
      in {\sl Subspaces} {\sl \#0}, {\sl \#1}, \& {\sl \#2} 
      and the one graph for the circular target trajectories 
      are illustrated respectively from the left to the right. 
      To help understanding of the meaning of this figure, 
      a block diagram of the whole network 
      and mapping relationships of the two {\sl Internetworks} 
      are drawn along with wide gray arrow lines in the upper part. 
    } 

\end{figure}

\vspace{3.00mm} 
\noindent
(2-3) Type \#1 Architecture

\vspace{3.00mm} 
\noindent
The experimental conditions in Fig. 20 
are basically the same as those in Fig. 17 
except that {\sl Forward Subnet \#1}'s mapping relationships 
differ from each other. 
Nevertheless, we can see at a glance that 
the aspects of the top row and mid row graphs in Fig. 20 
are very different from those in Fig. 17 in many ways.

Even when input to {\sl Backward Subnet \#1} is not very large, 
output from it becomes a certain level 
owing to the nature of expansive {\sl Internetwork \#1}. 
This property produces an effect such that 
it makes the difference between feedback signals and input ones 
in {\sl Subspace \#1} decrease in {\sl Subspace \#0}; 
in other words, it generates an effect that 
brings an output of {\sl Subspace \#1} 
close to the value of an input from the outside. 
Let us compare the top row graphs in Fig. 20 with those in Fig. 17; 
both of them are for Type \#1 architecture with Fast Input. 
Owing to the above-mentioned effect, 
the radii of periodic output trajectories in {\tt 20-S4-Fast-V<0>} 
are larger than those in {\tt 17-S2-Fast-V<0>}. 
The simulation results in which 
the radii of periodic output trajectories in {\tt 20-S4-Fast-V<1>} 
are bigger than those in {\tt 17-S2-Fast-V<1>}, as expected, 
stem from the dissimilarity as to 
whether {\sl Forward Subnet \#1} is expansive or contractive. 
From a different perspective, 
it is significant to compare {\tt 20-S4-Fast-V<1>} here 
with {\tt 19-S4-Fast-V<0>} in Fig. 19. 
The radii of periodic trajectories in the former 
are obviously larger than those in the latter. 
This suggests that 
the tracking property of activity dynamics 
based on the detoured feedback loop 
through expansive {\sl Internetwork \#1} 
is better than that according to the direct feedback loop 
only inside {\sl Subspace \#0}. 
The output trajectories in {\sl Subspace \#1} 
are expansively mapped by {\sl Forward Subnet \#2} 
in a static manner to {\sl Subspace \#2}, 
where output trajectories in the shape of a larger rounded square 
are drawn ({\tt 20-S4-Fast-V<2>}).

The mid row graphs show the case with Medium Speed Input. 
%
%
Comparing {\tt 20-S4-Medium -V<1>} here 
with {\tt 20-S4-Fast-V<1>} right above, 
the input-to-output tracking characteristics in the former 
get better than those in the latter, 
and the output of {\sl Subspace \#1} 
becomes largish periodic trajectories. 
Along with this behavior in {\sl Subspace \#1}, 
output of {\sl Subspace \#0} 
becomes larger trajectories ({\tt 20-S4-Medium-V<0>}) 
in comparison with the upper graph. 
The mapping relationship of {\sl Forward Subnet \#1} here 
is two-dimensionally expansive in a square on a coordinate plane. 
Therefore, an output trajectory 
near the diagonal line \( V^{<1>}_{1} = V^{<1>}_{0} \) 
or \( V^{<1>}_{1} = - V^{<1>}_{0} + 1.0 \) 
tends to pass further away from the center, 
compared with that near the straight line 
\( V^{<1>}_{1} = 0.5 \) or \( V^{<1>}_{0} = 0.5 \). 
The output of {\sl Forward Subnet \#1} is conveyed 
through {\sl Backward Subnet \#1} to {\sl Subspace \#0} 
after the input from the outside has been subtracted 
in {\sl Subspace \#1}; 
this procedure is repeated within 
the detoured feedback loop including {\sl Internetwork \#1}. 
Thus, output trajectories in {\sl Subspaces} {\sl \#0} \& {\sl \#1} 
are determined based on the trade-off relationship between 
``the time constant of dynamical neurons in {\sl Subspace \#0}" and 
``the frequency of input signals from the outside in {\sl Subspace \#1}." 
It should be noted that, as a result, 
the output of {\sl Subspace \#0} is shaped in the form of 
not a circle but a little bit rounded square 
and the output of {\sl Subspace \#1} 
clearly becomes a largish rounded square, 
even though the trajectory of an input signal 
applied from the outside is full-orbed. 
On the other hand, 
output of {\sl Subspace \#2} draws a more squared trajectory 
({\tt 20-S4-Medium-V<2>}) as a statically extended version 
of the output of {\sl Subspace \#1} 
due to expansive {\sl Forward Subnet \#2}.

In the bottom row graphs with Slow Input, 
output trajectories in {\sl Subspace \#1} 
well follow the full-orbed target signals 
as shown in {\tt 20-S4-Slow-V<1>}; 
this behavior is similar to {\tt 17-S2-Slow-V<1>} in Fig. 17. 
In contrast to the situation in {\sl Subspace \#1}, 
output trajectories in {\sl Subspace \#0} are determined 
so that those in {\sl Subspace \#1} get full-orbed 
owing to the cooperation of 
``dynamical neurons in {\sl Subspace \#0}" and 
``the detoured feedback loop via {\sl Internetwork \#1} 
between {\sl Subspaces} {\sl \#0} \& {\sl \#1}." 
Let us pay attention to the following two points 
on anyone of the full-orbed trajectories in {\sl Subspace \#1}: 
One is the point that crosses the diagonal line 
\( V^{<1>}_{1} = V^{<1>}_{0} \) 
or \( V^{<1>}_{1} = - V^{<1>}_{0} + 1.0 \). 
The other is the one that goes across 
the straight line \( V^{<1>}_{1} = 0.5 \) or \( V^{<1>}_{0} = 0.5 \). 
In {\sl Subspace \#0}, an output trajectory at the point 
corresponding to the former point in {\sl Subspace \#1} 
must pass more inside than 
the one at the point 
that corresponds to the latter point in {\sl Subspace \#1} 
because of the property of expansive {\sl Forward Subnet \#1}. 
As a result, the output trajectories in {\sl Subspace \#0} 
become a little smaller rounded squares rotated by \( 45^{\circ} \). 
Note here that 
the relation between 
{\tt 20-S4-Slow-V<0>} and {\tt 20-S4-Slow-V<1>} in Fig. 20 
is the same as the one between 
{\tt 16-S2-Slow-V<1>} and {\tt 16-S2-Slow-V<2>} in Fig. 16 
or 
the one between 
{\tt 18-S2-Slow-V<1>} and {\tt 18-S2-Slow-V<2>} in Fig. 18, 
since the architectures are different but the mapping relationships 
of the corresponding {\sl Internetworks} are simply identical. 
Output trajectories in {\sl Subspace \#2} 
draw largish rounded squares 
based on expansive {\sl Forward Subnet \#2}. 
In addition, the relation between 
{\tt 20-S4-Slow-V<1>} and {\tt 20-S4-Slow-V<2>} here 
is the same as the one between 
{\tt 17-S2-Slow-V<1>} and {\tt 17-S2-Slow-V<2>} in Fig. 17, 
and is also identical with the one between 
{\tt 19-S4-Slow-V<0>} and {\tt 19-S4-Slow-V<1>} in Fig. 19. 
In any cases, it should be noted that 
there are quite small signals or no ones in principle 
propagating through {\sl Backward Subnets} 
and output trajectories in each {\sl Subspace} are yielded 
only based on static mapping relationships of {\sl Forward Subnets}.

\vspace{3.00mm} 
\noindent
(2-4) Type \#2 Architecture

\vspace{3.00mm} 
\noindent
Figure 21 shows activity dynamics of Type \#2 architecture, 
which are generated based on the widely detoured feedback loop 
with {\sl Internetworks} {\sl \#1} \& {\sl \#2} 
in the same way as the case of Fig. 18.

Since {\sl Forward Subnet \#2} here 
is not contractive but expansive, 
output of the corresponding {\sl Backward Subnet \#2} 
does not become a very small value 
even if input to it is not large. 
The output of {\sl Backward Subnet \#2} 
is further sent to the input part of {\sl Backward Subnet \#1}, 
where the same computation as that in {\sl Backward Subnet \#2} 
is carried out. 
Consequently, even if the difference 
between feedback signals and input ones 
in {\sl Subspace \#2} is not so great, 
output of {\sl Backward Subnet \#1} reaches a certain level 
and it produces an effect that makes such difference 
in {\sl Subspace \#2} diminish in {\sl Subspace \#0}. 
In fact, regarding the top row graphs with Fast Input, 
output trajectories in {\sl Subspace \#0} here 
({\tt 21-S4-Fast-V<0>})
are slightly larger than those in the same {\sl Subspace} of Fig. 19 
({\tt 19-S4-Fast-V<0>}). 
The output trajectories in {\sl Subspace \#0} 
are expansively sent in turn 
to {\sl Subspaces} {\sl \#1} \& {\sl \#2} 
based on the widely detoured feedback loop 
with {\sl Internetworks} {\sl \#1} \& {\sl \#2}
that are both expansive; 
eventually, output trajectories in {\sl Subspace \#2} 
become fairly large rounded squares 
as indicated in {\tt 21-S4-Fast-V<2>}. 
From a different standpoint, 
it is of great interest to make a comparison between 
{\tt 19-S4-Fast-V<0>} 
(output of {\sl Subspace \#0} in Type \#0 architecture), 
{\tt 20-S4-Fast-V<1>} 
(output of {\sl Subspace \#1} in Type \#1 architecture), 
and {\tt 21-S4-Fast-V<2>} 
(output of {\sl Subspace \#2} in Type \#2 architecture). 
Looking at these three graphs with Fast Input, 
their output trajectories are larger in the order given. 
This is because greater also in this order is an effect 
that makes the difference between feedback signals and input ones 
(in the {\sl Subspace} to which input signals are applied from the outside)
lessen by means of a direct feedback loop, a detoured feedback loop, 
or a widely detoured feedback loop.

Comparing output for the case with Fast Input 
in Fig. 19 (Type \#0 architecture) 
with that in Fig. 20 (Type \#1 architecture) 
this time by each {\sl Subspace}, 
the circular trajectories in the latter 
are bigger than those in the former 
throughout every {\sl Subspace}. 
Next, let us make a comparison between 
output for the case with Fast Input in Fig. 20 (Type \#1 architecture) 
and that in Fig. 21 (Type \#2 architecture) 
also by each {\sl Subspace}. 
The circular trajectories in the latter 
are almost the same as or slightly smaller than those in the former 
throughout every {\sl Subspace}. 
In this way, variation from Fig. 19 through Fig. 20 to Fig. 21 
is not monotonous, 
so these relationships may seem strange at first glance. 
However, the associative dynamics are determined, 
through repeated pushing and pulling, 
comprehensively by various factors 
such as the relations between 
``the frequency of target signals applied from the outside" 
and ``the time constant of a dynamical neuron," 
the distinctions in architectures 
(i.e., the difference of signals flowing through {\sl Backward Subnets}), 
and the differences in mapping relationships of {\sl Forward Subnets}. 
Conversely, such situations are thought to suggest that, 
just by lowering the frequency of input target signals 
applied from the outside for instance, 
the differences in associative dynamics depending on architectures 
could become more clear and straightforward.

In the mid row graphs with Medium Speed Input, 
the tracking property gets better overall 
as shown in {\tt 21-S4-Medium-V<0>}, 
{\tt 21-S4-Medium-V<1>}, and {\tt 21-S4-Medium-V<2>}, 
and output trajectories in every {\sl Subspace} 
represent larger periodic ones 
compared with the top row graphs. 
Here, let us compare 
between output for the case with Medium Speed Input 
in Fig. 19 (Type \#0 architecture), 
that in Fig. 20 (Type \#1 architecture), 
and that in Fig. 21 (Type \#2 architecture) 
by each {\sl Subspace}. 
Quite interestingly, we can see that 
output trajectories in each {\sl Subspace} become smaller 
in the order of Type \#0, Type \#1, and Type \#2 architectures. 
In this way, if the frequency of input target signals 
is lowered up to that at Medium Speed Input, 
variation from Fig. 19 through Fig. 20 to Fig. 21 becomes monotonous; 
these results provide evidence to support the suggestion 
mentioned above for the case with Fast Input. 
Specifically, in a {\sl Subspace} 
to which input signals are given from the outside, 
an output trajectory is going to get closer to a full-orbed one 
corresponding to the input signals. 
Considering ``output trajectories in the {\sl Subspace} 
to which input signals are applied from the outside" 
as a basis for comparison, 
output trajectories in its right (outer) {\sl Subspaces} (if they exist) 
will get more expansive, 
and those in its left (inner) {\sl Subspaces} (if they exist) 
will become more contractive. 
Regarding the mid row graphs with Medium Speed Input 
in Figs. 19, 20, and 21, 
an output trajectory in every {\sl Subspace} 
reflects the complex interaction 
between various factors mentioned above.

In the bottom row graphs with Slow Input, 
output trajectories in {\sl Subspace \#2} 
well track the full-orbed target signals, 
as shown in {\tt 21-S4-Slow-V<2>}. 
{\sl Forward Subnet \#2} here is two-dimensionally expansive 
in a square on a coordinate plane. 
Accordingly, 
when an output trajectory in {\sl Subspace \#2} passes 
near the diagonal line \( V^{<2>}_{1} = V^{<2>}_{0} \) 
or \( V^{<2>}_{1} = - V^{<2>}_{0} + 1.0 \), 
the corresponding output one in {\sl Subspace \#1} 
has to get closer to the center, 
compared with the case in which 
an output trajectory in {\sl Subspace \#2} 
goes near the straight line 
\( V^{<2>}_{1} = 0.5 \) or \( V^{<2>}_{0} = 0.5 \). 
Output trajectories in {\sl Subspace \#1} 
eventually become smallish rounded squares 
rotated by \( 45^{\circ} \). 
Also in this case, the following facts are worthy of notice: 
The relation between output trajectories in {\sl Subspace \#1} 
and those in {\sl Subspace \#2}, 
i.e., the relation between 
{\tt 21-S4-Slow-V<1>} and {\tt 21-S4-Slow-V<2>}, 
is identical with that between 
{\tt 16-S2-Slow-V<1>} and {\tt 16-S2-Slow-V<2>} in Fig. 16 
or that between 
{\tt 18-S2-Slow-V<1>} and {\tt 18-S2-Slow-V<2>} in Fig. 18. 
Additionally, it is the same as the relation between 
{\tt 20-S4-Slow-V<0>} and {\tt 20-S4-Slow-V<1>} in Fig. 20. 
On the other hand, an output trajectory in {\sl Subspace \#0} 
emerges as what the output in {\sl Subspace \#1} was contracted to 
accompanied with emphasized peaks and troughs, 
and it becomes a smaller periodic trajectory 
whose shape is like a four-point Ninja Star 
as shown in {\tt 21-S4-Slow-V<0>}.

\subsection{Comparison of Simulation Results}

\vspace{3.00mm}
\noindent
{\large (1) Simulation Results}

\vspace{3.00mm}
\noindent
In the previous subsection, 
we chose two different combinations of mapping relationships 
for {\sl Internetworks} {\sl \#1} \& {\sl \#2}, 
and examined the associative activity dynamics 
for Type \#0, Type \#1, and Type \#2 architectures 
with those {\sl Internetworks} 
when various periodic target signals were applied 
to the corresponding input port from the outside. 
We then confirmed that 
their tracking behaviors were markedly different from each other 
depending on the architectures, 
the mapping relationships of {\sl Internetworks}, 
and the frequency of periodic inputs. 
In the case with Slow Input, 
we further found out that output trajectories 
determined only by {\sl Internetworks}' static mapping relationships 
appeared in each {\sl Subspace}.

Limiting the discussion to the cases 
with Slow Input in this subsection, 
we append another two different combinations 
of mapping relationships for {\sl Internetworks} {\sl \#1} \& {\sl \#2} 
to the previous two ones adopted in Subsection 5.2 
(i.e., we employ totally four different combinations), 
and compare dynamics of the networks 
with their sets of {\sl Internetworks} 
in the Constrained Association Mode. 
Figures 22, 23, and 24 respectively illustrate  
the simulation results for Type \#0, Type \#1, and Type \#2 architectures; 
these figures are individually composed of four rows, each of which 
is based on a different combination of mapping relationships 
for {\sl Forward Subnets}. 
For the horizontal direction in each row, 
the leftmost graph shows output of {\sl Subspace \#0}, 
its right one is the mapping relationship of {\sl Forward Subnet \#1}, 
the central one depicts output of {\sl Subspace \#1}, 
the next right one is the mapping relationship of {\sl Forward Subnet \#2}, 
and the rightmost one illustrates output of {\sl Subspace \#2}. 
The separated graph at the upper right in each figure 
indicates periodic input signals applied from the outside, 
and it is identical with the rightmost graphs 
at the bottom rows of Figs. 16-21. 
For better understanding, we appended a block diagram 
of the corresponding architecture 
at the upper part of each figure 
along with wide gray auxiliary arrows. 
The graphs for output responses 
at the second and fourth rows in Figs. 22, 23, and 24 
respectively correspond to 
those at the bottom rows in Figs. 16-21; 
we repeatedly show them for comparison purposes. 
What these simulation results have in common 
is that output responses in a {\sl Subspace} 
to which input signals are applied from the outside 
perfectly track the input ones 
regarding every architecture in the Association Mode, 
since the frequency of the sinusoidal target signals 
is sufficiently low 
compared with the time constant of a dynamical neuron. 
The total dynamics eventually progress 
with very small signals or no signals 
propagating through {\sl Backward Subnets}. 
With respect to Figs. 22, 23, and 24, 
it is extremely important that 
the mutual relation between output trajectories 
in adjacent {\sl Subspaces} is fundamentally determined 
only according to a ``static" mapping relationship 
of the corresponding {\sl Forward Subnet}.

\vspace{4.00mm}
\noindent
{\large (2) Comparison A}

\vspace{3.00mm}
\noindent
Comparing the graphs in the uppermost row of Fig. 22 
with those in the bottom one of Fig. 24, 
we notice that 
{\tt 22-S1-V<0>} and {\tt 24-S4-V<2>}, 
{\tt 22-S1-V<1>} and {\tt 24-S4-V<1>}, and 
{\tt 22-S1-V<2>} and {\tt 24-S4-V<0>} 
are respectively identical with each other. 
Let us look at the top row of Fig. 22 in detail. 
Output trajectories in {\sl Subspace \#0}, 
which are the same as the full-orbed target signals, 
are sent to outer {\sl Subspace \#1} 
through contractive {\sl Forward Subnet \#1}, 
and they become output ones 
in the shape of a \( 45^{\circ} \) slanted and smaller rounded square. 
The output trajectories there are further transmitted 
through contractive {\sl Forward Subnet \#2} to outer {\sl Subspace \#2}, 
where those in the shape similar to a four-point Ninja Star appear. 
The lowermost row of Fig. 24 is in contrast to this. 
Output trajectories 
in the shape of a \( 45^{\circ} \) slanted and smaller rounded square 
are produced in {\sl Subspace \#1}, 
in such a manner that 
full-orbed output ones in {\sl Subspace \#2} 
can be realized through expansive {\sl Forward Subnet \#2}. 
In {\sl Subspace \#0}, 
output trajectories are generated 
in the shape similar to a four-point Ninja Star, 
so that those in the shape of 
a \( 45^{\circ} \) slanted and smaller rounded square 
can be produced in {\sl Subspace \#1} 
through expansive {\sl Forward Subnet \#1}. 
From the other perspective, 
we can grasp these results in the following way: 
The output of {\sl Subspace \#1} ({\tt 24-S4-V<1>}) is 
what the one of {\sl Subspace \#2} ({\tt 24-S4-V<2>}) 
was converted to 
based on a contractive mapping relationship 
which is the ``inverse" of actual expansive {\sl Forward Subnet \#2}'s. 
In addition, 
the output of {\sl Subspace \#0} ({\tt 24-S4-V<0>}) becomes 
what the one of {\sl Subspace \#1} ({\tt 24-S4-V<1>}) 
was further transformed to 
according to a contractive mapping relationship 
which is the ``inverse" of real expansive {\sl Forward Subnet \#1}'s.

Thus, we can understand the relation between 
the top row graphs of Fig. 22 and 
the lowermost row ones of Fig. 24 in the following way: 
In the former graphs, 
full-orbed output trajectories in {\sl Subspace \#0} 
are sent through two-staged contractive {\sl Forward Subnets} 
to {\sl Subspace \#2}, 
where output ones are produced 
in the shape similar to a four-point Ninja Star. 
Contrary to this, in the latter graphs, 
full-orbed output trajectories in {\sl Subspace \#2} 
are transmitted in the opposite direction 
through two-staged ``imaginary" static layered networks 
with contractive mapping relationships 
(each of which is the ``inverse" of existing expansive 
{\sl Forward Subnet} {\sl \#1}'s or {\sl \#2}'s) 
to {\sl Subspace \#0}, 
where output ones are generated 
in the shape similar to a four-point Ninja Star.

\begin{figure}[p]

    \vspace*{-2.00mm}
    \hspace*{-1.00mm}
    \includegraphics[scale=0.78]{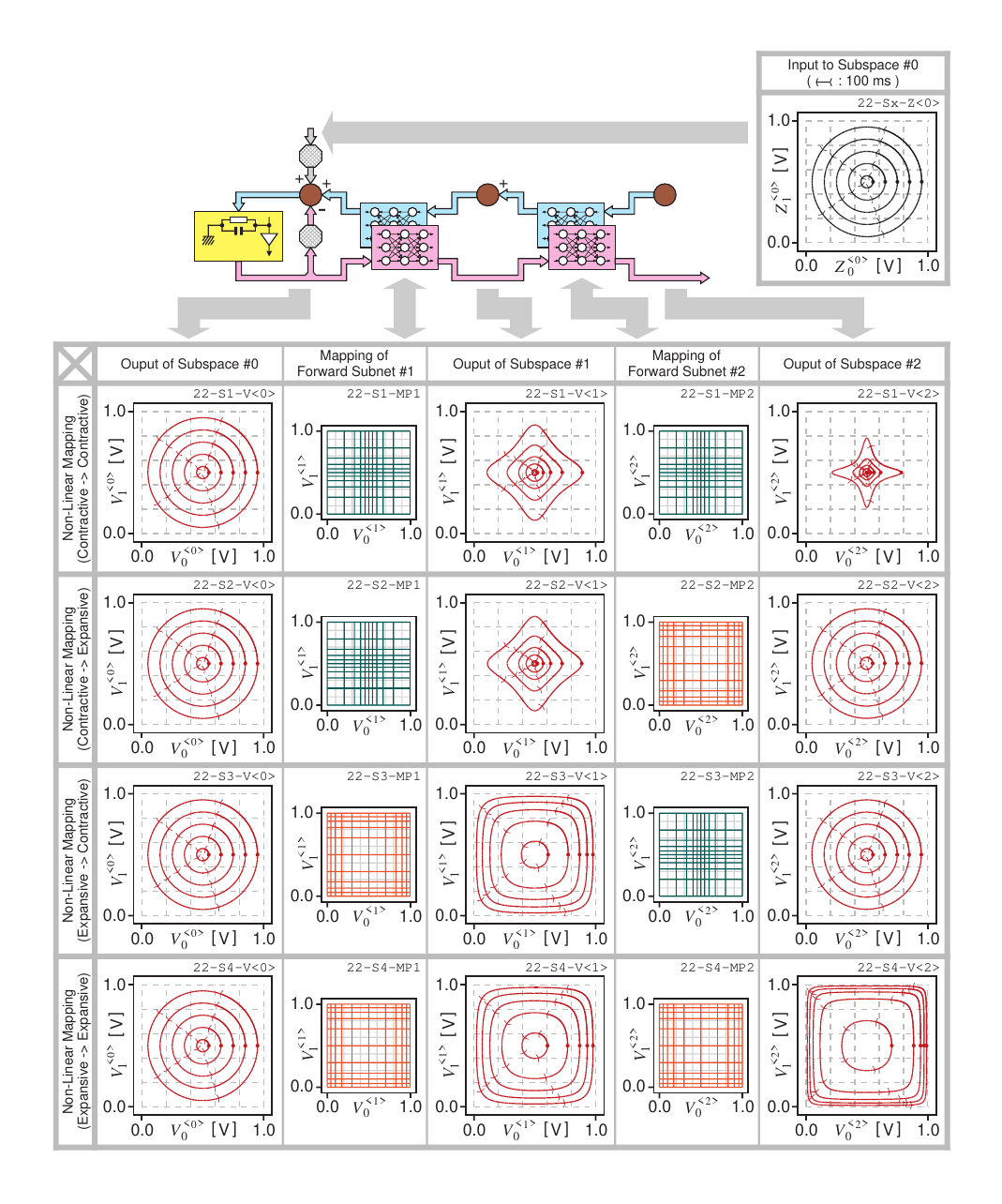}

    \vspace*{-4.00mm}
    \caption{ 
      Summary of simulation results for Type \#0 architecture 
      in the Constrained Associative Mode 
      of the two-dimensional model with \( P = 2 \) 
      particularly under Slow (Low Speed) Input indicated in Fig. 15. 
      The main part of the figure is composed of four rows, 
      each of which is based on a different combination 
      of mapping relationships 
      for {\sl Internetworks} {\#1} \& {\#2}. 
      For the horizontal direction in each row, 
      output of {\sl Subspace \#0}, 
      the mapping relationship of {\sl Forward Subnet \#1}, 
      output of {\sl Subspace \#1}, 
      the mapping relationship of {\sl Forward Subnet \#2}, 
      and output of {\sl Subspace \#2} 
      are illustrated respectively from the left to the right. 
      The separated graph at the upper right 
      specifically shows the periodic input signals 
      applied from the outside; 
      it is the same as the rightmost graphs 
      at the bottom rows of Figs. 16-21. 
      To aid understanding of the relation between the graphs, 
      a block diagram of the whole network 
      with wide gray arrow lines is drawn on the top. 
      Note that the graphs in the second row from the top 
      and those in the bottom row here are identical 
      respectively with those in the bottom row of Fig. 16 
      and those in the bottom row of Fig. 19. 
    } 

\end{figure}

\begin{figure}[p]

    \vspace*{-2.00mm}
    \hspace*{-1.00mm}
    \includegraphics[scale=0.78]{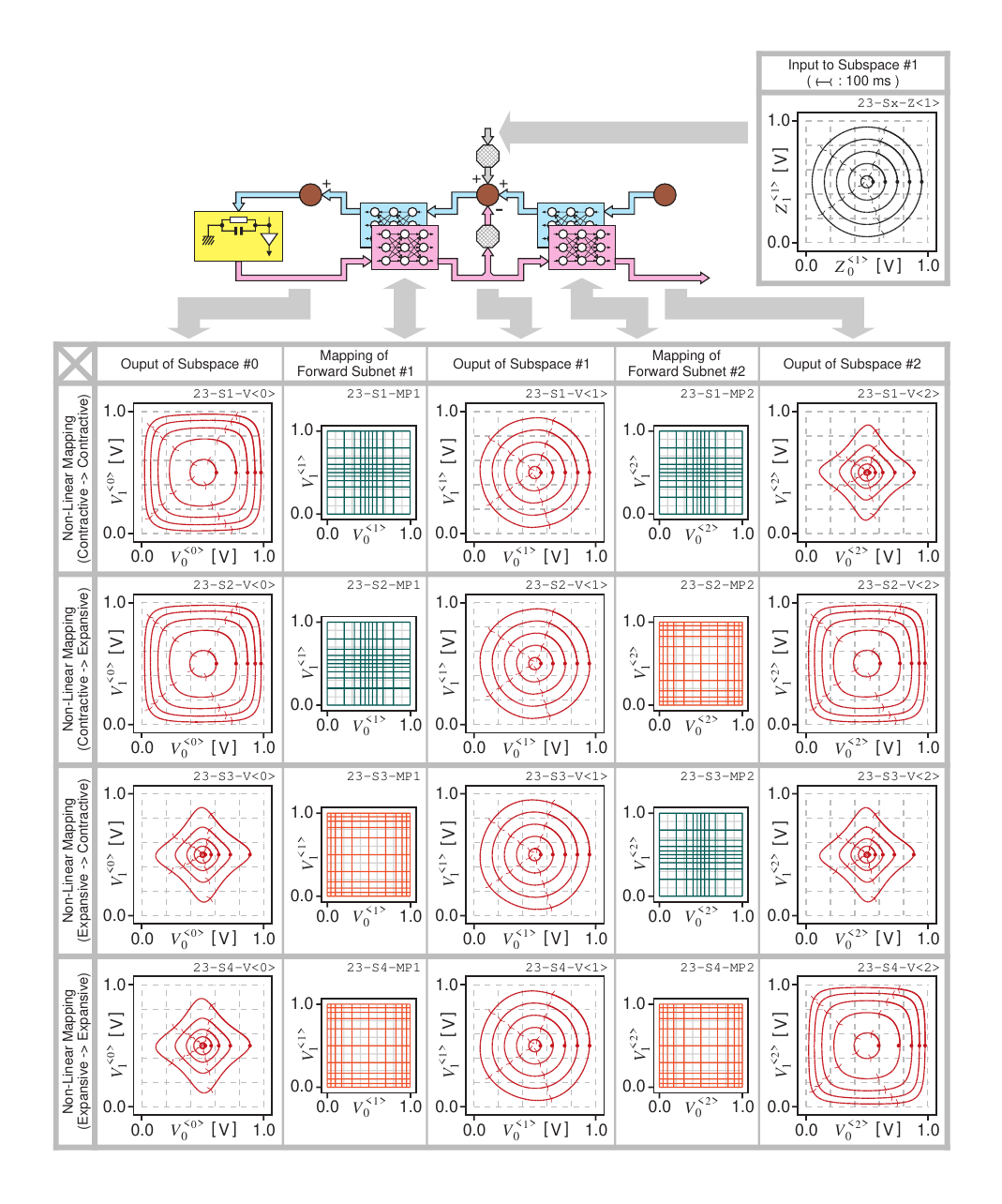}

    \vspace*{-4.00mm}
    \caption{ 
      Summary of simulation results for Type \#1 architecture 
      in the Constrained Associative Mode 
      of the two-dimensional model with \( P = 2 \) 
      particularly under Slow (Low Speed) Input indicated in Fig. 15. 
      The main part of the figure is composed of four rows, 
      each of which is based on a different combination 
      of mapping relationships 
      for {\sl Internetworks} {\#1} \& {\#2}. 
      For the horizontal direction in each row, 
      output of {\sl Subspace \#0}, 
      the mapping relationship of {\sl Forward Subnet \#1}, 
      output of {\sl Subspace \#1}, 
      the mapping relationship of {\sl Forward Subnet \#2}, 
      and output of {\sl Subspace \#2} 
      are illustrated respectively from the left to the right. 
      The separated graph at the upper right 
      specifically shows the periodic input signals 
      applied from the outside; 
      it is the same as the rightmost graphs 
      at the bottom rows of Figs. 16-21. 
      To aid understanding of the relation between the graphs, 
      a block diagram of the whole network 
      with wide gray arrow lines is drawn on the top. 
      Note that the graphs in the second row from the top 
      and those in the bottom row here are identical 
      respectively with those in the bottom row of Fig. 17 
      and those in the bottom row of Fig. 20. 
    } 

\end{figure}

\begin{figure}[p]

    \vspace*{-2.00mm}
    \hspace*{-1.00mm}
    \includegraphics[scale=0.78]{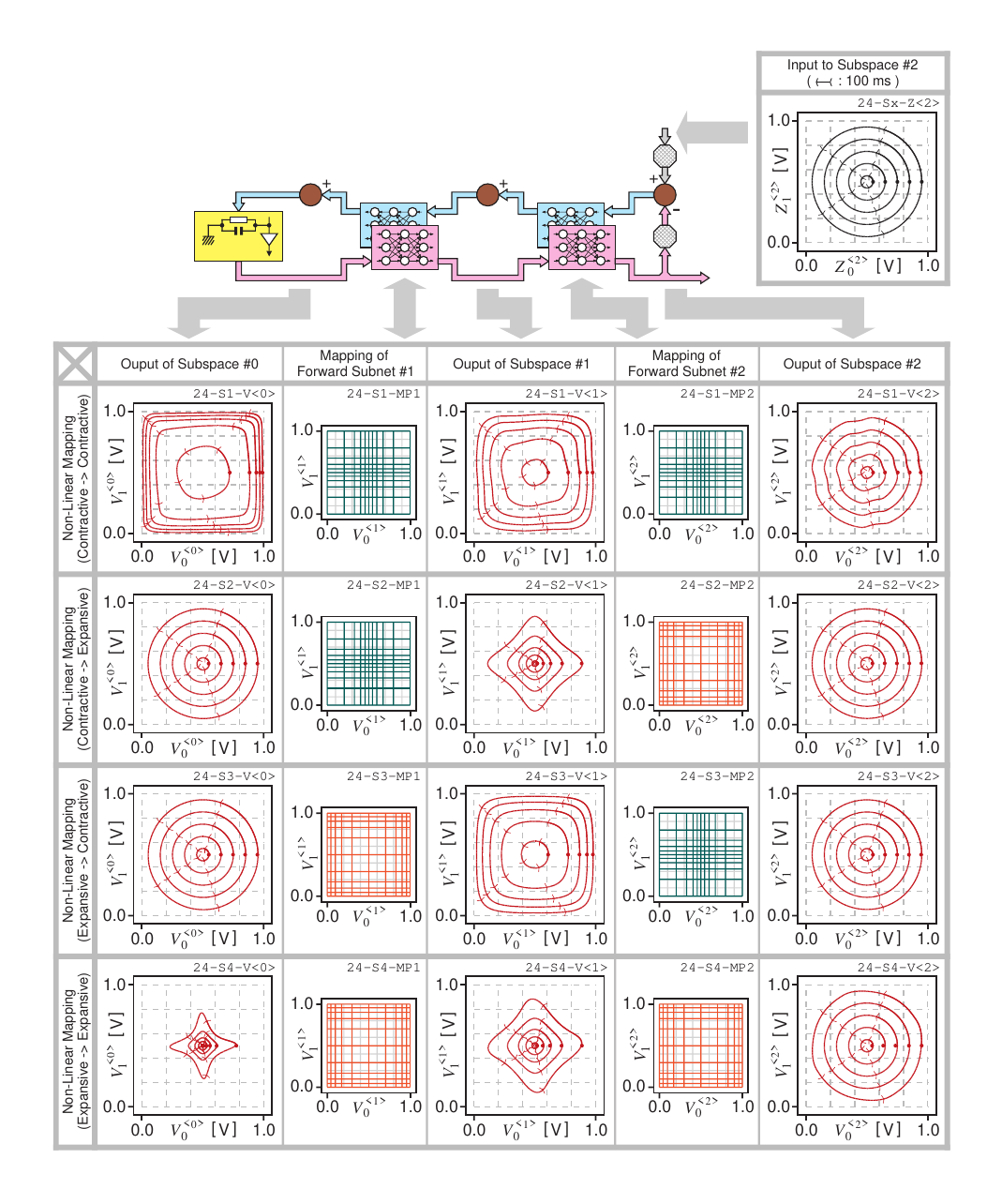}

    \vspace*{-4.00mm}
    \caption{ 
      Summary of simulation results for Type \#2 architecture 
      in the Constrained Associative Mode 
      of the two-dimensional model with \( P = 2 \) 
      particularly under Slow (Low Speed) Input indicated in Fig. 15. 
      The main part of the figure is composed of four rows, 
      each of which is based on a different combination 
      of mapping relationships 
      for {\sl Internetworks} {\#1} \& {\#2}. 
      For the horizontal direction in each row, 
      output of {\sl Subspace \#0}, 
      the mapping relationship of {\sl Forward Subnet \#1}, 
      output of {\sl Subspace \#1}, 
      the mapping relationship of {\sl Forward Subnet \#2}, 
      and output of {\sl Subspace \#2} 
      are illustrated respectively from the left to the right. 
      The separated graph at the upper right 
      specifically shows the periodic input signals 
      applied from the outside; 
      it is the same as the rightmost graphs 
      at the bottom rows of Figs. 16-21. 
      To aid understanding of the relation between the graphs, 
      a block diagram of the whole network 
      with wide gray arrow lines is drawn on the top. 
      Note that the graphs in the second row from the top 
      and those in the bottom row here are identical 
      respectively with those in the bottom row of Fig. 18 
      and those in the bottom row of Fig. 21. 
    } 

\end{figure}

\vspace{4.00mm}
\noindent
{\large (3) Comparison B}

\vspace{3.00mm}
\noindent
Next contrasting the graphs in the lowermost row of Fig. 22 
with those in the top one of Fig. 24, 
we realize that 
{\tt 22-S4-V<0>} and {\tt 24-S1-V<2>}, 
{\tt 22-S4-V<1>} and {\tt 24-S1-V<1>}, and 
{\tt 22-S4-V<2>} and {\tt 24-S1-V<0>} 
are respectively the same as each other, 
although the output trajectories in {\sl Subspace \#2} 
as shown in the rightmost graph at the uppermost row of Fig. 24 
({\tt 24-S1-V<2>}) 
are not perfectly circular but slightly distorted 
on account of the limits of 
{\sl Internetwork}'s generalization capability. 
Regarding the graphs in the bottom row of Fig. 22, 
full-orbed output trajectories generated in {\sl Subspace \#0} 
are converted by two-staged expansive {\sl Forward Subnets} 
to larger ones in the shape of a rounded square, 
which appear in {\sl Subspace \#2}. 
In contrast to this, 
for the graphs in the uppermost row of Fig. 24, 
output trajectories in the shape of a bigger rounded square 
are produced in {\sl Subspace \#0} 
so that full-orbed ones appear in {\sl Subspace \#2} 
through two-staged contractive {\sl Forward Subnets}. 
Viewing this result in a different light, 
we can understand them as follows: 
The output of {\sl Subspace \#1} ({\tt 24-S1-V<1>}) is 
what the output of {\sl Subspace \#2} ({\tt 24-S1-V<2>}) was converted to 
according to an expansive mapping relationship 
which is the ``inverse" of real contractive {\sl Forward Subnet \#2}'s. 
Then, the output of {\sl Subspace \#0} ({\tt 24-S1-V<0>}) 
emerges as what the one of {\sl Subspace \#1} ({\tt 24-S1-V<1>}) 
was transformed to based on an expansive mapping relationship 
which is the ``inverse" of actual contractive {\sl Forward Subnet \#1}'s.

Thus, we can summarize the relation between 
the bottom row graphs in Fig. 22 and the uppermost ones in Fig. 24 
in the following way: 
In the former graphs, 
full-orbed output trajectories in {\sl Subspace \#0} 
are straightforwardly transmitted in a static manner 
through two-staged expansive {\sl Forward Subnets} 
to {\sl Subspace \#2}, 
where output ones are yielded 
in the shape of a larger rounded square. 
As for the latter graphs, by contrast, 
full-orbed output trajectories in {\sl Subspace \#2} 
are sent in the opposite direction 
through two-staged ``imaginary" static layered networks 
with expansive mapping relationships 
(each of which is the ``inverse" of existing contractive 
{\sl Forward Subnet} {\sl \#1}'s or {\sl \#2}'s) 
to {\sl Subspace \#0}, 
where output ones are generated 
in the shape of a bigger rounded square.

\vspace{4.00mm}
\noindent
{\large (4) Comparison C}

\vspace{3.00mm}
\noindent
Let us compare the graphs in the second row of Fig. 22 
with those in the second row of Fig. 24. 
Also focus on the relation between the graphs 
in the third rows of Figs. 22 and 24. 
At a glance, they are respectively the same as each other. 
As inferred from the discussion in Subsection 4.3, 
in both the Unconstrained/Constrained Association Modes 
of Type \#2 architecture 
(i.e, regardless of the quality of input signals from the outside), 
the deformation components of signals 
going through {\sl Backward Subnet \#2} 
are canceled out by passing through {\sl Backward Subnet \#1}, 
under the condition that 
signals on two axes have no correlation 
in the two dimensional model 
and {\sl Forward Subnets} {\sl \#1} \& {\sl \#2} 
have mutually inverse mapping relationships. 
That is to say, under such conditions, 
output trajectories in Type \#0 architecture 
are identical with those in Type \#2 architecture 
by each {\sl Subspace}.

In the same way as the viewpoints 
stated in (2) and (3) of this subsection, 
we can comprehend the results 
in the second and third rows of Fig. 24 as follows: 
The full-orbed output trajectories in {\sl Subspace \#2} 
({\tt 24-S2-V<2>} and {\tt 24-S3-V<2>}) 
are sent in the opposite direction to {\sl Subspace \#1} 
through an ``imaginary" static layered network 
with the ``inverse" mapping relationship 
to real {\sl Forward Subnet \#2}'s. 
The output trajectories in {\sl Subspace \#1} 
({\tt 24-S2-V<1>} and {\tt 24-S3-V<1>}) 
are further converted to output ones in {\sl Subspace \#0} 
({\tt 24-S2-V<0>} and {\tt 24-S3-V<0>}) 
by an ``imaginary" static layered network 
with the ``inverse" mapping relationship 
to actual {\sl Forward Subnet \#1}'s.

\vspace{4.00mm}
\noindent
{\large (5) Comparison D}

\vspace{3.00mm}
\noindent
Let us look at Fig. 23 for Type \#1 architecture in the end. 
In {\sl Subspace \#1} to which 
input signals are applied from the outside, 
output trajectories become full-orbed 
as an outcome of well-tracking to those target signals. 
In contrast to the situation in {\sl Subspace \#1}, 
we can see that output trajectories 
in {\sl Subspaces} {\sl \#0} \& {\sl \#2} 
are established based only on the static mapping relationships 
of {\sl Forward Subnets} {\sl \#1} \& {\sl \#2}. 
As for each row, taking a look at the network responses 
in the direction from the inner {\sl Subspace} to the outer one, 
output trajectories in each {\sl Subspace} 
are straightforwardly determined 
according to the static mapping relationships 
of real {\sl Forward Subnets}. 
On the contrary, viewing the network responses 
in the direction from the outer {\sl Subspace} to the inner one, 
output trajectories in each {\sl Subspace} 
are sequentially converted in that direction 
based on ``imaginary" static layered networks 
with the ``inverse" mapping relationships 
to existing {\sl Forward Subnets}'.

In fact, it is for this reason that 
{\tt 23-S1-V<0>}, {\tt 23-S2-V<0>}, 
{\tt 23-S2-V<2>}, and {\tt 23-S4-V<2>} are all the same, 
even though {\sl Forward Subnet \#1} in the uppermost row 
and that in the second row are both contractive, 
and {\sl Forward Subnet \#2} in the second row 
and that in the bottom row are both expansive. 
Contrastively, 
{\sl Forward Subnet \#1} in the third row 
and that in the bottom row are both expansive, 
and {\sl Forward Subnet \#2} in the uppermost row 
and the one in the third row are both contractive. 
Under these circumstances, 
{\tt 23-S3-V<0>}, {\tt 23-S4-V<0>}, 
{\tt 23-S1-V<2>}, and {\tt 23-S3-V<2>} 
are all identical. 
These matches are thought to be for the same reason 
as the above ones.

\vspace{1.00mm} 
\section{Conclusions}

\vspace{2.00mm} 
\noindent 
In this paper, 
we propose a dynamical neural network model 
with a hierarchical and modular structure 
composed of two kinds of neurons with different time constants 
(one is dynamic and the other is regarded as static), 
and thoroughly examined how the model dynamically behaves. 
We summarize the obtained results item by item in the following.

\vspace{4.00mm} 
\noindent 
\( \diamondsuit \) Model Proposal

\begin{itemize}

  \item Supposing that output variables of a simple dynamical neural network 
        with direct feedback connections were converted 
        in turn to other output ones 
        through ``mapping functions" or ``multi-layered neural networks 
        composed of neurons with a tiny time constant 
        that could be deemed to be static," 
        we designed a particular energy function 
        to which the energy function defined in our basic model 
        was expanded as a generalized version 
        employing all of those output variables. 

  \item The architecture of the proposed network can be derived 
        through the process of minimizing the newly-designed energy function. 
        The finally obtained network is of ladder-shaped structure 
        with the arbitrary number of {\sl Subspaces}, 
        each of which has a pair of ``a feedback path" and ``an input port." 

  \item Adjacent {\sl Subspaces} in the model are related with 
        a static neural network named ``{\sl Internetwork}," 
        which further consists of a complementary pair 
        of a ``{\sl Forward Subnet}" and a ``{\sl Backward Subnet}." 

  \item In the whole network, 
        there exist some kinds of synaptic connections; 
        we suppose that synaptic connections 
        only inside {\sl Internetworks} have plasticity 
        and the other ones are all fixed. 
  \item Both ``activity dynamics based on the dynamical neurons" 
        and ``learning dynamics for the modifiable connections 
        in {\sl Internetworks}" 
        progress at the same time, 
        and they determine the total time-course behavior of the model. 

  \item Signals flowing through a {\sl Backward Subnet} essentially affect 
        learning dynamics in an {\sl Internetwork} composed of a pair 
        of a {\sl Forward Subnet} and a {\sl Backward Subnet}, 
        and they also influence activity dynamics of the whole network. 

  \item We consider this dynamical neural network as a model, 
        and assume two operational modes, 
        the Learning Mode and the Association Mode, 
        depending on the presence or absence 
        of a pair of a feedback path and an input port 
        in a {\sl Subspace}. 

  \item The Association Mode is further divided into 
        the ``Unconstrained" Association Mode and 
        the ``Constrained" Association Mode. 

  \item In the Learning Mode, we presume a unique architecture 
        in which every {\sl Subspace} has 
        a pair of a feedback path and an input port 
        regardless of the model size (i.e., the number of {\sl Subspaces}), 
        only considering a situation such that 
        all of {\sl Internetworks} are trained in a lump. 
        Whereas in the (Unconstrained/Constrained) Association Mode, 
        we assume \( P + 1 \) types of architectures, 
        each of which has a pair of a feedback path and an input port 
        only in one {\sl Subspace}, 
        defining \( P \) as the number of {\sl Internetworks} in the model. 

\end{itemize}

\vspace{3.00mm} 
\noindent 
\( \diamondsuit \) Learning Mode

\begin{itemize}

  \item On the basis of the results for our basic model 
        with only one {\sl Internetwork}, 
        we examined the Learning Mode of the generalized network model 
        in which there was no limitation 
        for the number of {\sl Internetworks} and {\sl Subspaces}. 

  \item When adequately applying sinusoidal periodic signals 
        to all input ports from the outside 
        in the Learning Mode of a one-dimensional model, 
        we confirmed that the {\sl Internetworks} 
        were able to automatically acquire 
        linear mapping relationships 
        between adjacent {\sl Subspaces}. 
        The periodic wave in this case is represented by 
        short-term average density of nerve impulses, 
        so it indicates repetitive bursts at the nerve impulse level. 

  \item We can expand the model to a two (or more) dimensional one. 
        In the two-dimensional model, 
        if the frequencies of two kinds of regular sinusoidal signals 
        (that are equally applied to every input port 
        with the horizontal and vertical axes) 
        properly differ from each other, 
        two-dimensional linear mapping relationships are formed 
        in all {\sl Internetworks}. 
        Then, preferable is the situation in which 
        the frequencies of those two periodic signals 
        have a relation such that 
        they are either slightly different (Fig. 9 and Fig. 10) 
        or very different (Fig. 11); 
        this mechanism is similar 
        to that of a well-known Lissajous curve. 
        Note that the frequency of input signals 
        for the horizontal or vertical axis 
        must be the same throughout all {\sl Subspaces}. 

  \item Regardless of a one-dimensional model or a two-dimensional model, 
        non-linear mapping relationships can successfully be obtained 
        in {\sl Internetworks} between {\sl Subspaces} 
        depending on how the periodic signals applied 
        from the outside to the input ports are deformed. 

  \item We also confirmed that 
        ``the variance in {\sl Internetwork}'s training speed" 
        and ``the synergistic effect of activity dynamics 
        and leaning dynamics" came out in various ways 
        according to circumstances of input signals 
        applied from the outside. 

  \item The Learning Mode works well even in the cases with \( P \geq 3 \), 
        although we only treated, in the present paper, 
        the case with \( P = 2 \) 
        that is the simplest one of the generalized network model 
        with \( P \geq 2 \). 

\end{itemize}

\vspace{1.50mm} 
\noindent 
\( \diamondsuit \) Unconstrained Association Mode

\begin{itemize}

  \item Let us look back the Unconstrained Association Mode 
        of our basic model with only one {\sl Internetwork}. 
        In terms of the one-dimensional model, 
        the convergence speed of an output in each space ({\sl Subspace}) 
        greatly varied depending on 
        how a mapping relationship in the {\sl Internetwork} warped. 
        As for the two-dimensional model, 
        an output trajectory in each space ({\sl Subspace}) 
        sometimes curved largely 
        because of the difference between the convergence speeds 
        in the horizontal and vertical axes. 

  \item In this paper, regarding the generalized network model 
        that is able to have arbitrary number of {\sl Internetworks}, 
        we precisely examined dynamics of Types \#0, \#1, and \#2 architectures 
        in the Unconstrained Association Mode. 
        Specifically, by limiting the model 
        to the two-dimensional one with \( P = 2 \) 
        and assuming no correlation between signal components 
        on the horizontal and vertical axes in {\sl Internetworks}, 
        we prepared five sets of the networks, 
        each of which owned two {\sl Internetworks} 
        with a different combination of mapping relationships 
        that were obtained beforehand in the Learning Mode, 
        and traced output trajectories in the three {\sl Subspaces} 
        when giving a fixed point 
        (corresponding to the center of a square on a coordinate plane) 
        to a unique input port in each network. 

  \item In Type \#0 architecture, a direct feedback circuit 
        composed of dynamical neurons in {\sl Subspace \#0} 
        determines fundamental behavior of the whole network. 
        Specifically when giving the fixed values (for the center) 
        to a unique input port in {\sl Subspace \#0}, 
        output of the dynamical neurons goes straightforwardly 
        to the fixed point. 
        The straight output trajectory of {\sl Subspace \#0} 
        is statically mapped in turn 
        by {\sl Forward Subnets} {\sl \#1} \& {\sl \#2} 
        to become output trajectories 
        of {\sl Subspaces} {\sl \#1} \& {\sl \#2}
        (not necessarily straight lines depending on 
        the mapping relationships of {\sl Forward Subnets}). 
        In this way, if we consider only the range within 
        {\sl Subspaces} {\sl \#0} \& {\sl \#1}, 
        the associative dynamical behaviors there 
        in Type \#0 architecture of the generalized model 
        are completely the same as those in Type \#0 architecture 
        of our basic model. 

  \item In Type \#1 architecture, 
        the difference between 
        ``output signals from {\sl Forward Subnet \#1}" 
        and 
        ``the fixed input signals (for the center) 
        applied in {\sl Subspace \#1}" 
        is large at the initial state, 
        so the signals corresponding to such difference 
        propagate through {\sl Backward Subnet \#1} 
        to {\sl Subspace \#0}. 
        Dynamical neurons there can be affected 
        by the output of {\sl Backward Subnet \#1}, 
        and their output signals are converted 
        by {\sl Forward Subnet \#1} 
        to become output of {\sl Subspace \#1}. 
        These processes are repeated;  
        thus, dynamics of Type \#1 architecture 
        within {\sl Subspaces} {\sl \#0} \& {\sl \#1} 
        come to be determined 
        based on the detoured feedback loop via {\sl Internetwork \#1}. 
        An output trajectory in {\sl Subspace \#1} of Type \#1 architecture 
        is capable of curving widely, 
        even though the direction and the magnitude of such bending 
        differ depending on a mapping relationship of {\sl Forward Subnet \#1}. 
        Then, note that an output trajectory in {\sl Subspace \#0} 
        may also bend instead of going straight. 
        As can easily be inferred from these characteristics, 
        the dynamics within {\sl Subspaces} {\sl \#0} \& {\sl \#1} 
        in Type \#1 architecture of the generalized model 
        are identical with those in the same architecture of our basic model. 
        An output trajectory in {\sl Subspace \#2} is produced 
        based only on a static mapping relationship of {\sl Forward Subnet \#2}; 
        this aspect is common to the one in Type \#0 architecture 
        as mentioned above. 

  \item In Type \#2 architecture, 
        the difference between 
        ``output signals from {\sl Forward Subnet \#2}" 
        and 
        ``the fixed input signals (for the center) 
        applied in {\sl Subspace \#2}" 
        is large at the initial state. 
        A certain level of signals based on such difference 
        flows through {\sl Backward Subnet \#2}, 
        whose output further goes through {\sl Backward Subnet \#1} 
        to {\sl Subspace \#0}. 
        Output of dynamical neurons there 
        can be influenced by signals passing through 
        {\sl Backward Subnets} {\sl \#2} \& {\sl \#1} in this order, 
        and then it becomes output of {\sl Subspace \#2} 
        via {\sl Forward Subnets} {\sl \#1} \& {\sl \#2}. 
        These processes are repeated 
        based on the widely detoured feedback loop 
        through the two {\sl Internetworks}. 
        Accordingly, if mapping relationships of 
        {\sl Forward Subnets} {\sl \#1} \& {\sl \#2} 
        are either both expansive or both contractive, 
        the curving of an output trajectory in {\sl Subspace \#2} 
        can be bigger in comparison to 
        that of any output trajectory generated in our basic model, 
        under the condition that its unique {\sl Internetwork} has one 
        of those two-staged mapping relationships in the generalized model. 

  \item When serially connected two {\sl Forward Subnets} 
        have mutually inverse mapping relationships 
        (e.g., from an expansive one to a contractive one, 
        or from a contractive one to an expansive one) 
        in Type \#2 architecture, 
        deformed components in signals 
        propagating through the corresponding {\sl Backward Subnets} 
        are interestingly canceled out. 
        Due to this, when setting the fixed values (for the center) 
        to an input port in {\sl Subspace \#2}, 
        an output trajectory in {\sl Subspace \#2} 
        draws a straight line, and at the same time, 
        it becomes identical with that in {\sl Subspace \#0}. 

  \item On the basis of the experimental results 
        for Types \#0, \#1, and \#2 architectures 
        in the Unconstrained Association Mode, 
        we can reasonably estimate associative dynamical behavior 
        of the generalized model with three or more {\sl Internetworks}. 

  \item Putting \( p = 1, 2, ..., P - 1 \), 
        let us suppose that 
        {\sl Internetworks} {\sl \#\( p \)} \& {\sl \#\( (p + 1) \)} 
        located next to each other 
        have mutually inverse mapping relationships
        and also that {\sl Subspace \#\( p \)} does not have 
        a pair of a feedback path and an input port. 
        Deformed components 
        in signals flowing through {\sl Backward Subnet \#\( (p + 1) \)} 
        are completely canceled out by the process 
        such that they further go through {\sl Backward Subnet \#\( p \)}, 
        regardless of whether 
        a pair of a feedback path and an input port exists or not 
        in {\sl Subspace \#\( (p - 1) \)} or {\sl Subspace \#\( (p + 1) \)}. 
        Eventually, output trajectories in {\sl Subspace \#\( (p - 1) \)} 
        and those in {\sl Subspace \#\( (p + 1) \)} 
        become identical with each other. 

\end{itemize}

\vspace{1.50mm} 
\noindent 
\( \diamondsuit \) Constrained Association Mode, 
Inverse Mapping, and Certainty/Uncertainty Relation

\begin{itemize}

  \item In our basic model with only one {\sl Internetwork}, 
        naming a particular Association Mode 
        in which a ``trajectory" instead of a fixed goal point 
        was applied to an input port 
        as the ``Constrained" Association Mode, 
        we precisely investigated its dynamical behavior 
        employing a straight target trajectory 
        from an initial point to a fixed goal point, 
        and found out various interesting phenomena. 
        In Type \#1 architecture of the Constrained Association Mode 
        with a pair of a feedback path and an input port 
        only in the external space, 
        if the speed of the target trajectory was sufficiently low 
        compared with the time constant of a dynamical neuron 
        located in the internal space, 
        signals propagating through the {\sl Backward Subnet} 
        became quite small due to the action of dynamical neurons, 
        and curvings of output trajectories 
        in both the internal and external spaces 
        eventually came to be determined 
        based solely on the static mapping relationship 
        realized by the {\sl Internetwork}. 
        That is to say, 
        if the mapping relationship of the {\sl Forward Subnet} 
        was regarded as \( M \), 
        generated in the internal space was an output trajectory 
        that was transformed by a mapper with \( M^{-1} \) 
        from a target trajectory applied to the input port in the external space, 
        although the mapper with \( M^{-1} \) did not exist 
        explicitly in the whole network. 

  \item In this paper, regarding the generalized network model 
        with arbitrary number of {\sl Internetworks}, 
        we also examined network dynamics 
        of the Constrained Association Mode in detail. 
        Specifically, in the two-dimensional model with \( P = 2 \), 
        preparing some combinations of mapping relationships 
        that were obtained for the two {\sl Internetworks}
        beforehand in the Learning Mode, 
        we analyzed dynamical behaviors 
        of Type \#0, Type \#1, and Type \#2 architectures 
        with those {\sl Internetworks}. 
        In each of these simulation studies, 
        circular periodic target signals 
        with three different frequencies 
        (Fast (High Speed) Input, Medium Speed Input, and Slow (Low Speed) Input) 
        and five kinds of amplitudes 
        were given from the outside to a unique input port. 

  \item Considering Type \#1 architecture 
        with expansive {\sl Forward Subnet \#1} for instance, 
        the level of signals 
        flowing through the corresponding {\sl Backward Subnet \#1} 
        tends to be larger 
        in comparison to the case with contractive {\sl Forward Subnet \#1}. 
        Therefore, compared with the latter case, 
        greater in the former case is an effect that 
        makes the difference between feedback signals and input ones 
        in {\sl Subspace \#1} 
        (to which the input signals are provided from the outside) 
        lessen in the inner {\sl Subspace}; 
        eventually, the tracking characteristics 
        for output of {\sl Subspace \#1} in the former case 
        become better than those in the latter case. 
        As is inferred from this example for Type \#1 architecture, 
        in general, output responses in a {\sl Subspace} 
        are able to vary greatly 
        depending on how the mapping relationship(s) 
        of the relevant {\sl Internetwork(s)} warps(warp). 
        In any cases, however, if applying input signals 
        with a very high frequency in particular, 
        circular output trajectories in every {\sl Subspace} 
        do not become large sufficiently 
        because of limited responsiveness of a dynamical neuron 
        in {\sl Subspace \#0}. 

  \item When the frequency of signals for a circular target trajectory 
        applied from the outside is really low 
        compared with the time constant of a dynamical neuron, 
        an output trajectory in the {\sl Subspace} 
        to which those input signals are applied 
        almost perfectly follows the input ones 
        in every architecture of the Constrained Association Mode; 
        then, signals flowing through {\sl Backward Subnets} 
        located in the very {\sl Subspace} and on its inner side 
        (if they exist) become quite small, 
        and there is no signal going through {\sl Backward Subnets} 
        located on the outer side of the very {\sl Subspace} 
        (even if they exist) in principle. 
        In the case with Slow Input, as a result, 
        an output trajectory in every {\sl Subspace} 
        is determined based only on 
        ``static" mapping relationships of {\sl Forward Subnets} 
        that were obtained in advance in the Learning Mode. 

  \item For instance, let us suppose a situation in which, 
        in Type \( \#P \) architecture, 
        very slowly varying signals for a circular trajectory 
        are applied to an input port in {\sl Subspace \#\( P \)}. 
        Considering the mapping relationships of 
        {\sl Forward Subnets} 
        {\sl \#1}, {\sl \#2}, ..., {\sl \#\( (P - 1) \)}, {\sl \#\( P \)} 
        as \( M^{<1>} \), \( M^{<2>} \), ..., \( M^{<P - 1>} \), \( M^{<P>} \) 
        respectively, 
        generated in {\sl Subspace \#0} is an output trajectory 
        that is converted from the one in {\sl Subspace \#\( P \)} 
        in turn by the ``imaginary" mappers 
        \( \{M^{<P>}\}^{-1} \), \( \{M^{<P - 1>}\}^{-1} \), ..., 
        \( \{M^{<2>}\}^{-1} \), \( \{M^{<1>}\}^{-1} \) 
        with the ``inverse" mapping relationships 
        to the really existing {\sl Forward Subnets}'. 

  \item In other words, this phenomenon suggests that 
        it is needed to strongly constrain the network dynamics from the outside 
        in order to ensure that output trajectories appear in inner {\sl Subspaces} 
        based more accurately on the ``inverse" mapping relationships 
        to actually existing {\sl Forward Subnets}'. 
        This also implies that 
        a certainty/uncertainty relation exists 
        between ``the speed of a target trajectory 
        applied from the outside to a unique input port" 
        and ``an output trajectory generated 
        by the inverse mapping relationships 
        to really existing {\sl Forward Subnets}'." 
        This is a crucial point that should not be overlooked 
        when considering dynamics of the generalized network model 
        that is proposed here. 

\end{itemize}

\vspace{2.0mm} 
Since the proposed model can include multiple {\sl Internetworks}, 
a method how to train those {\sl Internetworks} in the Learning Mode 
is not limited to the lumped training that was employed in Section 3. 
Essentially, we can consider a step-by-step training method
in which each {\sl Internetwork} is separately trained. 
If signals propagating through {\sl Backward Subnets} 
are not adequate in such a case, however, 
the training might end halfway and/or at an undesired state. 
It is of great interest to study other training methods 
beyond the lumped training by focusing particularly on signals 
passing through {\sl Backward Subnets}. 
In the Association Mode, 
we treated the selected architectures, each of which has 
a pair of a feedback path and an input port only in one {\sl Subspace}. 
It is interesting to examine associative dynamics 
of other architectures in which plural {\sl Subspaces} 
have a pair of a feedback path and an input port. 
In both the Learning and Association Modes, 
there is a need to widely discuss the relation between 
``the location(s) of a pair (pairs) of a feedback path and an input port" 
and ``the waveforms of input signals given from the outside." 
In the present paper, the Learning Mode and 
the two kinds of Association Modes (Unconstrained / Constrained) 
are artificially classified 
on the basis of their structural differences, 
and the ON/OFF operation of the learning dynamics given by Eq. (43) 
is consistently selected. 
``The process of synaptic modification in {\sl Internetworks}" 
and ``quality \& quantity of signals propagating back and forth 
through the {\sl Internetworks}" 
are closely connected with each other. 
It might be possible, by analyzing those circumstances in detail, 
to construct a model in which such modes can be switched automatically 
according to the waveforms of input signals given from the outside.

The dynamical neural network model 
with highly hierarchical and modular structure proposed here 
offers a framework in which 
state variables inside the network relax severally in warped spaces, 
each of which is deformed as convenient 
to a (or some) state variable(s); 
to put it differently, in the Association Mode of the model, 
various constraints described as terms in an energy function 
can be optimized one by one in warped spaces, 
each of which is designed 
as favorable for a (or plural) constraint(s). 
This framework is thought to have the validity as a brain model 
into which integrated are some of fundamental principles 
that have been proposed so far 
as models of brain functions and neural networks. 
At the same time, it has great potential 
to various practical applications, 
such as advanced associative memory tasks, optimization problems, 
and robotic control issues in which various constraints must be minimized 
respectively in the corresponding non-linear subspace. 
This paper deals with specific examples 
only for the one-dimensional and two-dimensional models. 
The methodology for expanding the model's dimension from one to two 
can be straightforwardly applied 
to the three-dimensional model and beyond. 
In terms of issues about control and robotics, 
a three-dimensional model will be important. 
For associative memory tasks and optimization problems, 
an ultra-high dimensional model should be considered. 
It is of extreme importance 
to investigate the model's characteristics 
with high dimensional models 
and also with elaborated simulation studies 
particularly aiming to improve the performance of such applications.

In the present paper, we mainly discussed 
depth-wise (hierarchical or vertical) modularity 
as for internetworking between {\sl Subspaces}. 
Regarding the proposed model 
with the large number of hierarchical modules, 
the whole of the serially connected {\sl Forward Subnets} 
constitutes a static neural network 
with the superlarge number of layers 
that is collecting a lot of attention in recent years
\cite{Hinton2012}. 
In contrast to this, 
with regard to spread-wise (horizontal) modularity 
in neural architecture, 
we have already started to discuss it, 
but not sufficiently the relation to depth-wise modularity
\cite{Tsutsumi1990}\cite{Ozawa1998}. 
From the viewpoint of considering 
how the modularity should be in neural architecture, 
the depth-wise (hierarchical or vertical) modularity 
and 
the spread-wise (horizontal) modularity 
seem to be both essential. 
Nevertheless, it is of great significance to compare those two 
by purposely separating them, 
in terms of 
the function of input applied from the outside, 
the role of output emitted in each {\sl Subspace}, 
and furthermore the fundamental significance of modularization 
in the first place.

Sinusoidal or quasi-sinusoidal input signals 
(corresponding to periodic bursts or semi-bursts 
in the nerve impulse level), 
which had been employed in our basic model presented previously, 
were given from the outside 
also in the Learning and Association Modes of this paper. 
From the perspective of expanding this situations, 
it is vital to discuss whether the proposed model 
(including its developed or partly-restricted version) 
has a capability to generate such oscillatory signals by itself or not; 
this extension must be also meaningful 
in order to improve our proposal's consistency 
as a neural network model and a brain one. 
Under these circumstances, 
it is crucial to analyze 
not only convergent dynamics toward equilibrium
but also divergent dynamics 
by altering a sign of each term in the energy function 
\cite{Tsutsumi1999a}\cite{Tsutsumi1999b}\cite{Tsutsumi2001}. 
In addition, it should be mandatory 
to study dynamics of a further enhanced model 
with extra interactions between distant {\sl Subspaces}.

\bibliographystyle{acm}
\bibliography{tsutsumi_niebur_02_references}

\vspace{2mm}
\noindent
{\bf Acknowledgements:} 
The authors are grateful to the deceased 
Professor Dr. Dagmar Niebur (Drexel University) 
for her helpful support and warm encouragement. 
This work was supported in part 
by Ryukoku University Fund for Overseas Research, 
Office of Naval Research Grant N00014-22-1-2699, 
and National Science Foundation grant 1835202.

\end{document}